\RequirePackage{fix-cm}
\documentclass[smallextended]{svjour3}       
\smartqed  
\usepackage{graphicx}
\usepackage[round]{natbib}
\usepackage{amssymb,amsmath}
\newcommand{\mnras}{MNRAS}
\newcommand{\nat}{Nature}
\newcommand{\aap}{A\&A}
\newcommand{\apj}{ApJ}
\newcommand{\aj}{AJ}
\newcommand{\apjl}{ApJL}
\newcommand{\apjs}{ApJS}
\newcommand{\jcap}{JCAP}
\newcommand{\physrep}{Phys.Rep.}
\newcommand{\prd}{Phys.Rev.D}
\newcommand{\araa}{ARA\&A}
\newcommand{\Ddt}{D_{\Delta{\rm t}}}
\newcommand{\Dd}{D_{\rm d}}
\newcommand{\Ds}{D_{\rm s}}
\newcommand{\Dds}{D_{\rm ds}}
\newcommand{\zd}{z_{\rm d}}
\newcommand{\zs}{z_{\rm s}}
\newcommand{\cospars}{\boldsymbol{\Omega}}
\newcommand{\Ok}{\Omega_{\rm k}}
\newcommand{\ODE}{\Omega_{\rm DE}}
\newcommand{\wDE}{w_0}
\newcommand{\x}{\boldsymbol{\theta}}
\newcommand{\y}{\boldsymbol{\beta}}
\newcommand{\grad}{\boldsymbol{\nabla}}
\newcommand{\deflectionangle}{\boldsymbol{\alpha}}
\begin{document}

\title{Time Delay Cosmography
}

\titlerunning{Time Delay Cosmography}        

\author{Tommaso Treu         \and
        Philip J. Marshall 
}


\institute{Tommaso~Treu \at
Department of Physics and Astronomy, \\
University of California,\\
Los Angeles, CA 90095, USA\\
\email{tt@astro.ucla.edu}           
           \and
Philip~J.~Marshall \at
Kavli Institute for Particle Astrophysics and Cosmology, \\
P.O. Box 20450, MS29, \\
Stanford, CA 94309, USA \\
}

\date{Received: date / Accepted: date}

\maketitle


\begin{abstract}

Gravitational time delays, observed in strong lens systems where the
variable background source is multiply-imaged by a massive galaxy in
the foreground, provide direct measurements of cosmological distance
that are very complementary to other cosmographic probes. The success
of the technique depends on the availability and size of a suitable
sample of lensed quasars or supernovae, precise measurements of the
time delays, accurate modeling of the gravitational potential of the
main deflector, and our ability to characterize the distribution of
mass along the line of sight to the source. We review the progress
made during the last 15 years, during which the first competitive
cosmological inferences with time delays were made, and look ahead to
the potential of significantly larger lens samples in the near future.
\end{abstract}


\section{Introduction}
\label{sec:intro}

The measurement of cosmic distances is central to our understanding of
cosmography, i.e. the description of the geometry and kinematics of
the universe. The discovery of the period luminosity relation for
Cepheids led to the realization that the universe is much bigger than
the Milky Way and that it is currently expanding. Relative distance
measurements based on supernova Ia light curves were the turning point
in the discovery of the acceleration of the universe
\citep{Riess:1998p21184,Per++99}.

In the two decades since the discovery of the acceleration of the
universe, distance measurements have improved steadily. For example,
the Hubble constant has now been measured to 2.4\% precision
\citep{Rie++16} while the distance to the last scattering surface of
the cosmic microwave backgrond is now known to approximately 0.5\%
precision \citep[depending on the assumed cosmological
model]{WMAP9,Pla15}. This precision is more than sufficient for all
purposes related to our understanding of phenomena occurring within
the universe, like galaxy evolution.

In spite of all this progress, the most fundamental question still
remains unanswered. What is causing the acceleration? Is this {\it
dark energy} something akin to Einstein's cosmological constant or is
it a dynamical component? Answering this question from an empirical
standpoint will require further improvements in the precision of
distance measurements \citep{Suy++12,Wei++13,Kim++15,Rie++16}.  In
practice, measuring the dark energy equation of state requires an
accurate model of the scale parameter of the universe as a function of
time, particularly when dark energy is dynamically most relevant,
i.e. below $z\sim1$.

Cosmic microwave background anisotropies primarily provide a
measurement of the angular distance to the last scattering surface,
obtained by comparing the angular scale of the acoustic peaks with the
sounds horizon at recombination. Therefore, the constraints set by CMB
anisotropy data on dark energy parametere are highly degenerate in a
generic cosmological model
\citep[e.g.,][]{Pla15}. Breaking the degeneracy requires strong
assumptions about the universe (e.g., flatness or dark energy being
the cosmological constant), or lower redshift distance
measurements. Many dedicated experiments are currently under way or
being planned with this goal in mind.

Precision, however, is not sufficient by itself. In addition to
controlling the known statistical uncertainties ({\it precision}),
modern day experiments need to control systematic errors ({\it
accuracy}) in order to fullfill their potential, including the
infamous unknown unknowns. The most direct way to demonstrate accuracy
is to compare independent measurements that have comparable
precision. An interesting, currently topical, and relevant case is
that of the 3$-\sigma$ tension between the local distance ladder
determination of the Hubble Constant H$_0$ by \citet{Rie++16} and that
inferred by the Planck satellite assuming a flat $\Lambda$CDM model
\citep{Pla15}. The tension could be due to an unknown source of
systematic errors in either or both of the two measurements, or it
could be indicative of new physics, for example an effective number of
relativistic species greater than three. Independent measurements with
comparable precision are the best way to make progress.  While
independent measurements of the same phenomenon, or reanalysis of the
same data \citep{Fre++12,Rig++15,Efs14,SFH15}, are certainly useful
and necessary, completely independent datasets based on different
physical phenomena provide qualitatively new information.

Ideally, the comparison between independent measurements should be
carried out blindly, so as to minimize experimenter bias. Two mutually
blind measurements agreeing that the equation of state parameter $w$
is not $-1$ would be a very convincing demonstration that the dark
energy is not the cosmological constant. Conversely, the significant
disagreement of two blind and independent measurements, could be the
first sign of new physics.

In this review we focus on strong lensing gravitational time delays as
a tool for cosmography. As we shall see, this probe provides a direct
and elegant way to measure absolute distances out to cosmological
redshift. When the line of sight to a distant source of light is
suitably well aligned with an intervening massive system, multiple
images appear to the observer. The arrival time of the images depends
on the interplay of the geometric and gravitational delays specific to
the configuration. If the emission from the source is variable in
time, the difference in arrival time is measurable, and can be
interpreted via a so-called ``time delay distance'' $\Ddt$.  In the
simplest case, this distance is just a multiplicative combination of
the three angular diameter distances between the observer, deflector
and source. $\Ddt$ is inversely proportional to $H_0$, and more weakly
dependent on other cosmological parameters. As several authors have
pointed out \citep{Hu05,Lin11,Suy++12,Wei++13}, achieving sub-percent
precision and accuracy on the measurement of the Hubble constant will
be a powerful addition to Stage III and IV dark energy
experiments. The independence of time delays from other traditional
probes of cosmology, makes them very valuable for precise and accurate
cosmology. For example, time delays yield an {\it absolute}
measurement of distance without relying on Cepheids or any other local
rung of the distance ladder, and because the relevant quantities are
angular diameter distances rather then luminosity distances, the
approach is insensitive to dust or other photometric errors.

This review is organized as follows. In Section~\ref{sec:intro} we
summarize the history of time delay cosmography up until the turn of
the millennium, in order to give a sense of the early challenges and
how they were overcome. In Section~\ref{sec:theory}, we review the
theoretical foundations of the method, in terms of the gravitational
optics version of Fermat's principle. In Section~\ref{sec:measurement}
we describe in some detail the elements of a modern time delay
distance measurement, emphasizing recent advances and remaining
challenges. In Section~\ref{sec:cosmo} we elucidate the connection
between time delay distance measurements and cosmological parameters,
discussing complementarity with other cosmological
probes. Section~\ref{sec:outlook} critically examines the future of
the method, discussing prospects for increasing the precision, testing
for accuracy, and synergy with other future probes of dark energy. A
brief summary is given in Section~\ref{sec:summary}.

Owing to space limitations, we could only present a selection of all
the beautiful work that has been published on this topic in the past
decades. We refer the readers to recent
\citep{Bar10,Ell10,Tre10,TMC12,Jackson:2013p30763,Jac15,T+E15}
and not-so-recent \citep{B+N92,CSS02,K+S04,Fal05,SKW06}
excellent reviews and textbooks \citep{SEF92} for additional
information and historical context.


\section{A brief history of time delay cosmography}
\label{sec:history}

\citet{Ref64} first suggested that strong lens time delays could be
used to measure absolute, cosmological distances, and
therefore the Hubble Constant to leading order. Unfortunately, no
strong lensing systems were known at that time, and therefore his
intuition remained purely theoretical for over a decade.

The prospects of using time delays for cosmography suddenly brightened
in the late seventies, with the discovery of the first strongly lensed
quasar \citep{WCW79}. Even though they were not the strongly lensed
supernovae that Refsdal had had in mind, quasar fluxes are sufficiently
variable \citep{Van82} that people were able to start to put Refsdal's
idea in practice \citep{Van89}.
The first multiply imaged supernova was discovered in 2014,
fifty years afer Refsdal's initial suggestion \citep{Kel++15}, lensed
by a foreground cluster of galaxies. The time delays are being
measured at the time of writing
\citep{Rod++16,Kel++16}; however, it is unclear at the moment whether the
cluster potential can be constrained with sufficient accuracy to yield
interesting cosmological information \citep{Tre++16}. In general, we
expect the more straightforwardly-modeled, more numerous galaxy-scale
time delay lenses to be the most useful systems for cosmography, with
supernovae competing for attention with quasars \citep{O+M10}.  

In this review, we will restrict our case to the hitherto much more
common and better understood case of variable active galactic nuclei
(AGN) being lensed by foreground elliptical galaxies.

Discovery and monitoring of lensed quasars continued in the eighties and
nineties, powered by heroic efforts. By the end of the millennium the
number of known strongly lensed systems was in double digits
\citep{CSS02}, and the first truly robust time delays were measured
\citep{Kun++97,Sch++97}.
The industrial detection of multiply imaged AGN finally took off at the
beginning of the current millennium, with the improvement of panoramic
search technology in dedicated or existing surveys
\citep{Bro++03,Oguri:2006p5865,Agn++15}.

\begin{figure*}
\includegraphics[width=0.48\textwidth]{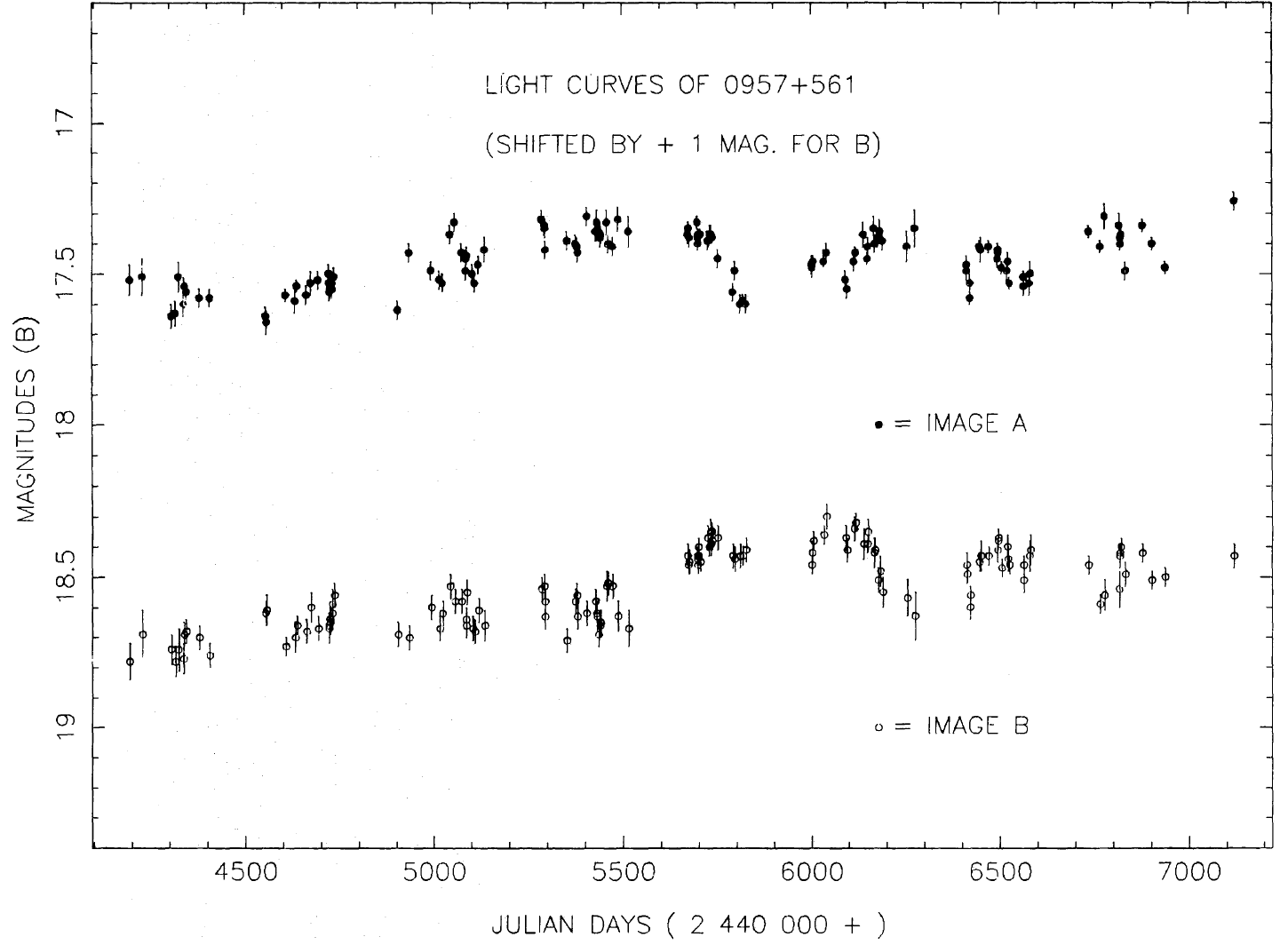}
\includegraphics[width=0.48\textwidth]{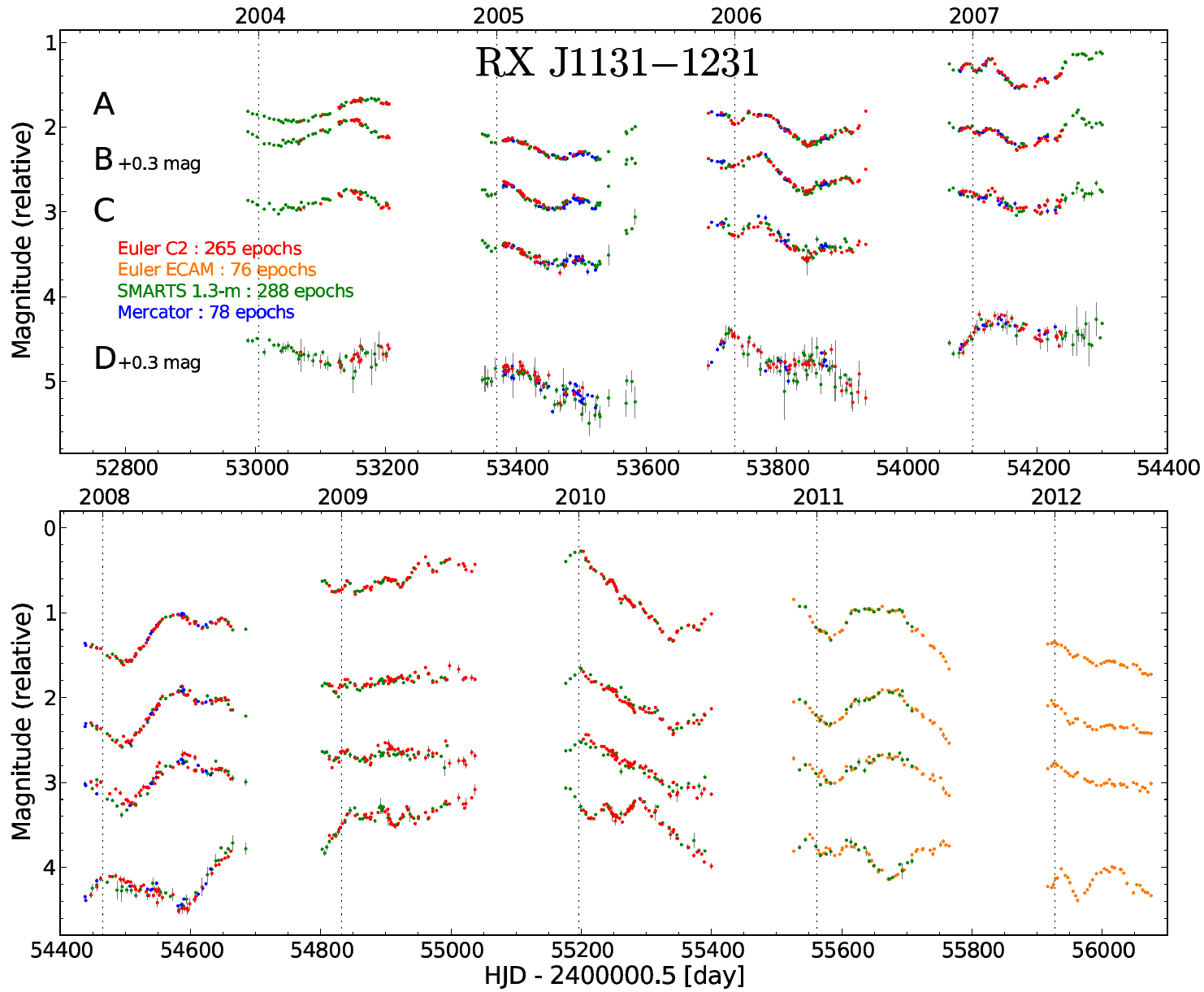}
\caption{Comparison between one of the early light curves \citep[left
panel, from][]{Van89}, and a modern light curve from COSMOGRAIL
\citep[right panel, from][]{Tew++13}. Note the improved photometric
precision, cadence, and duration of the light curves, allowing for
unambiguous determination of the time delay to within 1-2\% precision.}
\label{fig:oldvsmoderndt}
\end{figure*}

The initial period of time delay cosmography was marred by controversies over
systematic errors.  The measurement of time delays was particularly
controversial during the nineties as the quality of the early data
allowed for multiple estimated values \citep{PRH92}, owing to the combined
effects of gaps in the data, and microlensing noise in the optical
light curves. This problem was solved definitively at the turn of the
millennium, with the beginning of modern monitoring campaigns,
characterized by high cadence, high precision, and long duration, both
at optical and radio wavelengths
\citep{Fas++99,Fas++02,Bur++02,Hjo++02,Jak++05,Eig++05}, as illustrated in
Figure~\ref{fig:oldvsmoderndt}. We discuss
modern monitoring campaigns in more detail in Section~\ref{ssec:timedelay}.

\begin{figure*}
\includegraphics[height=3.5cm]{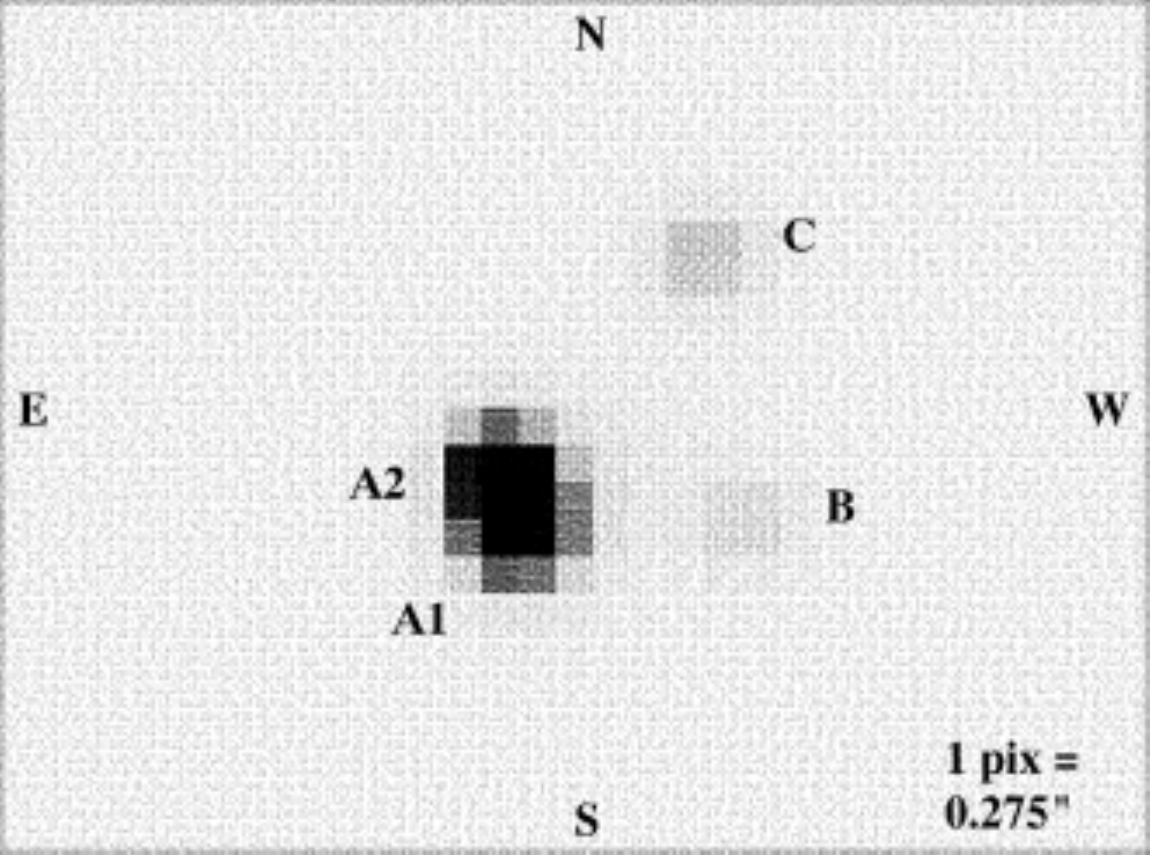}
\includegraphics[height=3.5cm]{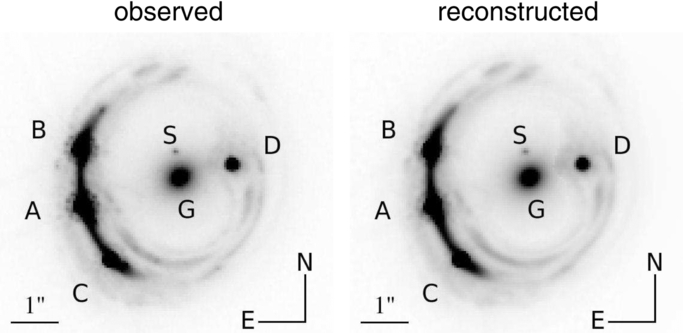}
\caption{Comparison between imaging data available in the nineties
\citep[left panel, from][]{Sch++97} and in the most recent studies
\citep[middle and right panels, from][]{Suy++14}). With modern data the
structure of the quasar host galaxy can be modeled in great detail,
providing thousands of constraints on the deflection angle, and thus on
the derivatives of the gravitational potential.}
\label{fig:oldvsmodernimage}
\end{figure*}

Finally, when robust time delays started to become available, the
focus of the controversy shifted to the modeling of the gravitational
potential of the lens. Typically, in the mid nineties, the only
constraints available to modelers were the quasar image positions,
time delays, and to lesser extent flux ratios (limited by
microlensing, variability and differential extinction). Thus, the best
one could do was to assume some simple form for the lens mass
distribution, such as a singular isothermal sphere \citep{K+F99},
and to neglect the
effects of structure along the line of sight. As a result of these necessary
but oversimplistic assumptions, the apparent random errors grossly underestimated
the total uncertainty, leading to measurements that were apparently
significantly inconsistent
between groups, or with those from other techniques
\citep{K+S04}. Since then, two methods have been pursued in order to
obtain realistic estimates of the uncertainties. One consists of using
large samples of systems with relatively weak priors
\citep{Ogu07b}. The other method consists of obtaining high quality data for
each lens system, including improved astrometry \citep{Cou++97}, detailed imaging of
the quasar host galaxy
\citep{Keeton:2000p241,KKM01,Koo++03,WBB04,Suy++06}, or non-lensing data like the deflector
stellar velocity dispersion \citep{T+K02b} and the properties of
galaxies along the line of sight \citep{K+Z04,Suy++10}. We discuss
these approaches in Section~\ref{ssec:lensmodel}. The astounding
improvement in data quality over the past two decades is illustrated
in Figure~\ref{fig:oldvsmodernimage}.

Ultimately, the controversies over systematic errors were essential to
spur the community to overcome the difficulties and find ways to
address them. This is a natural and probably inevitable part of the
scientific process. However, the bitterness of some of those
controversies during the ninenties and early noughties still resonates
today: unfortunately, some of the scientists that followed the field
with excitement at that time are still under the impression that
strong lensing time delays are inherently inaccurate and imprecise. As
we have briefly described here, and we will show in detail in the
next sections, in the last twenty years the field has moved forward
considerably implementing many solutions to the lessons learned the
hard way.


\section{Theoretical background}
\label{sec:theory}

In this section we provide a brief summary of the theory of
gravitational lens time delays. We have distilled much of the content
of this section from the excellent exposition of Schneider and
Kochanek \citep{SKW06}, as well as the various key papers we cite.

Fermat's Principle of Least Time holds for the propagation of light rays
through curved spacetime. The light travel time through a single, isolated, thin
gravitational lens is given by
\begin{align}
    \tau(\x) &= \frac{\Ddt}{c} \cdot \Phi(\x,\y), \\
    \text{where\;\;} \Phi(\x) &= \frac{1}{2}\left(\x - \y\right)^2 - \psi(\x).
\end{align}
Here, $\x$ denotes the light source's apparent position on the sky, and
$\y$ is the position of the unlensed source. The difference between the
observable position~$\x$ and the unobservable position~$\y$ is the
scaled deflection angle~$\deflectionangle({\x})$, which is typically
$\sim1$~arcsecond in a galaxy-scale strong gravitational lens system.
$\psi(\x)$ is the scaled gravitational potential of the lensing object,
projected onto the lens plane. Both $\deflectionangle(\x)$ and $\psi(\x)$ can be
predicted given a model for the mass distribution of the lens.

\begin{figure*}[!t]
\centering\includegraphics[width=0.96\textwidth]{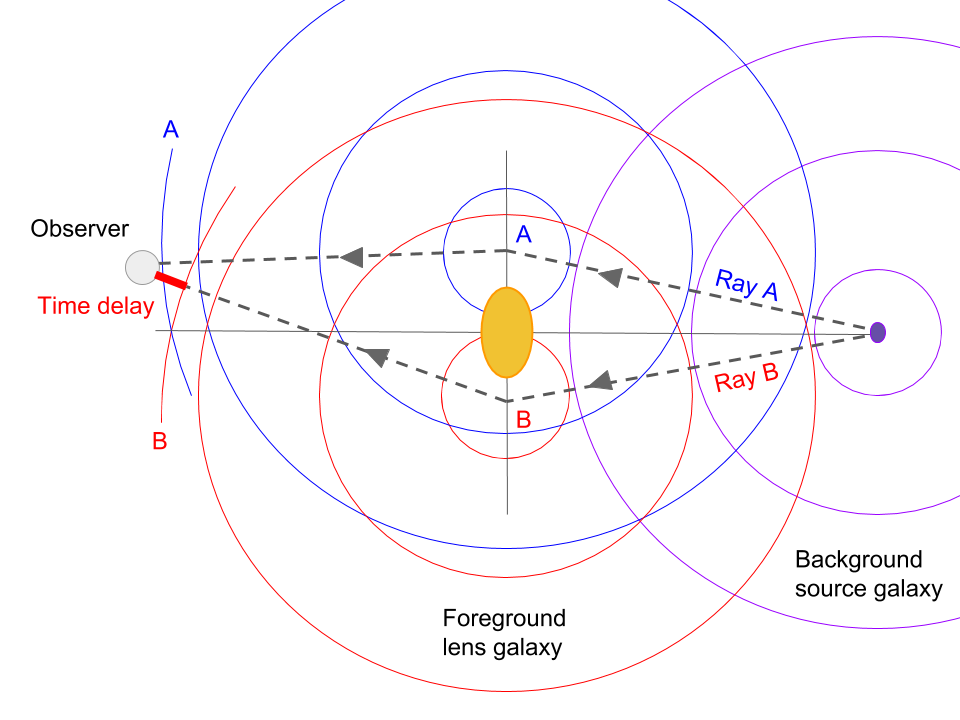}
\caption{Schematic diagram
illustrating the origin of the gometric component of the time delay.}
\label{fig:timedelaycartoon}
\end{figure*}

\begin{figure*}[!ht]
\centering\includegraphics[width=0.96\textwidth]{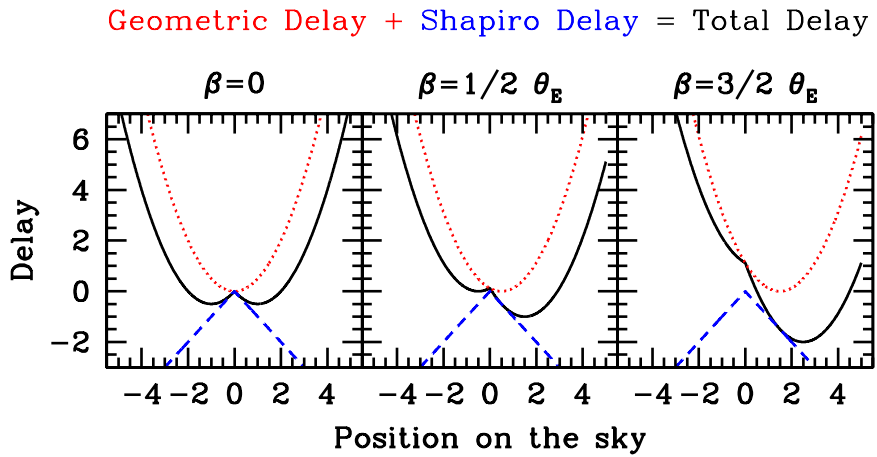}
\caption{Geometric and general relativistic (Shapiro) contributions
to the lens time delay, from \citep{T+E15}. Images form
at minima and saddle points of the delay surface, shown here in
cross-section. Different source positions result in different
geometrical delays as well as shifted image positions.}
\label{fig:delays}
\end{figure*}

Images form at the stationary points of the light travel time, where $\grad
\tau(\x) = \grad \Phi(\x) = 0$ \citep{Schneider1985}. For this reason,
$\Phi(\x)$ is known as the ``Fermat potential.'' This quantity can also
be thought of as the spatially-varying refractive index of the lens. The
arrival time itself is not observable, but differences in arrival time
between multiple images are. In the above approximation, the  ``time delay'' $\Delta \tau_{\rm AB}$
between image A and
image B can be predicted via
\begin{equation}
    \Delta \tau_{\rm AB} = \frac{\Ddt}{c} \Delta \Phi_{\rm AB} \label{eq:timedelay}
\end{equation}
where $\Delta \Phi_{\rm AB}$ is the Fermat potential difference
between the two image positions.  Figures~\ref{fig:timedelaycartoon}
and~\ref{fig:delays} illustrate the origin of the time delay between
the images in a simple gravitational lens system. The small magnitude
of the fractional time delay (typically $\Delta\tau \sim 10$~days out
of $\Ddt/c \sim 10^{12}$ days light travel time) is commensurate with
the square of the deflection angle (typically
$|\deflectionangle|\sim1$~arcsecond, or $\sim 5\times10^{-6}$
radians). Two characteristic scales are the critical surface mass
density $\Sigma_c$, and the Einstein Radius R$_{\rm Ein}$. The former
is given by a combination of angular diameter distances between the
source ($s$), deflector ($d$) and observer, $\Sigma_c=4c^2\Ds/4\pi
\Dd\Dds$, and it is used to define the dimensionless surface mass
density or convergence $\kappa=\Sigma/\Sigma_c$. The latter can be
defined, for axisymmetric mass distributions, as the radius of the
circle within which the mean convergence $\langle \kappa \rangle=1$.

We see from Equation~\ref{eq:timedelay} that given a mass
model that predicts $\Delta \Phi_{\rm AB}$, we can infer the ``time
delay distance'' $\Ddt$ from a measured time delay $\Delta \tau_{\rm AB}^{\rm obs}$.
This distance is actually a combination of angular diameter
distances:
%
\begin{equation}
    \Ddt = (1+\zd) \frac{\Dd \Ds}{\Dds}  \label{eq:ddt}
\end{equation}
These angular diameter distances can be predicted given the redshifts
of the lens and source, $\zd$ and $\zs$, and an assumed world model with
cosmological parameters~$\cospars$. The time delay distance is primarily
sensitive to the Hubble constant, since $\Ddt \propto H_0^{-1}$.
All the above formalism pertains to the simple model where
all the deflecting mass is arranged on a single lens plane. The multiple
lens plane case is more complex, but quantities like $\Ddt$ appear
throughout the equations that predict the time delays, capturing the
distances between the lens planes and preserving approximately the
same dependence on cosmological parameters \citep{Petters2001,McCullyEtal2014}.\footnote{Additional
distance dependences appear in the multi-plane formalism, but always as dimensionless
ratios with weaker cosmological dependence. The inverse proportionality
to the Hubble Constant is the same as in the single plane case.}


Knowledge of the lens mass distribution is of vital importance to the
success of this cosmological inference: Equation~\ref{eq:timedelay}
shows that the time delay distance is likely to be comparably
sensitive to uncertainty in
the predicted Fermat potential as it is to the measured time delay itself.
More concentrated mass distributions with steeper density
profiles produce longer time delays leading to shorter inferred time
delay distances, and thus larger inferred values of $H_0$ \citep{Wuc02,Koc02,Suyu12}.


Moreover, there is significant risk of systematic error when modeling
lens mass distributions. While image positions remain invariant under
the ``mass sheet transformation'' \citep{FGS85,S+S13}
\citep[and its generalization, the source-position transformation][]{SPT}, the time delays
predicted by the model can change significantly. The mass sheet transformation
and its effect on the time delay is as follows:
\begin{align}
    \kappa(\x) \rightarrow \kappa(\x)' &= (1 - \lambda) + \lambda \kappa(\x)  \label{eq:mst} \\
    \Delta \tau \rightarrow \Delta \tau' &= \lambda \Delta \tau.
\end{align}
This means that if we allow our model the freedom to generate both the
$\kappa(\x)$ and $\kappa(\x)'$ mass distributions, our image position
data will not favor one over the other: they will be equally likely
given the data. This model degeneracy can be broken by additional
information.

Perhaps the best sources of additional information are independent
measurements of the mass distribution: stellar kinematics is the
obvious choice \citep{Koo++03}. 
Another way to break degeneracy is to obtain non-lensing information
about the lensed source absolute size \citep{SBL11} or luminosity
\citep{Hol01}. This way requires special circumstances, and therefore
we focus on deflector kinematics in the remainder of this review.

\citet{JeeKomatsuSuyu2015} provide a derivation of the resulting
cosmological dependence of kinematics-constrained power-law lens
galaxy mass models with isotropic orbits, showing that were its
density profile and velocity isotropy to be known exactly, a time
delay lens would provide a measurement of the angular diameter
distance to the lens, $\Dd$, in addition to the time delay distance of
Equation~\ref{eq:ddt}. The reason given is that the velocity
dispersion and the time delay are both proportional to the enclosed
mass of the lens, but depend differently on galactocentric radius:
combining the measured velocity dispersion and time delay gives a
characteristic physical scale of the lens galaxy.  The image
separation provides a corresponding angular scale, allowing the
angular diameter distance to be probed \citep[see earlier work
by][]{GLB08}.

In practice, the same mass model must be used to predict all of the
measured velocity dispersion, Einstein ring appearance, and time
delay data, self-consistently. In the limit of low precision in the
velocity dispersion, the profile slope is weakly constrained by the
ring image alone, and the combination of time delay and lens mass
model provides information on $\Ddt$ but none on $\Dd$. As the
velocity dispersion precision increases, we expect the profile slope
to be pinned down, and the angular diameter distance to be constrained
{\it as well}. In the context of a cosmological model, the two
distances are not independent: the angular diameter distance
information provided by the velocity dispersion measurement should
translate into higher precision inference of the cosmological
parameters \citep{JeeEtal2016}.  We return to this in
Section~\ref{ssec:precision} below.

Another way to break degeneracy in the lens model is to include prior
knowledge of the lens mass distribution from measurements of other
galaxies similar to the lens, or perhaps from numerical
simulations. This type of information is typically encoded as a
simply-parametrized model, such as an elliptically-symmetric mass
distribution with power law density profile (as opposed to a free form
density map; see discussion in Section~\ref{ssec:lensmodel}). Assuming
a specific density profile partially breaks the mass sheet degeneracy:
how much systematic error in the time delay that assumption introduces
is an important topic for research.


The form of the mass sheet transformation given by
Equation~\ref{eq:mst} is a rescaling plus an offset. One way to
achieve such a transformation is therefore to change the overall mass
of the lens (by a factor of $\lambda$), and at the same time add a
``mass sheet,'' a constant convergence~$(1-\lambda)$.  Both these
variations are possible in nature: lens galaxies come in a range of
masses, and the combined gravitational lensing effect of all the other
galaxies, groups and filaments along the line of sight to the source
can, in the weak lensing limit, be approximated by a constant
``external convergence'' (which is associated with an ``external shear'',
capable of further distorting the lensed images). However, these
physical effects only complicate the modeling problem, as one is not
allowed to assume that the mass density profile of the deflector
should vanish exactly at large radii.  The physical effect should not
be confused with the mathematical degeneracy between lens model
parameters that is associated with the mass sheet transformation, and
which would be present regardless of any external weak lensing
effects. Having said that, any additional external physical mass
component must also be taken into account when modeling the lens.

In summary, independent information about the physical mass of the deflector
galaxy, such as the kinematics of its stars, can play an important role
in breaking the degeneracy in the mass model, which must be able to
predict self-consistently the strong lensing effects (image distortions and time
delays) and the internal dynamics of the lens galaxy, and take into
account the weak lensing effects of structures along the line of sight.
\citet{S+S13} provide demonstrations of the scale of this problem: very
good data (both imaging and spectroscopic), as well as physically
meaningful assumptions and careful treatment of the models used, will
be needed to obtain accurate results.  In Section~\ref{ssec:lensmodel}
we review the recent choices and approximations that have been made
when constructing such models.



\section{Modern time delay distance measurement}
\label{sec:measurement}

Since 2010, it has been recognized that accurate cosmography with
individual lens systems involves the following key analysis steps.

\begin{description}
    \item{\bf Time delay estimation} The light curve extracted from
    monitoring observations is used as input to an inference of the
    time delay between the multiple images.
    \item{\bf Lens galaxy mass modeling} High resolution imaging
    and spectroscopic data are used to
    constrain a model for the lens galaxy mass distribution, which can be used
    to predict Fermat potential differences. Both the Einstein ring
    image and the stellar velocity dispersion are important.
    \item{\bf Environment and line of sight modeling} Additional observational
    information about the field of view around the lens system is used
    to account for the weak lensing effects due to massive structures in
    the lens plane and along the line of sight.
\end{description}

Cosmological parameter inference can then proceed -- although in
practice the  separation between this final step and the ones above is
not clean. Practitioners aspire to a joint inference of lens, source,
environment and cosmological parameters from all the data
simultaneously, but have to date broken the problem down  into the above
steps. In the next three sections we describe current state of the art,
limitations, and principal sources of systematic error of these three
key measurement parts of the problem.


\subsection{Measuring time delays}
\label{ssec:timedelay}

The measurement of gravitational time delays involves two steps:  taking
observations to monitor the system over a period of several years,
and then inferring the time delays between the multiple images from
these data.


\subsubsection{Monitoring observations and results}

Active Galactic Nuclei (AGN) show intrinsic time variability on many
scales, with the variability amplitude increasing with timescale. Long
and regular
monitoring campaigns can build up high statistical significance as
more and more light curve features can be brought into play.  However,
such long campaigns are difficult to carry out in practice, because a
large number of
guaranteed, evenly spaced
observing nights are required (even if the
total exposure time is modest). Scheduling such a program has proved
difficult in traditional time allocation schemes, due to the competing
demands of the rest of the astronomy community and the long duration
requirements of lens monitoring. The highest precision time delays
have come from monitoring campaigns carried out with dedicated
facilities so far, i.e. observatories that were either able to commit
to the long term monitoring proposal submitted, or that were actually
operated in part by the monitoring collaboration.

Monitoring of the CLASS lens B1608$+$656 in the radio with the Very
Large Array enabled the breakthrough  time delay measurements of
\citet{Fas++02}. In its first season, this program  yielded measurements
of all three time delays in this quadruple image system with precision
of 6--10\% \citep{Fas++99}; with the variability of the source
increasing over the subsequent two seasons, \citet{Fas++02} were able
to reduce this uncertainty to 2--5\%. Such high precision was the
result of a dedicated campaign which consisted of 8-month seasons,
with a mean observation spacing of around 3 days. The light curves
were calibrated to 0.6\% accuracy.

While time delays had previously been measured in ten other lens
systems, this was the first time that all the delays in a quad had
been obtained; moreover, it brought the time delay uncertainty below
the systematic uncertainty due to the lens model, prompting new
efforts in this direction beyond what \citet{K+F99} needed to do.

While B1608$+$656 is not the only radio lens with measured time
delays, a combination of factors led the observational focus to shift
towards monitoring in the optical. With the sample of known, bright
lensed quasars increasing in size, networks of 1-2m class optical
telescopes began to be investigated. The variability in these systems
is somewhat more reliable, and while microlensing and image resolution
present observational challenges, the access to data was found to be
less restrictive. The COSMOGRAIL project \citep{Cou++05} took on the
task of measuring lens time delays with few-percent precision in this
way:
\citet{Eig++05} showed that microlensing was likely not to be an
insurmountable task, and \citet{Vui++07} provided the proof of concept
with a 4\% precision time delay measurement in SDSS\ J1650$+$4251.

One of the keys to the success of this program has been the simultaneous
deconvolution of the individual frames in the imaging dataset, using  a
mixture model to describe the point-like quasar images and extended lens
and AGN host galaxies \citep{MCS98}.  Another is the dedicated nature of
the network of telescopes employed, and the  careful calibration of the
photometry across this distributed system. Seasons of 8--12 months
duration over campaigns of up to 9 years have been achieved, with
typical mean observation gaps of around 3--4 days.

The COSMOGRAIL team and their collaborators have now published high
precision time delays in WFI\,J2033$-$4723 \citep[][3.8\%]{Vui++08},
HE\,0435$-$1223
\citep[][5.6\%]{Cou++11}, SDSS\,J1206$+$4332 \citep[][2.7\%]{Eul++13}
and  RX\,J1131$-$1231 \citep[][1.5\%]{Tew++13}, and SDSS\,J1001$+$5027
\citep[][2.8\%]{RK++13}, with more due to follow.  Typically multiple
years of monitoring are needed to obtain an accurate  time delay, as the
variability fluctuates and the reliability of the  measurement converges
\citep[see the discussion in e.g.\ ][]{Tew++13}. High precision optical time delays are also being obtained by other groups \citep{Poi++07a,Foh++07,Dah++15} using similar strategies on different telescopes.

A consistent picture seems to emerge from modern monitoring projects:
high precision gravitational time delay measurement requires campaigns
consisting of multiple, long seasons, with around 3-day cadence. The
baseline observing strategy for the Large Synoptic Survey Telescope
(LSST) is somewhat different to this, with seasons expected to be
around 4--5 months in length, and gaps between observation nights only
reaching 4--5 days when images in all filters are taken into
account. The ``Time Delay Challenge'' project was designed to test the
measurability of lens time delays with such light curves
\citep{DoblerEtal2015}, in a blind test offered to the astronomical
community. From the ten algorithms entered by seven teams, it was
concluded that time delay estimates of the precision and accuracy
needed for time delay cosmography would indeed be possible, in
$\sim$400 LSST lensed quasar systems \citep{LiaoEtal2015}. This result
came with two caveats: 1) the single filter light curve data presented
in the challenge is representative of the multi-filter data we
actually expect, and 2) that ``outliers'' (catastrophic time delay
mis-estimates) will be able to be caught during the measurement
process.  A second challenge to test these assumptions is in
preparation.


\subsubsection{Lightcurve analysis methods}

How were the time delays surveyed in the previous section derived from
the light curve data? Interest in this particular inference problem
has been high since the controversies of the late
1990's. \citet{Fas++99} used the ``dispersion method'' of
\citet{Pelt++96}, a technique that involves shifting one observed
light curve relative to another (both in time and in amplitude) and
minimizing the dispersion between adjacent points in the resulting
composite curve. Uncertainties were estimated by Monte Carlo
resampling of the data, assuming the minimum dispersion time delay and
magnification ratio to be true. In order to take into account the
slowly varying incoherent microlensing signals present in their
optical light curve data, the COSMOGRAIL team have investigated three
analysis techniques that all involve interpolation of the light curves
in some way \citep{TCM13}: free-knot splines, Gaussian processes and
simple linear interpolation have all been tested, within a common
``python curve-shifting'' (PyCS) framework.\footnote{The COSMOGRAIL
curve shifting analysis code is available from
\texttt{http://cosmograil.org}} These agree with each other given
light curves of sufficient length, providing an argument for
multiple-season monitoring campaigns.

The time delay challenge prompted seven analysis teams to develop and
test algorithms for time delay estimation. These are outlined in the
TDC1 analysis paper of \citet{LiaoEtal2015}, but we give a very brief
summary here as well, along with updated references.
The PyCS team tried a two-step approach (visual inspection and
interactive curve shifting, followed by automated analyses based on
spline model regressions for the AGN variability and the microlensing),
and submitted an entry after each step
\citep{BonvinEtal2016}. Two other teams applied similar
curve-shifting approaches: both \citet{A+S2015} and \citet{RK++2015}
devised smoothing and cross-correlation schemes that they find to be
both fast and reliable. Jackson applied the dispersion method of
\citet{Pelt++96}, but carefully supervised via visual inspection to
check for  catastrophic failures.  The three remaining teams used
Gaussian Processes (GPs) to model the light curves. \citet{TakEtal2016}
used a custom Gibbs sampler to infer the hyper-parameters describing the
GP for the AGN variability and polynomials for the microlensing signals,
although they ignored microlensing during the challenge itself.
Romero-Wolf \& Moustakas implemented a very similar model, also ignored
microlensing, and used a freely-available ensemble sampler for the
inference. \citet{H+L2014} used GPs for both the AGN and microlensing
variability, and marginalized over their hyper-parameters when focusing
on the time delay.

Two factors were important in the minimisation of catastrophic time
delay mis-estimation: explicitly including microlensing in the model,
and  visual inspection of the results. An additional promising avenue
for future challenges ought to be ensemble analysis, to exploit 1) the
intrinsic correlations between, for example, AGN variability, color and
brightness, and 2) the fact that  the cosmological parameters are common
to all lens systems.


\subsection{Modeling the lens mass distribution}
\label{ssec:lensmodel}

In addition to time delays, the second main ingredient entering the
determination of time delay distances is the mass model of the main
deflector. In the early days of time delay cosmography one could only
rely on the relative positions of the multiple images as constraints
(since in general the flux ratios are affected by micro and
millilensing, variability, and differential dust extinction, and are
therefore highly uncertain). Even for a quadruply imaged quasars, the
five positional constraints and three independent delays are
insufficient to determine Fermat potential differences to the desired
level of precision and accuracy.

There are two classes of solution to the problem of underconstrained
lens models. One is to analyze large samples of lenses with physically
motivated priors and exploit the fact that cosmological parameters are
the same for all lenses to remove model degeneracies. A number of
attempts along these lines have been made \citep{Ogu07b,RK++2015}, and
it is easy to imagine that this solution will be popular in the
future, when large samples of lenses with measured time delays will be
available.

The alternative solution is to increase dramatically the number of
emprical constraints per lens system by means of dedicated high
resolution imaging and spectroscopic observations
\citep{Suy++10,Suy++13,Suy++14}. We describe this approach in detail
below.

For simplicity, in this section we describe only the case of a single
deflector in a single plane, leaving line of sight and environmental
effects for a later section. For clarity, we describe each step
corresponding to a different dataset individually. Ideally, all the
data, including the time delays, should be modeled holistically at the
same time---although in practice the problem has, to date, been broken up
into parts to make it more tractable.


\subsubsection{High resolution imaging observations}

Lensed quasars reside in a host galaxy. For typical redshifts of lens
and source, the host galaxy apparent size is of order
arcseconds. Images with sufficient depth and resolution to isolate the
bright point source and detect the lower surface brightness host
galaxy often reveal extended lensed features connecting the
point-like images themselves (e.g. Figure~\ref{fig:oldvsmodernimage}).

In the best conditions these images cover hundreds if not thousands of
resolution elements. The distortion of the detailed features of the
lensed images are a direct measurement of the variation of the
deflection angle between the images.  In principle, for data with
infinite signal-to-noise ratio and resolution one could imagine
integrating the gradient of the deflection angle along a path between a
pair of images to obtain the difference in Fermat potential,
up to a mass sheet transformation (Section~\ref{sec:theory}).
In practice, in the presence of noisy data and limited resolution,
forward modeling approaches have been the most successful so far, as
discussed below. From an observational point of view, it has been
demonstrated that images with $0.1''-0.2''$ FWHM resolution provide
good results, provided that the point spread function can be
appropriately modeled or reconstructed as part of the lens model
itself. The Hubble Space Telescope in the optical/near infrared
\citep{Suy++10,Suy++13,Suy++14,BirrerEtal2015} and the Very Large Baseline
Interferometer in the radio \citep{WBB04} have been the main sources
of images for this application. Recent progress in adaptive optics
imaging at the 10m W.M.~Keck telescope \citep{Che++16}, the beautiful data being
obtained for lensed source by ALMA \citep{Hez++13a}, and the many
facilities currently being constructed or planned \citep{Men++15},
indicate that the prospects to scale up the number of systems with
available high resolution images are bright.


\subsubsection{Lens modeling techniques}

Conceptually, a detailed model of a lensed quasar and its host galaxy
needs to describe three different physical components: i) the surface
brightness of the source; ii) the surface brightness of the deflector;
iii) the gravitational potential of the deflector. It is useful to
conceptualize the problem in this way, in order to understand where
the information needed to break the degeneracy in the intrepretation
of the data comes from. Lensing is achromatic and preserves surface
brightness so any feature that belongs to the source \cite[including
in line of sight velocity][]{Hez++13} should appear in all the
multiple images (appropriately distorted). Likewise, the deflector is
typically a massive early-type galaxy with smooth surface brightness
distribution and approximately uniform colors \cite[except for dust,
see, e.g.][]{Suy++10}.

Each of the three components is typically described in terms of one or
both of the following choices:
i) simply parametrized functions such as a Sersic profile
for the surface brightness of the lens or the source, and a singular
isothermal ellipsoid for the gravitational potential of the deflector
\citep[e.g.][]{Mar++07,Kne++11,Kee11}; ii) as combinations of basis sets like surface brightness
pixels, lens potential pixel values, or Gauss-Hermite (``shapelet'')
functions
\citep[e.g.][]{Col08, BirrerEtal2015, Nig++15, TagoreAndJackson2016}.
Very flexible models require regularization to avoid overfitting the
noise in the data.\footnote{In the case of the shapelet basis set,
regularization can effectively be achieved through choosing the number of basis
functions to use as well as the scale of the underlying Gaussian. Most
analyses using shapelets having taken this approach to date, with
\citet{TagoreAndJackson2016} being a notable exception. A
promising alternative scheme would be to assign a less physically-motivated prior for the
shapelet coefficients.}
Hybrid approaches have been proposed where the parametrization of some of
the components is simple and others are complex
\citep{W+D03,T+K04,BrewerAndLewis2006,Suy++06,S+H10}, or where flexibly-parametrized
``corrections'' are added to simply parametrized models
\citep{Koo05,V+K09,Suy++09,BirrerEtal2015}.  The variety of approaches
in the literature reflect the inevitable tensions between the need to
impose as many physically motivated assumptions as possible, while
retaining sufficient flexibility to obtain a realistic estimate of the
uncertainties and avoid introducing biases by asserting incorrect
simplistic models. If the model is too constrained by the assumption
it will lead to underestimated errors, if it is more flexible than
necessary it will lead to a loss of precision.

Once the choice of modeling parametrization is set, exploring the
posterior PDF for the parameters is numerically non-trivial, often requiring weeks to months
of computing time. Fortunately, there are techniques to speed up the
calculations by limiting the number of non-linear parameters. For
example, for a given lens model, the transformation between source and
image plane can be described as a linear operation, or the pixellated
corrections to the potential can be found by linearizing the lens
equation (see references above).

Ideally, modeling choices should be explored systematically as well,
since they can potentially introduce systematic errors. This is
currently being done in the most advanced studies, at great expense in
term of computing time and investigator time. As we discuss in
Section~\ref{sec:outlook}, speeding up the modeling phase and reducing
the investigator time per system will be key to analyzing the large
statistical samples expected in the future.


\subsubsection{The role of stellar kinematics}

As introduced in Section~\ref{sec:theory}, stellar kinematics provide
a qualitatively different input and are therefore very valuable in
breaking degeneracies in the interpretation of lensing data
\citep[e.g., the mass-sheet degeneracy][]{Koo++03}, and in estimating systematic
uncertainties. Of course, translating kinematic data into estimates of
gravitational potential has its own uncertainties and degeneracies
(e.g. the mass anisotropy degeneracy for pressure supported systems),
but the combination of the two datasets in the context of a single
mass model has been proven to be very effective
\citep{T+K02a,T+K04}. Even a single measurement of stellar velocity dispersion,
interpreted via simple spherical Jeans modeling, has been shown to
substantially reduce modeling uncertainties
\citep{T+K02b,Koo++03,Suy++14}. It is clear that getting spatially
resolved kinematic data will enable breaking the mass-anistropy
degeneracy \citep[see, e.g.,][and references therein]{Cou++14} and
thus better constraints on the lens model and consequent
cosmological inference (Agnello et al. 2016, in prep).


\subsection{Lens environments and line of sight effects}
\label{ssec:los}

The analysis of B1608$+$656 by \citet{Suy++10} explicitly took into
account the weak lensing effects of external structures.  Such a
correction had been suggested by \citet{Fas++06b}, who identified 4
galaxy groups along the line of sight in a spectroscopic survey of the
B1608$+$656 field; the authors estimated that these groups could,
if left unaccounted
for, bias any inferred Hubble constant high by around 5\%, an amount
consistent with more general theoretical predictions \citep{Bar96,K+Z04}.
Further surveys have quantified the environments and line of sight
density structures of many more systems \citep{Mom++06,Aug++07,Won++11,Mom++15}.
Exactly how
to model the weak lensing contamination of strong lens signals
has been the topic of a number of papers since
2010: the problem is how to incorporate our knowledge of where the
galaxies are along the line of sight without introducing additional
bias due to the necessary assumptions about how their (dark) mass is
distributed, and how the rest of the mass budget in the field adds up.

\citet{Suy++10} attempted to solve these problems by comparing the
B1608$+$656 field with a large number of fields with similar  galaxy
number overdensity drawn from the Millennium Simulation, modeling the
line of sight effects with a single external convergence parameter and
accepting a somewhat broad prior distribution for it, in return for not
having to make  strong assumptions about the structure of the galaxy
groups in the field. The external convergence in the simulated fields
was calculated by ray-tracing by \citet{Hil++09}, and the
comparison in galaxy overdensity was enabled by the analysis of galaxy
number counts in archival HST images by \citet{FKW11}, who found that
the B1608$+$656 field was overdense by a factor of two. The resulting
prior PDF for the $\kappa_{\rm ext}$ parameter had median $0.10$ with
the 68\% credible interval spanning 0.05 to 0.18.
In the analysis of RXJ1131, \citet{Suy++13} also took into account the
inferred external shear from the lens model when deriving the prior for
$\kappa_{\rm ext}$, noting a significant improvement in precision (as
well as a marked shift in the PDF centroid).

Since these initial analyses, a number of improvements have been
suggested and investigated. All have in common the desire to bring more
information  to bear on the problem, in order to increase the precision
(while  continuing to avoid introducing bias). \citet{Gre++13} showed
that weighting the galaxy counts by distance, photometric redshift and
stellar mass  can significantly reduce the uncertainty in $\kappa_{\rm
ext}$, by up to 50\%. \citet{CollettEtal2013} claim an additional 30\%
improvement  by including knowledge of the stellar mass to halo mass
relation in galaxies,  and modeling each galaxy halo's contribution to
$\kappa_{\rm ext}$ individually in a  3-D reconstruction of the mass in
the field which is then calibrated to simulations in something like the
high resolution limit of the number counts approach.
\citet{McCullyEtal2014} showed how to compute the weak lensing
contamination accurately, using a full multi-plane lensing formalism
\cite[see also][]{Schneider2014}
but with fast approximations for less important
structures \citep{McCullyEtal2016}.

While research into these methods continues, one problem in particular
remains outstanding. The methods that involve calibration to numerical
simulations are dependent on the cosmological parameters assumed in
that simulation, while all methods involve modeling line of sight
structures at various distances as part of an evolving universe, whose
dynamics depend on cosmological parameters. We face two options:
either treat these cosmological parameters self-consistently as
hyperparameters in a joint analysis of the time delays and the lens
environments, or demonstrate that they can be decoupled via various
simplifying assumptions that introduce sub-dominant systematic
error. At the moment, comparison between the results of independent
systems \citep{Suy++13} seem to suggest that this source of
uncertainty is smaller than the estimated random uncertainty. However,
this issue will have to be addressed in detail as the sample sizes
increase and the random uncertainty decreases.


\section{From time delay distances to cosmological parameters}
\label{sec:cosmo}

Early approaches to inferring cosmological parameters from time delay
lens observations focused on measuring the Hubble constant in a
Friedman-Robertson-Walker model with asserted (fixed) density
parameters.\footnote{The original investigation by \citet{Ref64}
involved the ``assumption that the linear distance--red-shift relation
is valid.''} With better data came the recognition that time delay
lenses were really probes of cosmological distance
\citep{Koo++03,Suy++10}, and the emphasis shifted to
inferring the set of cosmological parameters that are needed to predict
the kinematics of the expansion of the Universe out to the redshift of
the source.
The parameter most strongly constrained is still the Hubble constant,
but as sample sizes increase we expect ensembles of lenses to support
the inference of several cosmological parameters (or combinations
thereof).

\begin{figure*}[!ht]
\centering\includegraphics[width=0.9\linewidth]{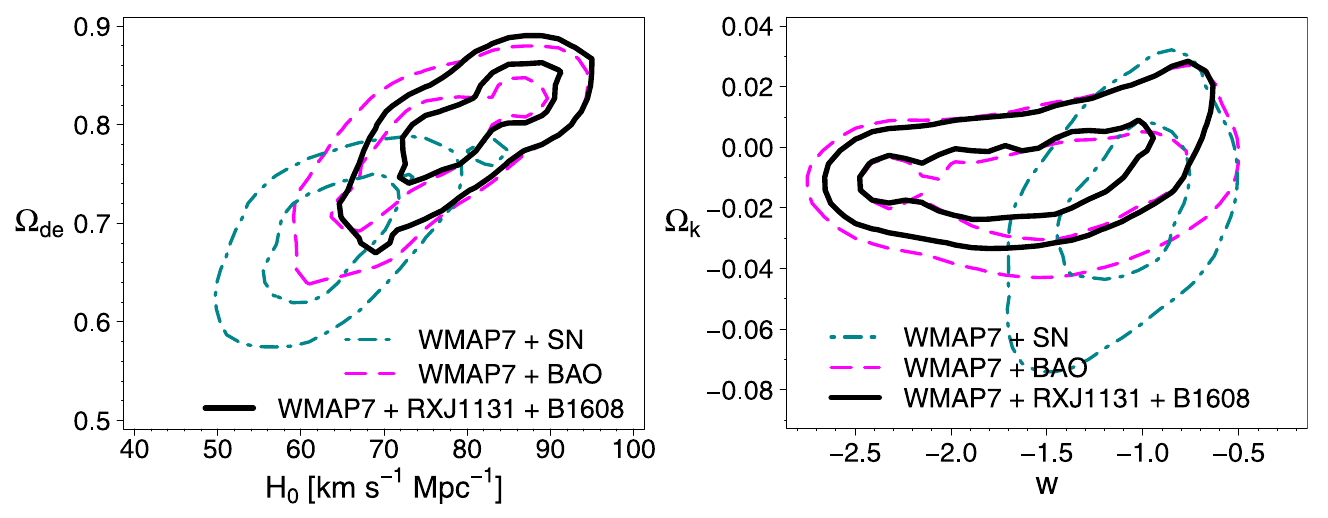}
\caption{Cosmological parameter constraints from time delay
lenses. The marginalized posterior PDFs, given the combined B1608$+$656
and RXJ1131 datasets and the assumption of an open CDM cosmology with
unknown dark energy equation of state, are shown in
two sets of two parameter dimensions,
and compared to those given contemporary BAO and Type Ia supernova data.
Figure reproduced from \citet{Suy++13}.}
\label{fig:current-constraints}
\end{figure*}

In Figure~\ref{fig:current-constraints} we reproduce the current
constraints on cosmological parameters, from the two best-measured
systems, B1608$+$656 and RXJ1131 \citep{Suy++14}. When this figure was
made, the available precision from just these two lenses was about the
same as that from SDSS DR7 Baryonic Acoustic Oscillations
\citep{PercivalEtal2010} or the ``Constitution'' set of Type Ia
supernovae \citep{HickenEtal2009}.  When all three of the curvature
density $\Ok$, Dark Energy density $\ODE$ and equation of state $\wDE$
parameters are allowed to vary, along with $H_0$, we see that the time
delay lenses provide similar constraints to BAO and complementary
constraints to the SNe: the time delays and the BAO signal depend on
angular diameter distances and $H_0$, while the supernovae probe
relative luminosity distances.

One important feature of the cosmological parameter inference carried
out in the RXJ1131 analysis of \citet{Suy++13} is that it was blinded.
Following the simple methodology suggested in the blind Type Ia
supernova analysis of \citet{Con++06}, all cosmological parameter PDFs
were plotted with centroids offset to the origin until the team agreed
(after notably lengthy discussions about systematic errors) to ``open
the box,'' just before publication.\footnote{Importantly, the authors
agreed to publish the unblinded results, no matter what.} Such
attempts to avoid ``unconscious experimenter bias'' introduced by
stopping systematics analysis when the ``right answer'' is obtained
have long been advocated in particle physics
\citep{klein2005blind}, and seem likely to become the standard in
cosmology as well \citep[e.g.][]{STEP,DESWL}. It is also crucial to
repeat the measurements using independent codes, assumptions, and
techniques, in order to quantify associated systematic
uncertainties. It is re-assuring that the independent analysis of
RXJ1131 carried by out by \citet{BAR16}, the one based on
more flexible models carried out by \citet{Suy++14}, and the one based
on ground based adaptive optics data by \citet{Che++16} find results that are
statistically consistent with the original blind analysis \citep{Suy++13}.

While the sample of very well-measured lenses was being painstakingly
expanded from zero to two, the exploration of statistical approaches
to dealing with large samples of lenses began.  Compressing the image
configuration and time delay in double image systems into a single
summary statistic, \citet{Ogu07b} derived a scaleable method for
measuring the Hubble constant (but not the other cosmological
parameters) from samples of lenses, finding
$H_0=68\pm6\,{\rm(stat.)}\,\pm8\,{\rm (syst.)}\,{\rm km s}^{-1}{\rm
Mpc}^{-1}$ from a sample of 16 lenses with measured time delays. The
systematic uncertainties associated with this result may be hard to
reduce given the approximations made: while the summary statistic is
model-independent, the interpretation is not.

An alternative approach is to work with more flexible lens models, fit
the data for each one, and combine the whole sample in a joint
inference.  This is the approach taken by \citet{Sah++06}, who found
$72^{+8}_{-11}\,{\rm km s}^{-1}{\rm Mpc}^{-1}$ from 10 lenses (again
assuming fixed curvature and dark energy parameters).
The amount of
information system used in this analysis was minimal: only the quasar
positions and time delays were taken as inputs. The high flexibility of the free
form models employed led to a likelihood that was effectively identically 1 or 0, and thus the
results are dominated by prior constraints set on the pixels and their
regularization \cite[see][for a description of pixel-lens in Bayesian
terms]{Col08}. Specifically, the known physical degeneracy between
density profile slope and predicted time delay
\citep{Wuc02,Suyu12} is broken by the choice of pixel value prior probability
distribution function. This assumption was tested by
\citet{Rea++07}, who used a hydrodynamic simulation of an elliptical
galaxy to generate mock image position and time delay data, and
confirm the accuracy of the previous study's Hubble constant
uncertainties. With improved time delay estimates in larger samples of
lenses, \cite{P+J10} and then \citet{RK++2015} reduced the random
uncertainty further.

While focused only on the Hubble constant and carried out unblind, and
with the lens environment and line of sight mass structures remain
unaccounted for and further tests on realistic simulated galaxies
warranted, these ensemble studies point the way towards a future of
considerably larger sample sizes. Our aspirations towards high accuracy
demand that we adopt more flexible mass models and then
cope with the degeneracies;
it is clear that such large
scale analysis will need careful consideration of the choice of the
priors, and ideally the ability to use more information than just the
image positions and time delays. We discuss these issues in detail in
the next section.


\section{Outlook}
\label{sec:outlook}

In this section we discuss the future of time delay cosmography, and
present a roadmap of how this measurement might be improved in the
next decade. In order to construct the roadmap
(Section~\ref{ssec:roadmap}), we will discuss in detail how to
decrease the random uncertainties (increasing the precision of the
method, Section~\ref{ssec:precision}), and the systematic
uncertainties that will need to be controlled as the random
uncertainties decrease (thus maintaining high accuracy,
Section~\ref{ssec:accuracy}).

However, before we lay out this roadmap, we first pause and
reflect on the broader context, and ask whether this is a worthy
endeavour.

This boils down to three simpler questions. The first question is
whether time delays contain valuable information {\it independent} of
other cosmological probes. As detailed in Section~\ref{sec:cosmo}, the
answer is a resounding yes: gravitational time delays are virtually
independent of the uncertainties affecting the other established
probes of dark energy, and provide valuable complementary information,
chiefly on the Hubble constant, which is commonly regarded as one of
the essential ingredients for interpreting other datasets such as the
cosmic microwave background \citep{Hu05,Suy++12,Wei++13,Rie++16}.  We
will expand on this topic in the remainder of this section by showing
cosmological forecasts for gravitational time delays by themselves and
in combination with other probes.

The second question is whether it is feasible to achieve an {\it
interesting} level of precision and accuracy in coming years. In this
mindset, {\it interesting} is defined as having total uncertainties
comparable to that of other contemporary probes. This will be
discussed in detail in Sections~\ref{ssec:precision}
and~\ref{ssec:accuracy} below.

The third and final question is what is the {\it cost} of pursuing
this roadmap, and how this cost compares to that of other probes. Our aim is not
to compute a full cost accounting, which will be almost impossible
considering that each probe involves facilities, observatories,
computing and brainpower, well beyond the boundaries of any individual
project, collaboration, or funding agency (not to mention that the
marginal cost of adding a technique to an existing program or
facility is very different from what the cost of building a facility
just for that purpose; for example, the cost of monitoring strongly
lensed quasars in LSST data is much less than building and operating
the LSST). Instead, we will aim to give an approximate sense of the
observational and human resources that will be needed to pursue the
roadmap.


\subsection{Precision}
\label{ssec:precision}

With considerable observational and data analysis effort, the
feasibility of reaching a {\it precision} of 6-7\% in time delay distance
per lens has been demonstrated. The contributions to this statistical
error budget
from the time delay measurement, mass model, and environment
correction are at present approximately equal, and somewhat larger than
the estimated systematic errors. In this situation it makes sense to
enlarge the sample of lenses, in order to beat down the statistical
uncertainties. We return to the question of how to reduce the residual
systematic errors in the next section.

\citet{C+M09b} made initial Fisher matrix forecasts of the likely
available precision on $H_0$ in large future surveys. They considered
several possible samples, concluding that 100 well-measured systems
(with 5\% distance precision each) should provide sub-percent precision
on the Hubble constant, and provide  dark energy parameter constraints
that are competitive with optimistic forecasts of other ``Stage IV''
cosmological probes. They also note that comparable constraints could be
available from a sample of 4000 time delay lens systems, each with only
photometric redshifts and simple image configuration model constraints
\citep[following][]{Ogu07b,P+J10}.  Continued investigation of both samples
seems warranted, keeping in mind that the size of such a photometric
sample would be set by the availability of time delays measured at the
few percent level.

\begin{figure*}[!ht]
\centering\includegraphics[width=0.9\linewidth]{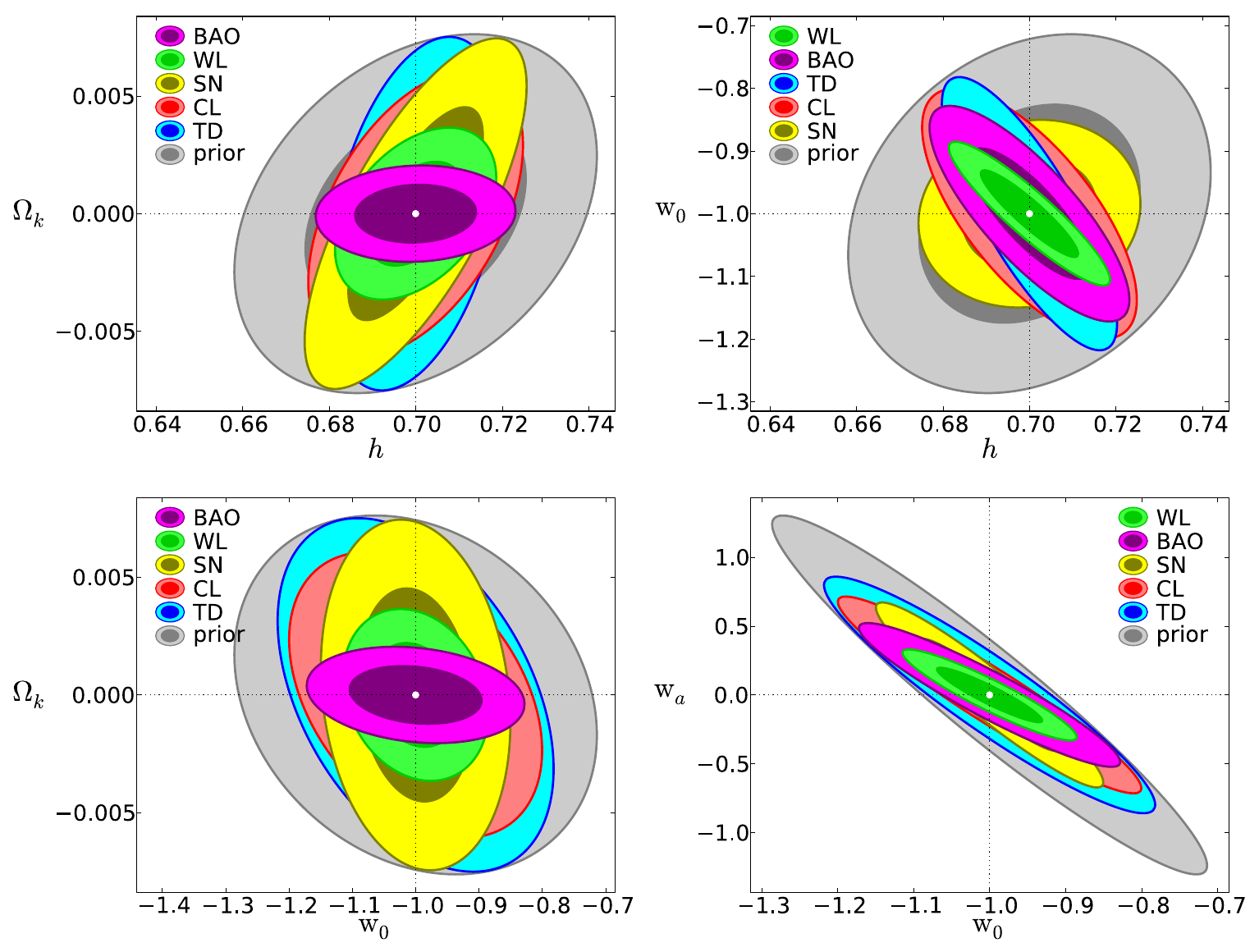}
\caption{Fisher matrix forecasts of cosmological parameters, based on
Dark Energy Task Force assumptions and having 5\% distance precision
for each of 100 time delay lenses. The Stage IV cosmological probes
being compared in an  open CDM cosmological model with time-variable
dark energy equation of state are weak lensing (WL), BAO, supernovae
(SN), cluster mass function (CL) and time delay cosmography (TD).
Figure reproduced from \citet{C+M09b}.}
\label{fig:fisher}
\end{figure*}

While Figure~\ref{fig:fisher} allows different cosmological probes to
be compared (and assessed for competitiveness), it does not show the
value of combining those probes. Indeed, \citet{Lin11} found that the
particular combination of a type Ia supernova dataset with a time  delay
lens dataset holds promise, with a sample of 150 time delay distances,
each measured to 5\% precision, improving the dark energy figure of
merit by a factor of about 5 over what could be  plausibly obtained with
a sample of about 1000 Stage~III supernovae and a Planck CMB prior alone.

\begin{figure*}[!t]
\begin{minipage}{0.48\linewidth}
    \centering\includegraphics[width=\linewidth]{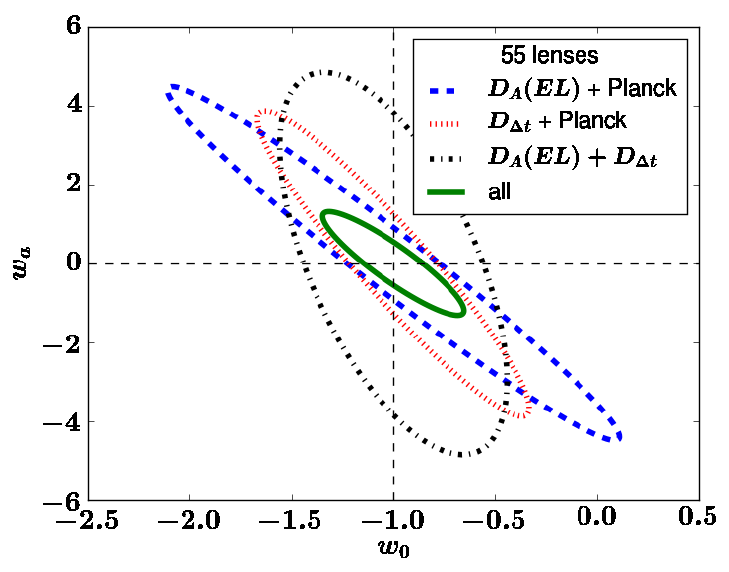}
\end{minipage}\hfill
\begin{minipage}{0.48\linewidth}
    \centering\includegraphics[width=\linewidth]{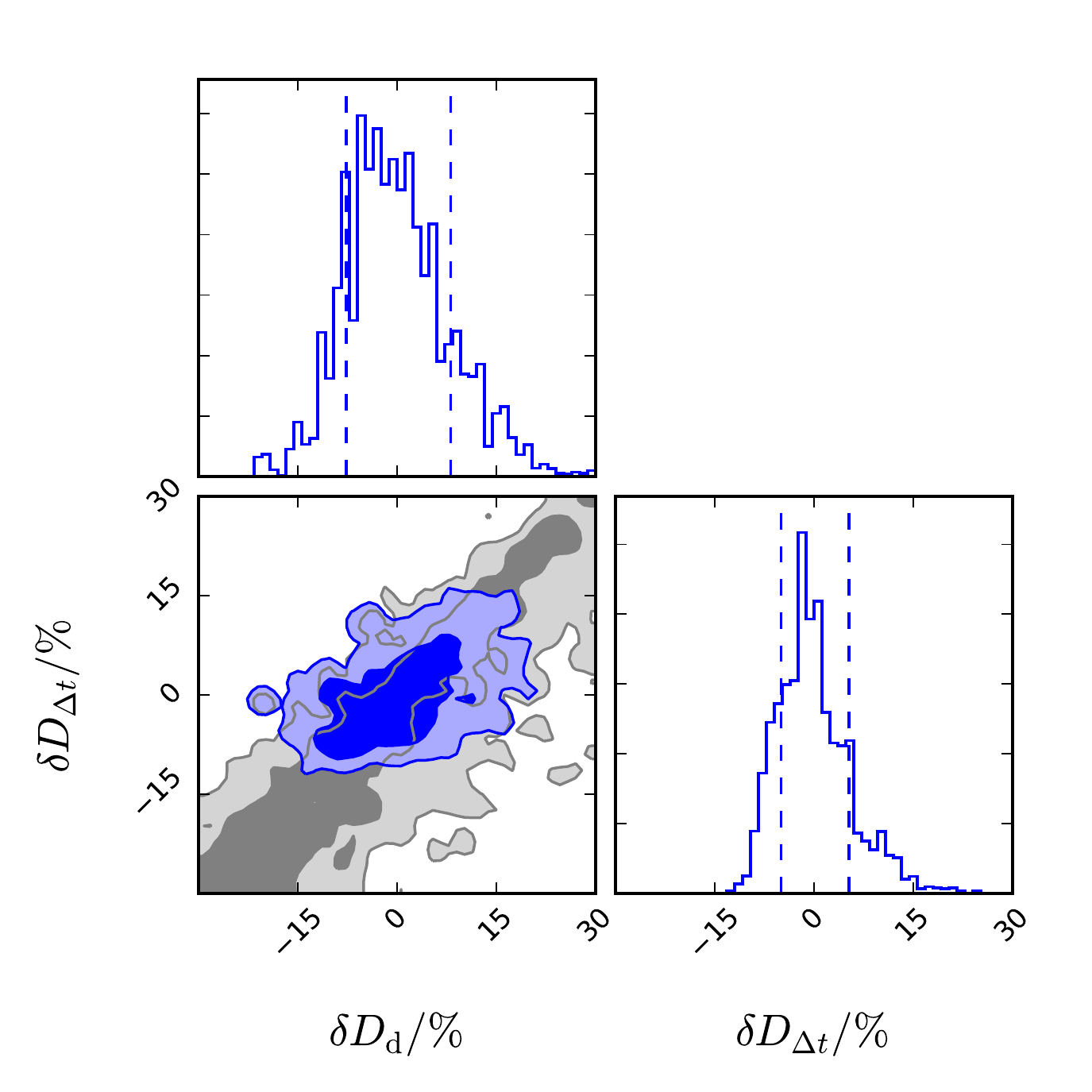}
\end{minipage}
\caption{Cosmological information from angular diameter distances
as well as time delay distances. Left: Fisher matrix
forecasts for a future time delay lens sample where 5\% precision on
each distance is assumed: the combination of distances gives a
significantly more powerful constraint than the time delay distances
would alone \citep[reproduced from][]{JeeEtal2016}. ``$D_{\rm A}(EL)$'' is
the angular diameter distance to the lens from Earth. Right: marginalized
prior (gray) and posterior (blue) PDFs for the two
distances in the B1608$+$656 system, assuming uninformative priors for the
cosmological parameters of a flat $\Lambda$CDM model and offset and
rescaled to reveal the implied percentage precisions of 5.7\% and 8.1\%
in $\Ddt$ and $\Dd$ respectively (as shown by the vertical dashed lines
enclosing the 68\% credible region). }
\label{fig:DdDdt}
\end{figure*}

More recently, \citet{JeeKomatsuSuyu2015} have pointed out that
cosmological parameter forecasts for time delay lens samples are
conservative, if each lens is assumed only to measure the time delay
distance. Including the angular diameter distance dependence as well can
have a marked effect on the projection, especially if the spectroscopic
constraints on the lens mass distribution are assumed to be very strong.
The reproduction of Figure 5 from
\citet{JeeEtal2016} in the lefthand panel of Figure~\ref{fig:DdDdt}
illustrates this. These authors find that a future sample of
55 lenses
with 5\% measurements of both time delay distance and angular diameter
distance would increase the figure of merit by a factor of two over that
provided by a Stage~III supernova, CMB, and BAO joint analysis. The righthand
panel of Figure~\ref{fig:DdDdt} puts such improvements in the  current
observational context. In the B1608$+$656 analysis, the  angular
diameter distance dependence {\it was} accounted for during the
calculation of the predicted time delay and velocity dispersion data,
but the constraints on the angular diameter distance were not strong:
assigning a uniform prior PDF for the cosmological parameters rather
than the distances introduced degeneracy between $\Dd$ and $\Ddt$,
which then seems to have been broken primarily by the time delay
information to yield a $5.7\%$ precision prediction for $\Ddt$, and
a corresponding $8.1\%$ precision prediction for $\Dd$.
With spatially resolved spectroscopy we should anticipate the angular
diameter distance becoming more important in future analyses, with
some work on simulated data needed to quantify this.

\subsection{Accuracy}
\label{ssec:accuracy}

While the precision available from Stage~III and Stage~IV samples
makes time delay lenses an interesting prospect for cosmology, they
will, like the other probes, be limited by systematic errors. As the
forecasts show, competitive contributions to joint dark energy
parameter inferences correspond to sub-percent precision in
characteristic distance, which implies that the residual systematic
error in distance needs to be well below 1\%. This residual systematic
may or may not be present at this level in every lens system -- what
matters is the bias in the overall measurement from the combined
sample. However, the term ``mean accuracy per lens'' is helpful, since
it reminds us that systematic errors can affect all members of a
sample in the same (or at least similar) way.  In this section we
revisit the primary sources of systematic error and assess the
prospects for this stringent requirement to be met.

\subsubsection{Time delay measurement}

\citet{LiaoEtal2015} showed that a
mean accuracy of 0.1\% per lens would already be achievable in
plausible samples of several hundred LSST lenses, were all images to
be taken in the same band.  While these results are encouraging,
questions about our ability to measure time delays from sparse,
multi-filter light curves extracted from realistic images remain
\citep{TCM13}.  Time delay measurement accuracy from LSST multi-filter
light curve could be tested by a second challenge; the success of the
analyses is likely to hinge on the treatment of quasar color
variability \citep[see e.g.\ ][and references
therein]{Sch++12,SunEtal2014} and chromatic microlensing
\citep[see e.g.][and references therein]{HainlineEtal2013}.
Joint inference from whole samples is likely to
be important, in mitigating against both outliers and also imprecision
arising from inappropriate uniform priors on  population parameters.
Insights into AGN physics and the stellar composition  of lens galaxies
would be welcome by-products of such an analysis. An alternative approach
could be to continue to pursue single-filter monitoring, but at
increased efficiency. Experiments with higher cadence campaigns,
exploiting the short (sub-day) timescale variability of AGN are in
progress \citep[][F.~Courbin, priv.\ comm.]{BorosonEtal2016}.

It is worth noting that time delay {\it perturbations},  due to small
scale structure in lens plane or along line of sight, will likely not be
a significant  additional source of systematic error, since they
primarily cause additional scatter on hour-long timescales
\citep{K+M09}.

\subsubsection{Lens mass modeling}

\citet{S+S13} have pointed out the possibility of systematic errors at
the twenty percent level due to modeling assumptions (and their
interaction with the mass sheet degeneracy) when fitting lensing data
alone. \citet{Suy++14} fitted the same two models used by
\citet{S+S13}  to the current state of the art Einstein ring imaging and
lens galaxy velocity dispersion data, and found that present
measurements of stellar kinematics reduce the error by a factor of
10, at least within their framework of pixelized source
reconstruction.  It is not clear how much of the residual 2\%
uncertainty is random, and how much residual bias there is: we do not
yet know how our mass modeling methods respond to the variety of lens
mass distributions we expect.

An important first step has been taken by \citet{XuEtal2016}, who
looked at the density profiles of a sample of mock galaxies from the
Illustris simulation, finding significant departures from simple power
law profiles. However, they note that at the mass scales typical of
galaxy scale lenses, where the total mass density profiles happen to
be well approximated by isothermal spheres
\citep{Koo++09,Aug++10}, the residual systematic uncertainty
in the Hubble constant could be restricted to a few percent.
Again, it remains to be verified how much of this averages out
given particular, large samples of systems.

As a result of these investigations, we know that 1) the dynamical
information is as important as the lensing data, 2) more complex
models than simple power law density profiles will likely be needed to
enable sub-percent accuracy to be reached, and 3) we now have
simulated galaxies that are {\it sufficiently realistic and suitably
complex} that we can carry out meaningful tests where the ground truth
is very different from our assumed analysis models. The acid test will
be whether we can recover the input cosmological parameters from
realistic simulated high resolution imaging and stellar spectroscopic
data made using numerical simulations that resolve massive
galaxies well \citep[e.g.,][]{Fia++16}.

As we look ahead to samples of dozens to hundreds of lenses in the
next decade, we can also consider the observational capabilities we
will have in that time period. Giant segmented mirror telescopes
(including the James Webb Space Telescope as well as planned 30m class
ground-based telescopes) will bring up to a factor of 10 increase in
angular resolution beyond what HST and today's 10-m class adaptive
optics-enabled telescopes can deliver, improving further the available
Einstein ring constraints. These facilities will be instrumented with
Integral Field Unit spectrographs that will provide
spatially-resolved spectroscopy of the lens galaxy stellar populations.
It is yet to be seen how accurately lens mass distributions can be
modeled with such data: extending the realistic lens galaxy data
simulation work to include them would seem to be very important. The
``crash test'' of \citet{Bar++09a} was an excellent start in this
direction, probing as it did the performance of a fully self-consistent
lensing and dynamics modeling code on a numerically simulated lens
galaxy mock dataset.

The upcoming increase in available precision per lens will support
significantly more flexible mass models, as reviewed in
Section~\ref{ssec:lensmodel}.  The opportunity here is to find a
flexible mass model whose parameters can be taken to have been drawn
from a relatively simple prior PDF, which could be derived from either
large samples of observed non-lens galaxies, plausible hydrodynamic
simulated galaxies, or both.

The mass sheet degeneracy, and indeed all model parameter degeneracies
are broken by incorporating more information, but this needs to be
done in such a way as to not introduce bias. Using flexible mass
models with reasonably broad but not uninformative priors is the first
step, but unless these priors are themselves movable, the introduced
bias might remain. The clear-cut solution is to learn the
hyper-parameters that govern the intrinsic distribution of mass model
parameters from the data as well. It is really the prior on these
``hyper-parameters'' that needs to come from simulations. An initial
attempts at this kind of ``hierarchical inference'' can be found in
the analysis of \citet{SonnenfeldEtal2015}, where the authors infer
the values of some 28 hyper-parameters assumed to govern the scaling
relations between massive galaxies, as well as the selection function
of the lens sample. As surveys yield larger and larger samples of
lenses, joint inferences from ensembles
of all lenses (time delay and otherwise) will bring in more
information about the density structure of somewhat self-similar
massive galaxies.

In addition to carrying out tests on simulated data, it is important
to continue empirical investigations of systematic errors. In addition
to the generally applicable strategy of comparing the cosmological
parameter estimates between individual systems or subsets of systems
to measure whether statistical uncertainties are under-estimated, we
must continue to look for other independent tests. An interesting
example is that of the multiply-imaged supernova ``Refsdal.'' Several teams
carried out modeling analyses of this lens system, in order
to predict in a truly blind fashion the
magnification, timing, and position of the next appearance of the
supernova \citep{Kel++15,Ogu15,S+J15,Jau++16,Tre++16,Kaw++16,Gri++16}. Even
though the deflector is a merging cluster, and thus significantly more
challenging to model than the typical relaxed elliptical galaxies used
in time-delay cosmography, several teams managed to predict the event
\citep{Tre++16,Kel++16} within the estimated model prediction uncertainties.
It is particularly re-assuring that the code {\sc GLEE} \citep{S+H10},
which was designed and used extensively to model time delay lenses for
cosmography \citep{Suy++10,Suy++13,Suy++14}, performed very well
\citep{Gri++16,Kel++16}, along with other methods based on similar assumptions \citep{Kaw++16}. As the precision of time delay cosmography increases
with sample size it will be important to seize any new opportunity to
carry out additional blind tests, e.g. by predicting time delays of
lens quasars before measuring them, and by actively searching for
multiply imaged supernovae in galaxy-scale lenses \citep{O+M10}.

\subsubsection{Environment and line of sight characterization}

The current methodology (Section~\ref{ssec:los}) includes dependencies
on both cosmological simulations and reference imaging surveys. The
Stage~III and~IV wide field surveys will help with the latter,
providing much larger, more homogeneous sets of control fields.
Systematic errors associated with calibrating against simulations is
the bigger problem, and both the number counts and 3D reconstruction
approaches that have been implemeted to date are affected.
\citet{CollettEtal2013} give some indication of the magnitude of the
issue, finding a bias of 3\% in the average inferred time delay distance
when assuming a stellar mass to halo mass relation that is incorrect
but still consistent with other observations. This reduces to 1-2\%
if the bright galaxies in the lens fields have spectroscopic redshifts,
suggesting that this kind of data will continue to be important.

The 3D reconstruction approach can, in principle, be made to be
independent from simulations \citep[indeed, this was a design feature
of][]{McCullyEtal2014}.  However, more information about mass in the
Universe will be required.  Both \citet{McCullyEtal2014} and
\citet{CollettEtal2013} use halo models, with very simplistic treatments
of the mass outside of halos, and the voids between them. Both weak
shear data and clustering information could be used to improve the
accuracy of these  models; statistical halo models are already well
constrained by summary statistics from these probes
\citep[e.g.][]{CouponEtal2015},  and these results are already
potentially useful (although the scatter in the model's relations
will likely need to be taken into account).  Covariance with cosmic
shear, galaxy clustering, and the halo mass function may then have to
be accounted for in any joint cosmological parameter inferences; this
may turn out to be negligible, once quantified.  Some
mitigation of the environment and line of sight systematics could be
achieved by selecting low density lines of sight
\citep{CollettEtal2013}, but this selection would have to be done with
some care, propagating all the uncertainties.

A key point made by
\citeauthor{McCullyEtal2014}~(\citeyear{McCullyEtal2014},~\citeyear{McCullyEtal2016})
is that the mass structures external to the primary lens should, in
principle, be included in the lens model itself.  Doing this would
allow the correct multiple lens plane formalism to be employed,
thereby reducing the systematic error introduced by the single lens
plane approximation. This approach is being actively pursued in the
ongoing analyses (K.~Wong, priv.\ comm.). An interesting feature of
multiple plane lensing is that the appearance of the lens galaxy is
also affected by weak lensing due to foreground mass structures: this
may well need to be taken into account when striving for high accuracy
modeling of the primary lens (R.Blandford \& S.~Birrer, priv.\ comm.).

Even when all the above systematics have been investigated and tested
for, others that are unforeseen may remain. Strategies for detecting
these ``unknown unknowns'' include jack-knife testing, which  will
become possible with larger samples. Other kinds of ``null tests'' may
also be possible when we are out of the small number stastistics regime:
research is needed on developing such tests. The ultimate test is
cosmological parameter consistency with other datasets, but for this
comparison to be meaningful the analysis of each dataset must be done
blindly, to avoid unconscious experimenter bias and the resulting
groupthink towards (or away from) concordance
\citep[see e.g.][]{Con++06,Suy++13}. In principle all
systematics tests need to be done before unblinding; as a result, end
to end tests on highly realistic mock data will become ever more
important. The time delay challenge was carried out blind;
similarly-designed lens modeling and environment characterization
challenges are called for too. Success at blind cosmological parameter
recovery from realistic mock samples is the surest way to generate
confidence in a probe's accuracy.


\subsection{Roadmap}
\label{ssec:roadmap}

We conclude the outlook section by proposing an ambitious, yet (in our
opinion) feasible roadmap for time delay cosmography in the next
decade. This roadmap aims to achieve $\sim0.5\%$ precision on time delay
distance\footnote{For simplicity, we refer to equivalent uncertainty
on an average time delay distance at the typical redshift of the
deflector and source. In practice of course, there will be a
distribution of redshifts and thus of individual distances. As the
way in which the time delay distance depends on cosmological parameters
varies slightly with redshift, the analysis of a real sample of
lenses will have the added benefit of breaking some of the
degeneracies between the cosmological parameters,
and reducing the uncertainties more rapidly than if all
the systems were at the same redshift.} by 2027 (by which time
the 5-year LSST light curves should be in hand),
building on the tools and techniques demonstrated in
the past 15 years and exploiting the large scale surveys that are
currently under way or planned. It is based on a specific strategy,
consisting of constructing the most precise and accurate models of
each lens based on rich datasets for each system (alternative
strategies are discussed at the end of this section). Specifically,
for each system one needs the following:

\begin{enumerate}
\item Time delays precise to better than $3\%$;
\item High resolution imaging (resolution much better than the Einstein
radius) with point spread function known well enough to reconstruct the
differences in Fermat potential to better than $3\%$ precision;
\item Spectroscopic redshifts of the deflector and the source;
\item Stellar velocity dispersion of the deflector to better than 10\%
precision, possibly spatially resolved;
\item Imaging and spectroscopic data sufficient to characterize the
weak lensing effects due to structure along the line of sight to better than
$3\%$ precision.
\end{enumerate}

These targets can be met with present technology, as it has already
been demonstrated for a few systems \citep{Tew++13,Suy++13}, and the
observational requirements have been investigated for a variety of
telescopes and configurations
\citep{Gre++13,CollettEtal2013,Men++15,Lin15}.

\begin{figure*}
\includegraphics[width=0.98\textwidth]{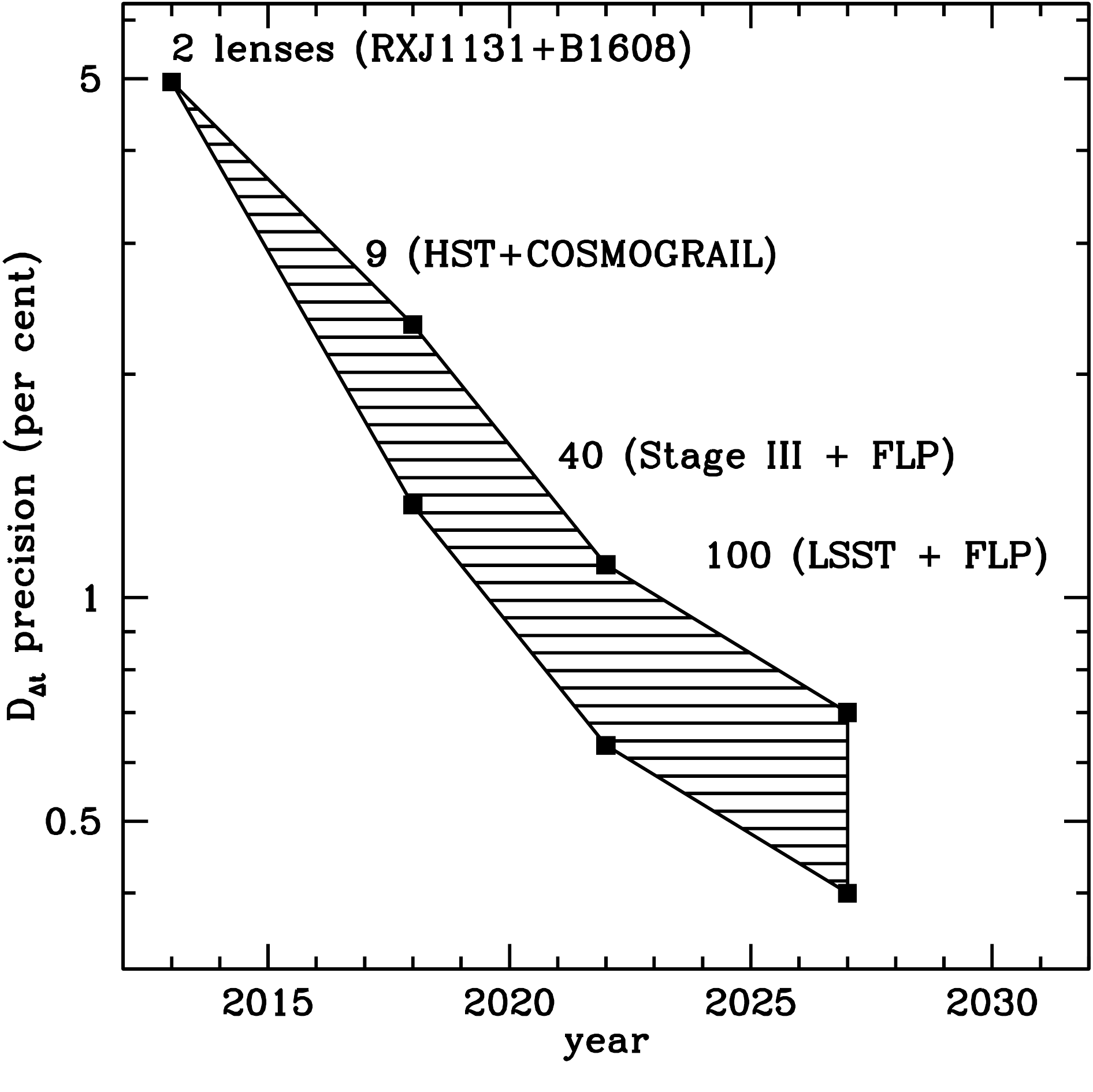}
\caption{Roadmap for time delay cosmography. The shaded region
represents the estimated range of uncertainty attainable on the
effective (ensemble) time delay distance $\Ddt$ as a
function of time, over the next ten years. Points on the roadmap are
labeled by the types of survey and
follow-up (FLP) observation required.}
\label{fig:roadmap}
\end{figure*}

The proposed roadmap is summarized in Figure~\ref{fig:roadmap}. The
shaded region represents an estimate of the ensemble precision attainable on
$\Ddt$,\footnote{The time delay distance referred to here is the same as
ensemble average quantity that \citet{C+M09b} call $\mathcal{\tau}_{\rm C}$.}
ranging from the most conservative to the most favorable
scenario. The most conservative case assumes only a central velocity
dispersion measurement for each system, while the most favorable
scenario involves spatially resolved stellar velocity dispersions,
obtained either from space or from the ground assisted by adaptive optics
(Agnello et al.\ in prep.). We neglect the additional independent
information that in principle can be obtained via the angular
diameter distance dependence \citep{JeeEtal2016}.  As
discussed in Section~\ref{ssec:precision}, this additional piece of
information would in principle improve the constraints on cosmological
parameters from time delay lenses, so this envelope should be regarded
as a conservative estimate.

The roadmap is divided into steps, whose timing is dictated by
available observational facilities.
The first step in the roadmap is the analysis
of two systems that was published in 2013, and was based on COSMOGRAIL time
delays, HST imaging, Keck spectroscopy, and other ancillary data from
a variety of sources.

The second step consists of the full analysis of nine lens systems for
which time delays have been measured by the COSMOGRAIL team, and for
which HST imaging is being completed this year (HST-GO-14254).
Completing the second step by the launch of the James Webb Space
Telescope at the end of 2018, and in doing so delivering $\sim2\%$
distance precision, would be ideal, so as to provide a useful
comparison for expected improvements in local distance ladder
measurements.

The third step will require the discovery of new lenses, in addition
to the usual follow-up (FLP) effort. Systematic searches are currently
under way based on large scale surveys such as the Dark Energy Survey
or Hyper-Suprime-Camera Survey \citep{Agn++15,Mor++16}, and should
discover hundreds of new lensed quasar systems by the end of the
decade \citep{O+M10}, while providing high quality photometric lens
environment information from the survey itself.  Focused follow-up of
a carefully selected sample of systems should be sufficient to reach
the goal for step 3, that is, $\sim1\%$ distance precision from 40
lenses in 5 years (i.e. by 2022). The selected sample should consist
of as many quads as possible, since they contain more cosmological
information and favor systems with time delays in the range 50-150
days, such that they are measurable at the $<3\%$ level in one
observing season with daily cadence. It is worth noting that in the
high-quality follow-up approach each individual system provides a
highly informative measurement of cosmology, and therefore this method
is very robust with respect to the precise selection function imposed
by the search and monitoring algorithms \citep{C+C16}, at the level of
accuracy required in this step . The observational bottle necks are
likely going to be the time delay measurements, which will require
dedicated monitoring on 1--4m class telescopes, and high resolution
imaging \citep{Tre++13}.

Scaling up to the fourth step will require a change of strategy, in
order to cope with the intrinsically fainter targets and larger sample
size. One natural strategy will consist of using time delays measured
from LSST light curves \citep{LiaoEtal2015}, perhaps supplemented in
part from 1--4m telescopes in order to increase the cadence, or potentially
from a dedicated lens monitoring satellite
\citep{Mou++08}. Unfortunately, LSST imaging will be insufficient for
detailed modeling, and higher resolution imaging will be required
\citep{Men++15}.
Planned surveys like Euclid and WFIRST will be excellent at
discovering new lenses, but probably will have insufficient depth and
resolution except for the brightest systems. Therefore targeted
follow-up will be required, achievable either with JWST from space, or
with improved adaptive optics systems on 8--10m class ground-based
telescopes \citep{Mar++07,Che++16,Rus++16}. Integral field
spectrographs on giant segmented mirror telescopes will be the ideal
complement to LSST, by providing the necessary high resolution imaging
and spectroscopy with relatively short exposure times
\citep[e.g.][]{Ski++15}.
Much of the information needed for lens environment characterization will again
come from the surveys themselves, although synergy with spectroscopic
surveys should be explored to increase the redshift accuracy.
This fourth step
aims to reach $\sim0.5\%$ precision, through follow-up of systems discovered
and monitored in the first five years of the LSST survey.

\begin{figure*}
\begin{center}
\includegraphics[height=5.5cm,clip]{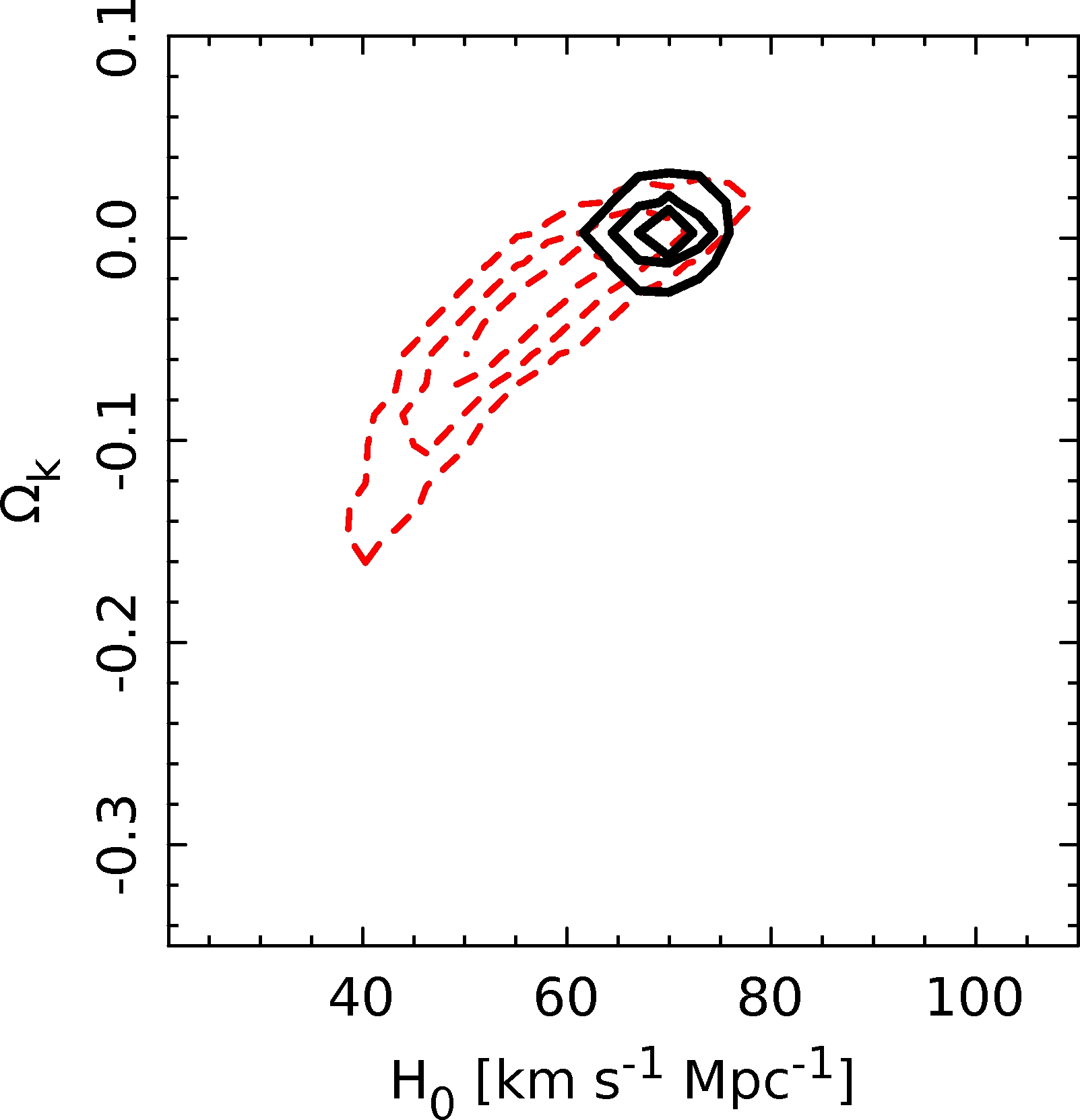}
\includegraphics[height=5.5cm,clip]{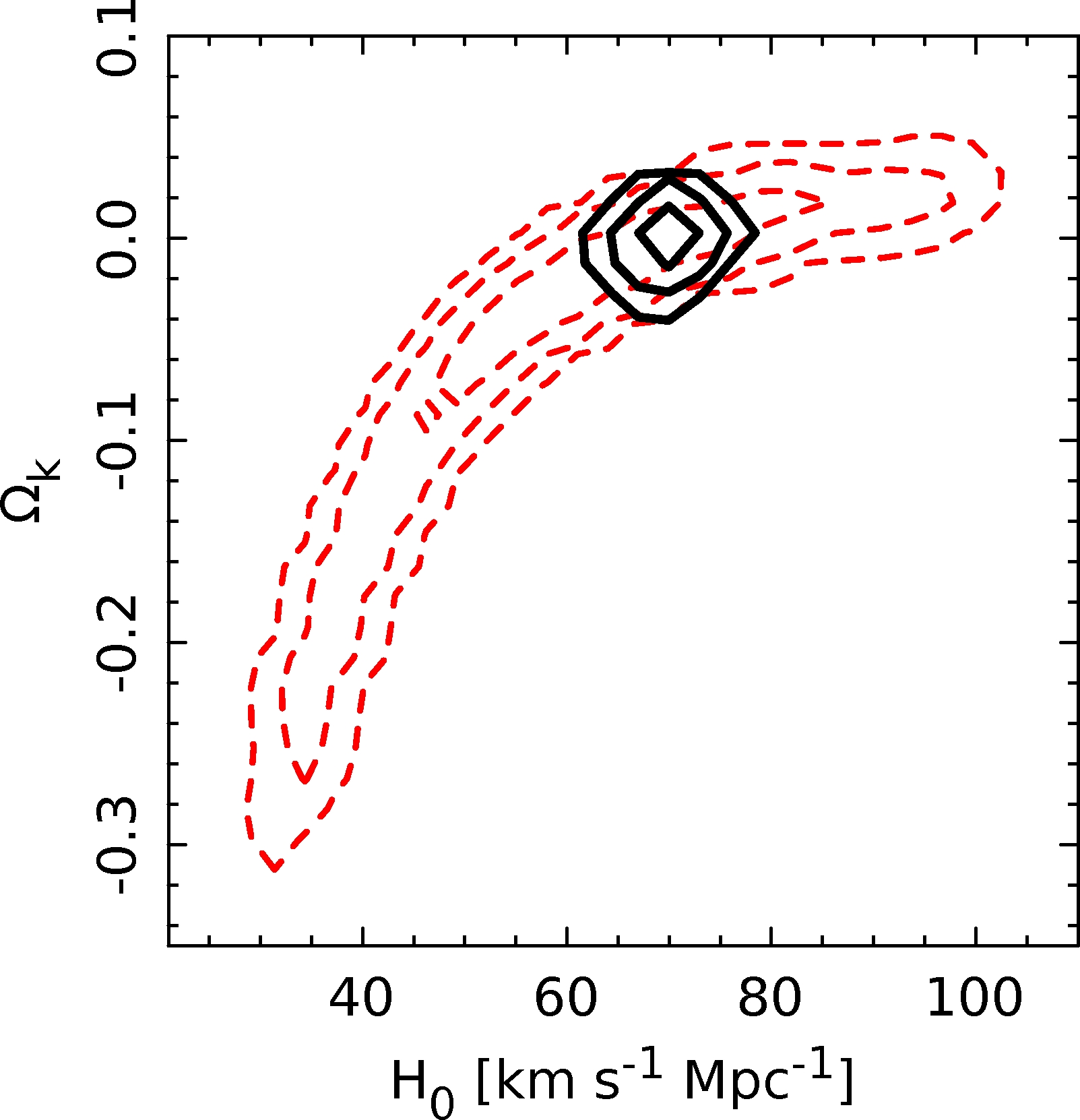}
\includegraphics[height=5.5cm,clip]{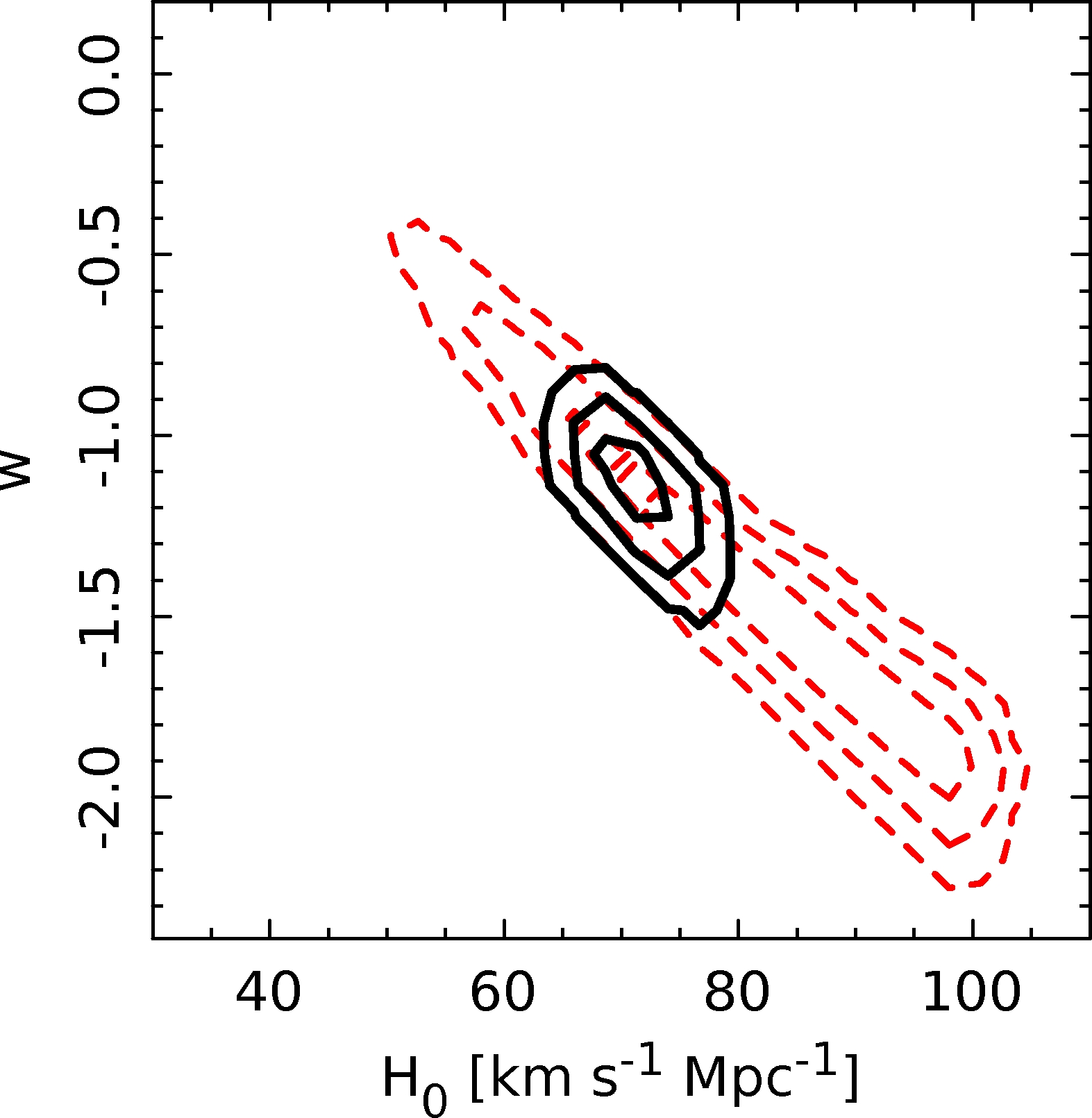}
\includegraphics[height=5.5cm,clip]{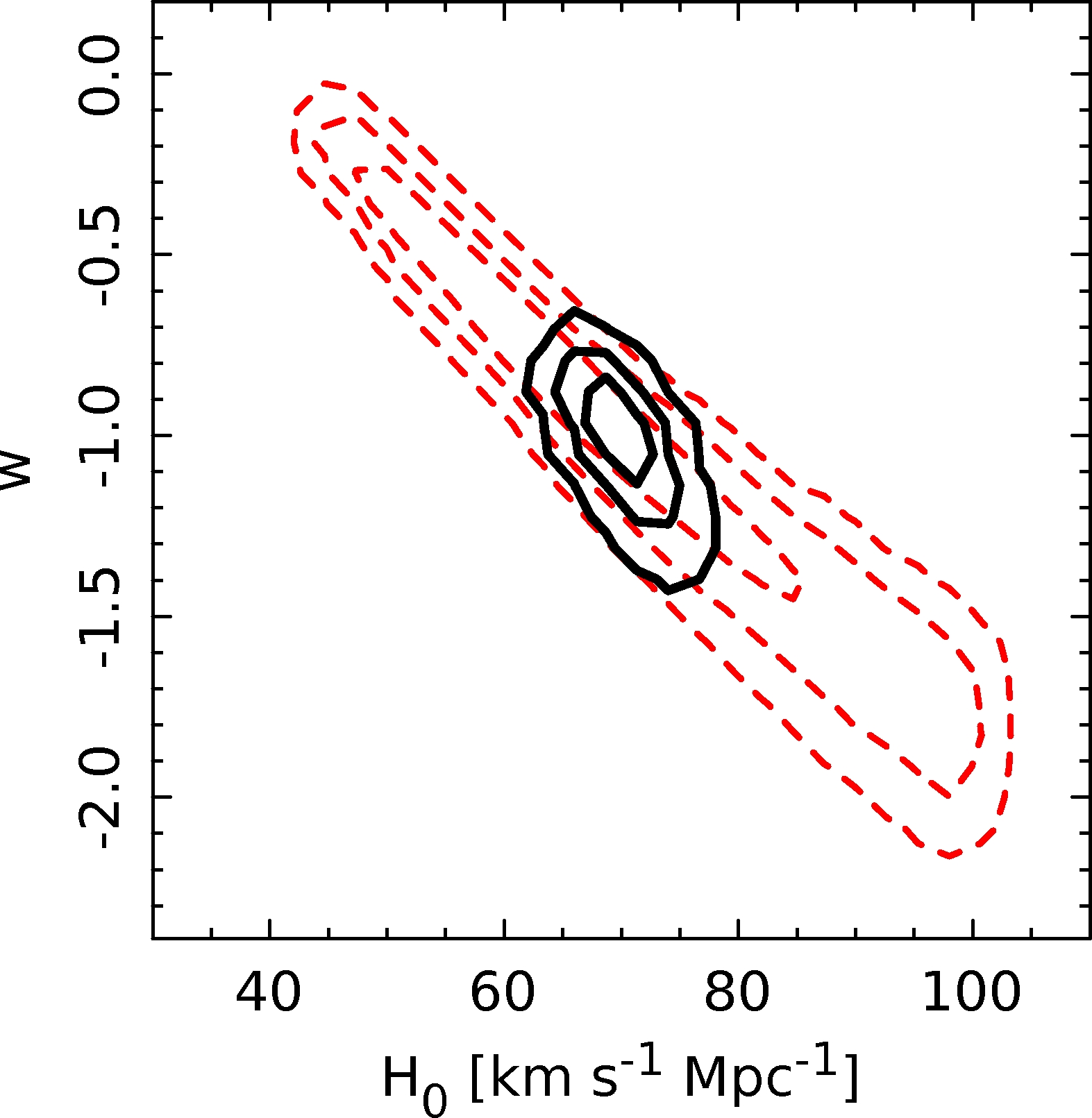}
\end{center}
\caption{Cosmological forecast (black solid lines, represent the 68\%,
95\%, and 99\% posterior probability contours) for step 2 in the roadmap
(see Figure~\ref{fig:roadmap} and Section~\ref{ssec:roadmap} for
details), assuming one parameter extensions of flat $\Lambda$CDM, and
Planck (left column) and WMAP9 (right column) priors (red dashed lines).
Figure courtesy of S.H.~Suyu.}
\label{fig:roadmap-step2}
\end{figure*}

Converting the uncertainty
in time delay distance $\Ddt$ to cosmological parameters requires
specific assumptions about the cosmological model and priors from
independent measurements. For step 1, the equivalent precision on H$_0$
using WMAP7 prior in one parameter extensions of $\Lambda$CDM is 4-5\%.
The forecast for step 2 with WMAP9 and Planck priors is shown in
Figure~\ref{fig:roadmap-step2}. For steps 3 and 4, the equivalent
precision on H$_0$ is in the range 1.1-1.3\% and 0.8\%-1.0\%
respectively, assuming ``Planck + Stage II'' priors \citep{C+M09b}.

In addition to the observational challenges, the main challenge in
pursuing this roadmap is likely to be the analysis cost. Lens modeling
is at present fairly labor intensive, requiring several months of work
per system by an expert modeler. This high labor cost is chiefly due
to code development. Up to now, the analysis of each single lens has
required the development and testing of new features (e.g. multi-plane
lensing, point spread function reconstruction; Suyu et al. 2016, in
prep.). In order to analyze the future large samples, the analysis
codes will have to transition from development to production, reducing
substantially the investigator time per system. Distributing the work
among a large team of modelers will likely be necessary to speed
things up and keep modeling uncertainties in check.

Naturally, this proposed roadmap is not the only possible way
forward. As discussed earlier in this section, several authors have
proposed the analysis of larger samples of lenses, each with fewer
ancillary data and thus lower precision per system. Alternatively, one
could imagine a hybrid strategy in which a subset of the lenses are
analyzed in great detail with lots of ancillary data, and the lessons
learned from that subset are propagated to a large sample through
judicious use of priors.


\section{Summary}
\label{sec:summary}

We reviewed gravitational time delays as a tool for measuring
cosmological parameters. In addition to giving a brief introduction to
the theoretical underpinnings of the method, we discussed the past
history of the field, before turning to present day accomplishments
and the challenges ahead. The main points of this review can be
summarized as follows:

\begin{enumerate}
\item From a theoretical point of view gravitational time delays are a
clean and well understood probe of cosmic acceleration. Conceptually,
each time delay measurement provides a one step measurement of
absolute distance. The typical redshifts of deflectors and sources
span the range between $z\sim2$ and today, covering the era of cosmic
time during which dark energy rose to prominence.
\item Even though the potential cosmological application of strongly
lensed, time-variable sources was recognized as early as 1964, it took
decades for the method to come to practical fruition. Two sets of
challenges have been overcome over the past 15 years. Observationally, the
main challenge has been organizing long term monitoring campaigns and
mustering the range of resources required to constrain accurate mass
models. Theoretically, the main challenge consisted of learning how to
exploit the available information to construct lens models with
realistic estimates of the uncertainties.
\item It has been demonstrated through blind measurements that each
individual system can deliver a measurement of absolute distance to
about 6-7\% total uncertainty, given current data quality.   The power of
the method is currently limited by the number of systems with
well-measured time delays and sufficient ancillary data to carry out
detailed modeling ($\lesssim10$ at the time of writing).
\item Systematic searches for strongly lensed quasars are under way and
should be able to increase the cosmographic sample size by more than
an order of magnitude in the next decade. With improvements in
follow-up image resolution and spatially resolved spectroscopy as
well, we can aspire to sub-percent precision in the Hubble constant by
the middle of the next decade.
\item Before LSST, dedicated monitoring campaigns will be required to
measure each time delay; LSST can potentially alter the landscape, if
it can deliver hundreds of time delays from the survey data themselves.
\item Throughout the next decade in order to deliver the
available precision it will be necessary to obtain a small amount of
high quality follow-up data in order to minimize the uncertainty per
system. These include: high resolution imaging from space or with
adaptive optics; redshifts; stellar velocity dispersions, and
spatially resolved kinematics of the deflectors. These data can be
obtained from the James Webb Space Telescope or large and extremely
large ground based telescopes with adaptive optics.
\item  
As samples increase, further work will be needed to
understand, quantify and mitigate against potential systematic errors
in the method. Extensive parameter recovery tests on realistic
simulated monitoring, high resolution imaging, spatially resolved
spectroscopy, and field weak lensing and photometry will be essential
to ensure that systematic errors are kept subdominant, the precision
of the method is realized, and an accurate cosmological measurement
is achieved.
\end{enumerate}

In gravitational time delays we have a theoretically sound,
experimentally competitive, and cost-effective cosmographic tool.
Like every other probe, a lot of hard work will be necessary to reach
the sub-percent level of precision and accuracy that is needed to make
progress. This effort seems well motivated, not only by the ultimate
goal of improving our understanding of the fundamental constituents of
the universe, but also by the opportunity to use lensed quasars to
learn about the astrophysics of dark matter
\citep{M+M01,D+K02,Metcalf:2005p1203,Xu++09,Veg++14,Nie++14}, active galactic
nuclei \citep{PMK08,Eig++08a,Eig++08b,Blackburne:2010p6600,Mac++15},
and stars
\citep{Sch++14}.



%
%


\begin{acknowledgements}
We are grateful to S.~Suyu and E.~Komatsu for insightful discussions
about the cosmological distance information content of time delay
lenses, and to S.~Suyu for making the B1608$+$656 MCMC chains
available for us to make Figure~\ref{fig:DdDdt}.
We thank A.~Agnello, S.~Birrer, V.~Bonvin, D.~Coe, T.~Collett,
F.~Courbin, I.~Jee, C.~Kochanek, E.~Linder, D.~Sluse, S.Suyu, and
M. Tewes for very valuable feedback on a draft of this review.
T.T. thanks the Packard Foundation for generous support through a
Packard Research Fellowship, the NSF for funding through NSF grant
AST-1450141, ``Collaborative Research: Accurate cosmology with strong
gravitational lens time delays''.
P.J.M. acknowledges support from the U.S.\ Department of Energy under
contract number DE-AC02-76SF00515.
\end{acknowledgements}



\begin{thebibliography}{174}
\providecommand{\natexlab}[1]{#1}
\providecommand{\url}[1]{{#1}}
\providecommand{\urlprefix}{URL }
\expandafter\ifx\csname urlstyle\endcsname\relax
  \providecommand{\doi}[1]{DOI~\discretionary{}{}{}#1}\else
  \providecommand{\doi}{DOI~\discretionary{}{}{}\begingroup
  \urlstyle{rm}\Url}\fi
\providecommand{\eprint}[2][]{\url{#2}}

\bibitem[{{Aghamousa} and {Shafieloo}(2015)}]{A+S2015}
{Aghamousa} A, {Shafieloo} A (2015) {Fast and Reliable Time Delay Estimation of
  Strong Lens Systems Using the Smoothing and Cross-correlation Methods}. \apj
  804:39, \doi{10.1088/0004-637X/804/1/39}, \eprint{1410.8122}

\bibitem[{{Agnello} et~al(2015){Agnello}, {Treu}, {Ostrovski}, {Schechter},
  {Buckley-Geer}, {Lin}, {Auger}, {Courbin}, {Fassnacht}, {Frieman},
  {Kuropatkin}, {Marshall}, {McMahon}, {Meylan}, {More}, {Suyu}, {Rusu},
  {Finley}, {Abbott}, {Abdalla}, {Allam}, {Annis}, {Banerji},
  {Benoit-L{\'e}vy}, {Bertin}, {Brooks}, {Burke}, {Rosell}, {Kind},
  {Carretero}, {Cunha}, {D'Andrea}, {da Costa}, {Desai}, {Diehl}, {Dietrich},
  {Doel}, {Eifler}, {Estrada}, {Neto}, {Flaugher}, {Fosalba}, {Gerdes},
  {Gruen}, {Gutierrez}, {Honscheid}, {James}, {Kuehn}, {Lahav}, {Lima}, {Maia},
  {March}, {Marshall}, {Martini}, {Melchior}, {Miller}, {Miquel}, {Nichol},
  {Ogando}, {Plazas}, {Reil}, {Romer}, {Roodman}, {Sako}, {Sanchez},
  {Santiago}, {Scarpine}, {Schubnell}, {Sevilla-Noarbe}, {Smith},
  {Soares-Santos}, {Sobreira}, {Suchyta}, {Swanson}, {Tarle}, {Thaler},
  {Tucker}, {Walker}, {Wechsler}, and {Zhang}}]{Agn++15}
{Agnello} A, {Treu} T, {Ostrovski} F, {Schechter} PL, {Buckley-Geer} EJ, {Lin}
  H, {Auger} MW, {Courbin} F, {Fassnacht} CD, {Frieman} J, {Kuropatkin} N,
  {Marshall} PJ, {McMahon} RG, {Meylan} G, {More} A, {Suyu} SH, {Rusu} CE,
  {Finley} D, {Abbott} T, {Abdalla} FB, {Allam} S, {Annis} J, {Banerji} M,
  {Benoit-L{\'e}vy} A, {Bertin} E, {Brooks} D, {Burke} DL, {Rosell} AC, {Kind}
  MC, {Carretero} J, {Cunha} CE, {D'Andrea} CB, {da Costa} LN, {Desai} S,
  {Diehl} HT, {Dietrich} JP, {Doel} P, {Eifler} TF, {Estrada} J, {Neto} AF,
  {Flaugher} B, {Fosalba} P, {Gerdes} DW, {Gruen} D, {Gutierrez} G, {Honscheid}
  K, {James} DJ, {Kuehn} K, {Lahav} O, {Lima} M, {Maia} MAG, {March} M,
  {Marshall} JL, {Martini} P, {Melchior} P, {Miller} CJ, {Miquel} R, {Nichol}
  RC, {Ogando} R, {Plazas} AA, {Reil} K, {Romer} AK, {Roodman} A, {Sako} M,
  {Sanchez} E, {Santiago} B, {Scarpine} V, {Schubnell} M, {Sevilla-Noarbe} I,
  {Smith} RC, {Soares-Santos} M, {Sobreira} F, {Suchyta} E, {Swanson} MEC,
  {Tarle} G, {Thaler} J, {Tucker} D, {Walker} AR, {Wechsler} RH, {Zhang} Y
  (2015) {Discovery of two gravitationally lensed quasars in the Dark Energy
  Survey}. \mnras 454:1260--1265, \doi{10.1093/mnras/stv2171},
  \eprint{1508.01203}

\bibitem[{{Auger} et~al(2007){Auger}, {Fassnacht}, {Abrahamse}, {Lubin}, and
  {Squires}}]{Aug++07}
{Auger} MW, {Fassnacht} CD, {Abrahamse} AL, {Lubin} LM, {Squires} GK (2007)
  {The Gravitational Lens-Galaxy Group Connection. II. Groups Associated with
  B2319+051 and B1600+434}. \aj 134:668--679, \doi{10.1086/519238},
  \eprint{arXiv:astro-ph/0603448}

\bibitem[{{Auger} et~al(2010){Auger}, {Treu}, {Bolton}, {Gavazzi}, {Koopmans},
  {Marshall}, {Moustakas}, and {Burles}}]{Aug++10}
{Auger} MW, {Treu} T, {Bolton} AS, {Gavazzi} R, {Koopmans} LVE, {Marshall} PJ,
  {Moustakas} LA, {Burles} S (2010) {The Sloan Lens ACS Survey. X. Stellar,
  Dynamical, and Total Mass Correlations of Massive Early-type Galaxies}. \apj
  724:511--525, \doi{10.1088/0004-637X/724/1/511}, \eprint{1007.2880}

\bibitem[{{Bar-Kana}(1996)}]{Bar96}
{Bar-Kana} R (1996) {Effect of Large-Scale Structure on Multiply Imaged
  Sources}. \apj 468:17, \doi{10.1086/177666}, \eprint{astro-ph/9511056}

\bibitem[{{Barnab{\`e}} et~al(2009){Barnab{\`e}}, {Nipoti}, {Koopmans},
  {Vegetti}, and {Ciotti}}]{Bar++09a}
{Barnab{\`e}} M, {Nipoti} C, {Koopmans} LVE, {Vegetti} S, {Ciotti} L (2009)
  {Crash-testing the CAULDRON code for joint lensing and dynamics analysis of
  early-type galaxies}. \mnras 393:1114--1126,
  \doi{10.1111/j.1365-2966.2008.14208.x}, \eprint{0808.3916}

\bibitem[{{Bartelmann}(2010)}]{Bar10}
{Bartelmann} M (2010) {TOPICAL REVIEW Gravitational lensing}. Classical and
  Quantum Gravity 27(23):233001, \doi{10.1088/0264-9381/27/23/233001},
  \eprint{1010.3829}

\bibitem[{{Bennett} et~al(2013){Bennett}, {Larson}, {Weiland}, {Jarosik},
  {Hinshaw}, {Odegard}, {Smith}, {Hill}, {Gold}, {Halpern}, {Komatsu}, {Nolta},
  {Page}, {Spergel}, {Wollack}, {Dunkley}, {Kogut}, {Limon}, {Meyer}, {Tucker},
  and {Wright}}]{WMAP9}
{Bennett} CL, {Larson} D, {Weiland} JL, {Jarosik} N, {Hinshaw} G, {Odegard} N,
  {Smith} KM, {Hill} RS, {Gold} B, {Halpern} M, {Komatsu} E, {Nolta} MR, {Page}
  L, {Spergel} DN, {Wollack} E, {Dunkley} J, {Kogut} A, {Limon} M, {Meyer} SS,
  {Tucker} GS, {Wright} EL (2013) {Nine-year Wilkinson Microwave Anisotropy
  Probe (WMAP) Observations: Final Maps and Results}. \apjs 208:20,
  \doi{10.1088/0067-0049/208/2/20}, \eprint{1212.5225}

\bibitem[{{Birrer} et~al(2015{\natexlab{a}}){Birrer}, {Amara}, and
  {Refregier}}]{BirrerEtal2015}
{Birrer} S, {Amara} A, {Refregier} A (2015{\natexlab{a}}) {Gravitational Lens
  Modeling with Basis Sets}. \apj 813:102, \doi{10.1088/0004-637X/813/2/102},
  \eprint{1504.07629}

\bibitem[{{Birrer} et~al(2015{\natexlab{b}}){Birrer}, {Amara}, and
  {Refregier}}]{BAR16}
{Birrer} S, {Amara} A, {Refregier} A (2015{\natexlab{b}}) {The mass-sheet
  degeneracy and time-delay cosmography: Analysis of the strong lens
  RXJ1131-1231}. JCAP, submitted \eprint{1511.03662}

\bibitem[{Blackburne et~al(2010)Blackburne, Pooley, Rappaport, and
  Schechter}]{Blackburne:2010p6600}
Blackburne JA, Pooley D, Rappaport S, Schechter PL (2010) Sizes and temperature
  profiles of quasar accretion disks from chromatic microlensing. arXiv
  astro-ph.CO, \urlprefix\url{http://arxiv.org/abs/1007.1665v1},
  \eprint{1007.1665v1}

\bibitem[{{Blandford} and {Narayan}(1992)}]{B+N92}
{Blandford} RD, {Narayan} R (1992) {Cosmological applications of gravitational
  lensing}. \araa 30:311--358, \doi{10.1146/annurev.aa.30.090192.001523}

\bibitem[{{Bonvin} et~al(2016){Bonvin}, {Tewes}, {Courbin}, {Kuntzer}, {Sluse},
  and {Meylan}}]{BonvinEtal2016}
{Bonvin} V, {Tewes} M, {Courbin} F, {Kuntzer} T, {Sluse} D, {Meylan} G (2016)
  {COSMOGRAIL: the COSmological MOnitoring of GRAvItational Lenses. XV.
  Assessing the achievability and precision of time-delay measurements}. \aap
  585:A88, \doi{10.1051/0004-6361/201526704}, \eprint{1506.07524}

\bibitem[{{Boroson} et~al(2016){Boroson}, {Moustakas}, {Romero-Wolf}, and
  {McCully}}]{BorosonEtal2016}
{Boroson} TA, {Moustakas} LA, {Romero-Wolf} A, {McCully} C (2016) {Using the
  LCOGT Network To Measure a High-Precision Time Delay in the Four-Image
  Gravitational Lens HE0435-1223}. In: American Astronomical Society Meeting
  Abstracts, American Astronomical Society Meeting Abstracts, vol 227, p 338.04

\bibitem[{{Brewer} and {Lewis}(2006)}]{BrewerAndLewis2006}
{Brewer} BJ, {Lewis} GF (2006) {The Einstein Ring 0047-2808 Revisited: A
  Bayesian Inversion}. \apj 651:8--13, \doi{10.1086/507475},
  \eprint{astro-ph/0606714}

\bibitem[{{Browne} et~al(2003)}]{Bro++03}
{Browne} IWA, et~al (2003) {The Cosmic Lens All-Sky Survey - II. Gravitational
  lens candidate selection and follow-up}. \mnras 341:13--32,
  \doi{10.1046/j.1365-8711.2003.06257.x}, \eprint{arXiv:astro-ph/0211069}

\bibitem[{{Burud} et~al(2002){Burud}, {Courbin}, {Magain}, {Lidman},
  {Hutsem{\'e}kers}, {Kneib}, {Hjorth}, {Brewer}, {Pompei}, {Germany},
  {Pritchard}, {Jaunsen}, {Letawe}, and {Meylan}}]{Bur++02}
{Burud} I, {Courbin} F, {Magain} P, {Lidman} C, {Hutsem{\'e}kers} D, {Kneib}
  JP, {Hjorth} J, {Brewer} J, {Pompei} E, {Germany} L, {Pritchard} J, {Jaunsen}
  AO, {Letawe} G, {Meylan} G (2002) {An optical time-delay for the lensed BAL
  quasar HE 2149-2745}. \aap 383:71--81, \doi{10.1051/0004-6361:20011731},
  \eprint{astro-ph/0112225}

\bibitem[{{Chen} et~al(2016){Chen}, {Suyu}, {Wong}, {Fassnacht}, {Chiueh},
  {Halkola}, {Hu}, {Auger}, {Koopmans}, {Lagattuta}, {McKean}, and
  {Vegetti}}]{Che++16}
{Chen} GCF, {Suyu} SH, {Wong} KC, {Fassnacht} CD, {Chiueh} T, {Halkola} A, {Hu}
  I, {Auger} MW, {Koopmans} LVE, {Lagattuta} DJ, {McKean} JP, {Vegetti} S
  (2016) {SHARP - III: First Use Of Adaptive Optics Imaging To Constrain
  Cosmology With Gravitational Lens Time Delays}. ArXiv e-prints
  \eprint{1601.01321}

\bibitem[{{Coe} and {Moustakas}(2009)}]{C+M09b}
{Coe} D, {Moustakas} L (2009) {Cosmological Constraints from Gravitational Lens
  Time Delays}. arXiv09064108 \eprint{0906.4108}

\bibitem[{{Coles}(2008)}]{Col08}
{Coles} J (2008) {A New Estimate of the Hubble Time with Improved Modeling of
  Gravitational Lenses}. \apj 679:17-24, \doi{10.1086/587635},
  \eprint{0802.3219}

\bibitem[{Collett and Cunnington(2016)}]{C+C16}
Collett TE, Cunnington S (2016) Selection biases in time-delay cosmography.
  Monthly Notices of the Royal Astronomical Society submitted

\bibitem[{Collett et~al(2013)Collett, Marshall, Auger, Hilbert, Suyu, Greene,
  Treu, Fassnacht, Koopmans, Brada{\v c}, and Blandford}]{CollettEtal2013}
Collett TE, Marshall PJ, Auger MW, Hilbert S, Suyu SH, Greene Z, Treu T,
  Fassnacht CD, Koopmans LVE, Brada{\v c} M, Blandford RD (2013) Reconstructing
  the lensing mass in the universe from photometric catalogue data. Monthly
  Notices of the Royal Astronomical Society 432:679, \doi{10.1093/mnras/stt504}

\bibitem[{{Conley} et~al(2006){Conley}, {Goldhaber}, {Wang}, {Aldering},
  {Amanullah}, {Commins}, {Fadeyev}, {Folatelli}, {Garavini}, {Gibbons},
  {Goobar}, {Groom}, {Hook}, {Howell}, {Kim}, {Knop}, {Kowalski}, {Kuznetsova},
  {Lidman}, {Nobili}, {Nugent}, {Pain}, {Perlmutter}, {Smith}, {Spadafora},
  {Stanishev}, {Strovink}, {Thomas}, {Wood-Vasey}, and {Supernova Cosmology
  Project}}]{Con++06}
{Conley} A, {Goldhaber} G, {Wang} L, {Aldering} G, {Amanullah} R, {Commins} ED,
  {Fadeyev} V, {Folatelli} G, {Garavini} G, {Gibbons} R, {Goobar} A, {Groom}
  DE, {Hook} I, {Howell} DA, {Kim} AG, {Knop} RA, {Kowalski} M, {Kuznetsova} N,
  {Lidman} C, {Nobili} S, {Nugent} PE, {Pain} R, {Perlmutter} S, {Smith} E,
  {Spadafora} AL, {Stanishev} V, {Strovink} M, {Thomas} RC, {Wood-Vasey} WM,
  {Supernova Cosmology Project} (2006) {Measurement of {$\Omega$}$_{m}$,
  {$\Omega$}$_{Λ}$ from a Blind Analysis of Type Ia Supernovae with CMAGIC:
  Using Color Information to Verify the Acceleration of the Universe}. \apj
  644:1--20, \doi{10.1086/503533}, \eprint{astro-ph/0602411}

\bibitem[{{Coupon} et~al(2015){Coupon}, {Arnouts}, {van Waerbeke}, {Moutard},
  {Ilbert}, {van Uitert}, {Erben}, {Garilli}, {Guzzo}, {Heymans},
  {Hildebrandt}, {Hoekstra}, {Kilbinger}, {Kitching}, {Mellier}, {Miller},
  {Scodeggio}, {Bonnett}, {Branchini}, {Davidzon}, {De Lucia}, {Fritz}, {Fu},
  {Hudelot}, {Hudson}, {Kuijken}, {Leauthaud}, {Le F{\`e}vre}, {McCracken},
  {Moscardini}, {Rowe}, {Schrabback}, {Semboloni}, and
  {Velander}}]{CouponEtal2015}
{Coupon} J, {Arnouts} S, {van Waerbeke} L, {Moutard} T, {Ilbert} O, {van
  Uitert} E, {Erben} T, {Garilli} B, {Guzzo} L, {Heymans} C, {Hildebrandt} H,
  {Hoekstra} H, {Kilbinger} M, {Kitching} T, {Mellier} Y, {Miller} L,
  {Scodeggio} M, {Bonnett} C, {Branchini} E, {Davidzon} I, {De Lucia} G,
  {Fritz} A, {Fu} L, {Hudelot} P, {Hudson} MJ, {Kuijken} K, {Leauthaud} A, {Le
  F{\`e}vre} O, {McCracken} HJ, {Moscardini} L, {Rowe} BTP, {Schrabback} T,
  {Semboloni} E, {Velander} M (2015) {The galaxy-halo connection from a joint
  lensing, clustering and abundance analysis in the CFHTLenS/VIPERS field}.
  \mnras 449:1352--1379, \doi{10.1093/mnras/stv276}, \eprint{1502.02867}

\bibitem[{{Courbin} et~al(1997){Courbin}, {Magain}, {Keeton}, {Kochanek},
  {Vanderriest}, {Jaunsen}, and {Hjorth}}]{Cou++97}
{Courbin} F, {Magain} P, {Keeton} CR, {Kochanek} CS, {Vanderriest} C, {Jaunsen}
  AO, {Hjorth} J (1997) {The geometry of the quadruply imaged quasar PG
  1115+080: implications for H\_0\_.} \aap 324:L1--L4,
  \eprint{astro-ph/9705093}

\bibitem[{{Courbin} et~al(2002){Courbin}, {Saha}, and {Schechter}}]{CSS02}
{Courbin} F, {Saha} P, {Schechter} PL (2002) {Quasar Lensing}. In: {Courbin} F,
  {Minniti} D (eds) Gravitational Lensing: An Astrophysical Tool, Lecture Notes
  in Physics, Berlin Springer Verlag, vol 608, p~1, \eprint{astro-ph/0208043}

\bibitem[{{Courbin} et~al(2005){Courbin}, {Eigenbrod}, {Vuissoz}, {Meylan}, and
  {Magain}}]{Cou++05}
{Courbin} F, {Eigenbrod} A, {Vuissoz} C, {Meylan} G, {Magain} P (2005)
  {COSMOGRAIL: the COSmological MOnitoring of GRAvItational Lenses}. In:
  {Mellier} Y, {Meylan} G (eds) Gravitational Lensing Impact on Cosmology, IAU
  Symposium, vol 225, pp 297--303, \doi{10.1017/S1743921305002097}

\bibitem[{{Courbin} et~al(2011){Courbin}, {Chantry}, {Revaz}, {Sluse}, {Faure},
  {Tewes}, {Eulaers}, {Koleva}, {Asfandiyarov}, {Dye}, {Magain}, {van Winckel},
  {Coles}, {Saha}, {Ibrahimov}, and {Meylan}}]{Cou++11}
{Courbin} F, {Chantry} V, {Revaz} Y, {Sluse} D, {Faure} C, {Tewes} M, {Eulaers}
  E, {Koleva} M, {Asfandiyarov} I, {Dye} S, {Magain} P, {van Winckel} H,
  {Coles} J, {Saha} P, {Ibrahimov} M, {Meylan} G (2011) {COSMOGRAIL: the
  COSmological MOnitoring of GRAvItational Lenses. IX. Time delays, lens
  dynamics and baryonic fraction in HE 0435-1223}. \aap 536:A53,
  \doi{10.1051/0004-6361/201015709}, \eprint{1009.1473}

\bibitem[{{Courteau} et~al(2014){Courteau}, {Cappellari}, {de Jong}, {Dutton},
  {Emsellem}, {Hoekstra}, {Koopmans}, {Mamon}, {Maraston}, {Treu}, and
  {Widrow}}]{Cou++14}
{Courteau} S, {Cappellari} M, {de Jong} RS, {Dutton} AA, {Emsellem} E,
  {Hoekstra} H, {Koopmans} LVE, {Mamon} GA, {Maraston} C, {Treu} T, {Widrow} LM
  (2014) {Galaxy masses}. Reviews of Modern Physics 86:47--119,
  \doi{10.1103/RevModPhys.86.47}, \eprint{1309.3276}

\bibitem[{{Dahle} et~al(2015){Dahle}, {Gladders}, {Sharon}, {Bayliss}, and
  {Rigby}}]{Dah++15}
{Dahle} H, {Gladders} MD, {Sharon} K, {Bayliss} MB, {Rigby} JR (2015) {Time
  Delay Measurements for the Cluster-lensed Sextuple Quasar SDSS J2222+2745}.
  \apj 813:67, \doi{10.1088/0004-637X/813/1/67}, \eprint{1505.06187}

\bibitem[{{Dalal} and {Kochanek}(2002)}]{D+K02}
{Dalal} N, {Kochanek} CS (2002) {Direct Detection of Cold Dark Matter
  Substructure}. \apj 572:25--33, \doi{10.1086/340303},
  \eprint{arXiv:astro-ph/0111456}

\bibitem[{{Dobler} et~al(2015){Dobler}, {Fassnacht}, {Treu}, {Marshall},
  {Liao}, {Hojjati}, {Linder}, and {Rumbaugh}}]{DoblerEtal2015}
{Dobler} G, {Fassnacht} CD, {Treu} T, {Marshall} P, {Liao} K, {Hojjati} A,
  {Linder} E, {Rumbaugh} N (2015) {Strong Lens Time Delay Challenge. I.
  Experimental Design}. \apj 799:168, \doi{10.1088/0004-637X/799/2/168}

\bibitem[{{Efstathiou}(2014)}]{Efs14}
{Efstathiou} G (2014) {H$_{0}$ revisited}. \mnras 440:1138--1152,
  \doi{10.1093/mnras/stu278}, \eprint{1311.3461}

\bibitem[{{Eigenbrod} et~al(2005){Eigenbrod}, {Courbin}, {Vuissoz}, {Meylan},
  {Saha}, and {Dye}}]{Eig++05}
{Eigenbrod} A, {Courbin} F, {Vuissoz} C, {Meylan} G, {Saha} P, {Dye} S (2005)
  {COSMOGRAIL: The COSmological MOnitoring of GRAvItational Lenses. I. How to
  sample the light curves of gravitationally lensed quasars to measure accurate
  time delays}. \aap 436:25--35, \doi{10.1051/0004-6361:20042422},
  \eprint{astro-ph/0503019}

\bibitem[{{Eigenbrod} et~al(2008{\natexlab{a}}){Eigenbrod}, {Courbin},
  {Meylan}, {Agol}, {Anguita}, {Schmidt}, and {Wambsganss}}]{Eig++08a}
{Eigenbrod} A, {Courbin} F, {Meylan} G, {Agol} E, {Anguita} T, {Schmidt} RW,
  {Wambsganss} J (2008{\natexlab{a}}) {Microlensing variability in the
  gravitationally lensed quasar QSO 2237+0305 {$\equiv$} the Einstein Cross.
  II. Energy profile of the accretion disk}. \aap 490:933--943,
  \doi{10.1051/0004-6361:200810729}, \eprint{0810.0011}

\bibitem[{{Eigenbrod} et~al(2008{\natexlab{b}}){Eigenbrod}, {Courbin}, {Sluse},
  {Meylan}, and {Agol}}]{Eig++08b}
{Eigenbrod} A, {Courbin} F, {Sluse} D, {Meylan} G, {Agol} E
  (2008{\natexlab{b}}) {Microlensing variability in the gravitationally lensed
  quasar QSO 2237+0305 {$\equiv$} the Einstein Cross . I. Spectrophotometric
  monitoring with the VLT}. \aap 480:647--661,
  \doi{10.1051/0004-6361:20078703}, \eprint{0709.2828}

\bibitem[{{Ellis}(2010)}]{Ell10}
{Ellis} RS (2010) {Gravitational lensing: a unique probe of dark matter and
  dark energy}. Philosophical Transactions of the Royal Society of London
  Series A 368:967--987, \doi{10.1098/rsta.2009.0209}

\bibitem[{{Eulaers} et~al(2013){Eulaers}, {Tewes}, {Magain}, {Courbin},
  {Asfandiyarov}, {Ehgamberdiev}, {Rathna Kumar}, {Stalin}, {Prabhu}, {Meylan},
  and {Van Winckel}}]{Eul++13}
{Eulaers} E, {Tewes} M, {Magain} P, {Courbin} F, {Asfandiyarov} I,
  {Ehgamberdiev} S, {Rathna Kumar} S, {Stalin} CS, {Prabhu} TP, {Meylan} G,
  {Van Winckel} H (2013) {COSMOGRAIL: the COSmological MOnitoring of
  GRAvItational Lenses. XII. Time delays of the doubly lensed quasars SDSS
  J1206+4332 and HS 2209+1914}. \aap 553:A121,
  \doi{10.1051/0004-6361/201321140}, \eprint{1304.4474}

\bibitem[{{Falco}(2005)}]{Fal05}
{Falco} EE (2005) {A most useful manifestation of relativity: gravitational
  lenses}. New Journal of Physics 7:200--+, \doi{10.1088/1367-2630/7/1/200}

\bibitem[{{Falco} et~al(1985){Falco}, {Gorenstein}, and {Shapiro}}]{FGS85}
{Falco} EE, {Gorenstein} MV, {Shapiro} II (1985) {On model-dependent bounds on
  H(0) from gravitational images Application of Q0957 + 561A,B}. \apjl
  289:L1--L4, \doi{10.1086/184422}

\bibitem[{{Fassnacht} et~al(1999){Fassnacht}, {Pearson}, {Readhead}, {Browne},
  {Koopmans}, {Myers}, and {Wilkinson}}]{Fas++99}
{Fassnacht} CD, {Pearson} TJ, {Readhead} ACS, {Browne} IWA, {Koopmans} LVE,
  {Myers} ST, {Wilkinson} PN (1999) {A Determination of $H_0$ with the CLASS
  Gravitational Lens B1608+656. I. Time Delay Measurements with the VLA}. \apj
  527:498--512, \doi{10.1086/308118}, \eprint{arXiv:astro-ph/9907257}

\bibitem[{{Fassnacht} et~al(2002){Fassnacht}, {Xanthopoulos}, {Koopmans}, and
  {Rusin}}]{Fas++02}
{Fassnacht} CD, {Xanthopoulos} E, {Koopmans} LVE, {Rusin} D (2002) {A
  Determination of $H_0$ with the CLASS Gravitational Lens B1608+656. III. A
  Significant Improvement in the Precision of the Time Delay Measurements}.
  \apj 581:823--835, \doi{10.1086/344368}, \eprint{arXiv:astro-ph/0208420}

\bibitem[{{Fassnacht} et~al(2006){Fassnacht}, {Gal}, {Lubin}, {McKean},
  {Squires}, and {Readhead}}]{Fas++06b}
{Fassnacht} CD, {Gal} RR, {Lubin} LM, {McKean} JP, {Squires} GK, {Readhead} ACS
  (2006) {Mass along the Line of Sight to the Gravitational Lens B1608+656:
  Galaxy Groups and Implications for $H_0$}. \apj 642:30--38,
  \doi{10.1086/500927}, \eprint{arXiv:astro-ph/0510728}

\bibitem[{{Fassnacht} et~al(2011){Fassnacht}, {Koopmans}, and {Wong}}]{FKW11}
{Fassnacht} CD, {Koopmans} LVE, {Wong} KC (2011) {Galaxy number counts and
  implications for strong lensing}. \mnras 410:2167--2179,
  \doi{10.1111/j.1365-2966.2010.17591.x}, \eprint{0909.4301}

\bibitem[{{Fiacconi} et~al(2016){Fiacconi}, {Madau}, {Potter}, and
  {Stadel}}]{Fia++16}
{Fiacconi} D, {Madau} P, {Potter} D, {Stadel} J (2016) {Cold Dark Matter
  Substructures in Early-Type Galaxy Halos}. ArXiv e-prints \eprint{1602.03526}

\bibitem[{{Fohlmeister} et~al(2007){Fohlmeister}, {Kochanek}, {Falco},
  {Wambsganss}, {Morgan}, {Morgan}, {Ofek}, {Maoz}, {Keeton}, {Barentine},
  {Dalton}, {Dembicky}, {Ketzeback}, {McMillan}, and {Peters}}]{Foh++07}
{Fohlmeister} J, {Kochanek} CS, {Falco} EE, {Wambsganss} J, {Morgan} N,
  {Morgan} CW, {Ofek} EO, {Maoz} D, {Keeton} CR, {Barentine} JC, {Dalton} G,
  {Dembicky} J, {Ketzeback} W, {McMillan} R, {Peters} CS (2007) {A Time Delay
  for the Cluster-Lensed Quasar SDSS J1004+4112}. \apj 662:62--71,
  \doi{10.1086/518018}

\bibitem[{{Freedman} et~al(2012){Freedman}, {Madore}, {Scowcroft}, {Burns},
  {Monson}, {Persson}, {Seibert}, and {Rigby}}]{Fre++12}
{Freedman} WL, {Madore} BF, {Scowcroft} V, {Burns} C, {Monson} A, {Persson} SE,
  {Seibert} M, {Rigby} J (2012) {Carnegie Hubble Program: A Mid-infrared
  Calibration of the Hubble Constant}. \apj 758:24,
  \doi{10.1088/0004-637X/758/1/24}, \eprint{1208.3281}

\bibitem[{{Greene} et~al(2013){Greene}, {Suyu}, {Treu}, {Hilbert}, {Auger},
  {Collett}, {Marshall}, {Fassnacht}, {Blandford}, {Brada{\v c}}, and
  {Koopmans}}]{Gre++13}
{Greene} ZS, {Suyu} SH, {Treu} T, {Hilbert} S, {Auger} MW, {Collett} TE,
  {Marshall} PJ, {Fassnacht} CD, {Blandford} RD, {Brada{\v c}} M, {Koopmans}
  LVE (2013) {Improving the Precision of Time-delay Cosmography with
  Observations of Galaxies along the Line of Sight}. \apj 768:39,
  \doi{10.1088/0004-637X/768/1/39}, \eprint{1303.3588}

\bibitem[{{Grillo} et~al(2008){Grillo}, {Lombardi}, and {Bertin}}]{GLB08}
{Grillo} C, {Lombardi} M, {Bertin} G (2008) {Cosmological parameters from
  strong gravitational lensing and stellar dynamics in elliptical galaxies}.
  \aap 477:397--406, \doi{10.1051/0004-6361:20077534}, \eprint{0711.0882}

\bibitem[{{Grillo} et~al(2016){Grillo}, {Karman}, {Suyu}, {Rosati}, {Balestra},
  {Mercurio}, {Lombardi}, {Treu}, {Caminha}, {Halkola}, {Rodney}, {Gavazzi},
  and {Caputi}}]{Gri++16}
{Grillo} C, {Karman} W, {Suyu} SH, {Rosati} P, {Balestra} I, {Mercurio} A,
  {Lombardi} M, {Treu} T, {Caminha} GB, {Halkola} A, {Rodney} SA, {Gavazzi} R,
  {Caputi} KI (2016) {The story of supernova 'Refsdal' told by MUSE}.
  151104093,in press \eprint{1511.04093}

\bibitem[{{Hainline} et~al(2013){Hainline}, {Morgan}, {MacLeod}, {Landaal},
  {Kochanek}, {Harris}, {Tilleman}, {Goicoechea}, {Shalyapin}, and
  {Falco}}]{HainlineEtal2013}
{Hainline} LJ, {Morgan} CW, {MacLeod} CL, {Landaal} ZD, {Kochanek} CS, {Harris}
  HC, {Tilleman} T, {Goicoechea} LJ, {Shalyapin} VN, {Falco} EE (2013) {Time
  Delay and Accretion Disk Size Measurements in the Lensed Quasar SBS 0909+532
  from Multiwavelength Microlensing Analysis}. \apj 774:69,
  \doi{10.1088/0004-637X/774/1/69}, \eprint{1307.3254}

\bibitem[{{Heymans} et~al(2006){Heymans}, {Van Waerbeke}, {Bacon}, {Berge},
  {Bernstein}, {Bertin}, {Bridle}, {Brown}, {Clowe}, {Dahle}, {Erben}, {Gray},
  {Hetterscheidt}, {Hoekstra}, {Hudelot}, {Jarvis}, {Kuijken}, {Margoniner},
  {Massey}, {Mellier}, {Nakajima}, {Refregier}, {Rhodes}, {Schrabback}, and
  {Wittman}}]{STEP}
{Heymans} C, {Van Waerbeke} L, {Bacon} D, {Berge} J, {Bernstein} G, {Bertin} E,
  {Bridle} S, {Brown} ML, {Clowe} D, {Dahle} H, {Erben} T, {Gray} M,
  {Hetterscheidt} M, {Hoekstra} H, {Hudelot} P, {Jarvis} M, {Kuijken} K,
  {Margoniner} V, {Massey} R, {Mellier} Y, {Nakajima} R, {Refregier} A,
  {Rhodes} J, {Schrabback} T, {Wittman} D (2006) {The Shear Testing Programme -
  I. Weak lensing analysis of simulated ground-based observations}. \mnras
  368:1323--1339, \doi{10.1111/j.1365-2966.2006.10198.x},
  \eprint{astro-ph/0506112}

\bibitem[{{Hezaveh} et~al(2013{\natexlab{a}}){Hezaveh}, {Dalal}, {Holder},
  {Kuhlen}, {Marrone}, {Murray}, and {Vieira}}]{Hez++13}
{Hezaveh} Y, {Dalal} N, {Holder} G, {Kuhlen} M, {Marrone} D, {Murray} N,
  {Vieira} J (2013{\natexlab{a}}) {Dark Matter Substructure Detection Using
  Spatially Resolved Spectroscopy of Lensed Dusty Galaxies}. \apj 767:9,
  \doi{10.1088/0004-637X/767/1/9}, \eprint{1210.4562}

\bibitem[{{Hezaveh} et~al(2013{\natexlab{b}}){Hezaveh}, {Marrone}, {Fassnacht},
  {Spilker}, {Vieira}, {Aguirre}, {Aird}, {Aravena}, {Ashby}, {Bayliss},
  {Benson}, {Bleem}, {Bothwell}, {Brodwin}, {Carlstrom}, {Chang}, {Chapman},
  {Crawford}, {Crites}, {De Breuck}, {de Haan}, {Dobbs}, {Fomalont}, {George},
  {Gladders}, {Gonzalez}, {Greve}, {Halverson}, {High}, {Holder}, {Holzapfel},
  {Hoover}, {Hrubes}, {Husband}, {Hunter}, {Keisler}, {Lee}, {Leitch},
  {Lueker}, {Luong-Van}, {Malkan}, {McIntyre}, {McMahon}, {Mehl}, {Menten},
  {Meyer}, {Mocanu}, {Murphy}, {Natoli}, {Padin}, {Plagge}, {Reichardt},
  {Rest}, {Ruel}, {Ruhl}, {Sharon}, {Schaffer}, {Shaw}, {Shirokoff}, {Stalder},
  {Staniszewski}, {Stark}, {Story}, {Vanderlinde}, {Wei{\ss}}, {Welikala}, and
  {Williamson}}]{Hez++13a}
{Hezaveh} YD, {Marrone} DP, {Fassnacht} CD, {Spilker} JS, {Vieira} JD,
  {Aguirre} JE, {Aird} KA, {Aravena} M, {Ashby} MLN, {Bayliss} M, {Benson} BA,
  {Bleem} LE, {Bothwell} M, {Brodwin} M, {Carlstrom} JE, {Chang} CL, {Chapman}
  SC, {Crawford} TM, {Crites} AT, {De Breuck} C, {de Haan} T, {Dobbs} MA,
  {Fomalont} EB, {George} EM, {Gladders} MD, {Gonzalez} AH, {Greve} TR,
  {Halverson} NW, {High} FW, {Holder} GP, {Holzapfel} WL, {Hoover} S, {Hrubes}
  JD, {Husband} K, {Hunter} TR, {Keisler} R, {Lee} AT, {Leitch} EM, {Lueker} M,
  {Luong-Van} D, {Malkan} M, {McIntyre} V, {McMahon} JJ, {Mehl} J, {Menten} KM,
  {Meyer} SS, {Mocanu} LM, {Murphy} EJ, {Natoli} T, {Padin} S, {Plagge} T,
  {Reichardt} CL, {Rest} A, {Ruel} J, {Ruhl} JE, {Sharon} K, {Schaffer} KK,
  {Shaw} L, {Shirokoff} E, {Stalder} B, {Staniszewski} Z, {Stark} AA, {Story}
  K, {Vanderlinde} K, {Wei{\ss}} A, {Welikala} N, {Williamson} R
  (2013{\natexlab{b}}) {ALMA Observations of SPT-discovered, Strongly Lensed,
  Dusty, Star-forming Galaxies}. \apj 767:132,
  \doi{10.1088/0004-637X/767/2/132}, \eprint{1303.2722}

\bibitem[{{Hicken} et~al(2009){Hicken}, {Wood-Vasey}, {Blondin}, {Challis},
  {Jha}, {Kelly}, {Rest}, and {Kirshner}}]{HickenEtal2009}
{Hicken} M, {Wood-Vasey} WM, {Blondin} S, {Challis} P, {Jha} S, {Kelly} PL,
  {Rest} A, {Kirshner} RP (2009) {Improved Dark Energy Constraints from \~{}100
  New CfA Supernova Type Ia Light Curves}. \apj 700:1097--1140,
  \doi{10.1088/0004-637X/700/2/1097}, \eprint{0901.4804}

\bibitem[{{Hilbert} et~al(2009){Hilbert}, {Hartlap}, {White}, and
  {Schneider}}]{Hil++09}
{Hilbert} S, {Hartlap} J, {White} SDM, {Schneider} P (2009) {Ray-tracing
  through the Millennium Simulation: Born corrections and lens-lens coupling in
  cosmic shear and galaxy-galaxy lensing}. \aap 499:31--43,
  \doi{10.1051/0004-6361/200811054}, \eprint{0809.5035}

\bibitem[{{Hjorth} et~al(2002){Hjorth}, {Burud}, {Jaunsen}, {Schechter},
  {Kneib}, {Andersen}, {Korhonen}, {Clasen}, {Kaas}, {{\O}stensen}, {Pelt}, and
  {Pijpers}}]{Hjo++02}
{Hjorth} J, {Burud} I, {Jaunsen} AO, {Schechter} PL, {Kneib} JP, {Andersen} MI,
  {Korhonen} H, {Clasen} JW, {Kaas} AA, {{\O}stensen} R, {Pelt} J, {Pijpers} FP
  (2002) {The Time Delay of the Quadruple Quasar RX J0911.4+0551}. \apjl
  572:L11--L14, \doi{10.1086/341603}, \eprint{astro-ph/0205124}

\bibitem[{{Hojjati} and {Linder}(2014)}]{H+L2014}
{Hojjati} A, {Linder} EV (2014) {Next generation strong lensing time delay
  estimation with Gaussian processes}. \prd 90(12):123501,
  \doi{10.1103/PhysRevD.90.123501}, \eprint{1408.5143}

\bibitem[{{Holz}(2001)}]{Hol01}
{Holz} DE (2001) {Seeing Double: Strong Gravitational Lensing of High-Redshift
  Supernovae}. \apjl 556:L71--L74, \doi{10.1086/322947},
  \eprint{astro-ph/0104440}

\bibitem[{{Hu}(2005)}]{Hu05}
{Hu} W (2005) {Dark Energy Probes in Light of the CMB}. In: {Wolff} SC, {Lauer}
  TR (eds) Observing Dark Energy, Astronomical Society of the Pacific
  Conference Series, vol 339, p 215, \eprint{astro-ph/0407158}

\bibitem[{Jackson(2013)}]{Jackson:2013p30763}
Jackson N (2013) Quasar lensing. eprint arXiv:13044172
  \urlprefix\url{http://adsabs.harvard.edu/abs/2013arXiv1304.4172J}

\bibitem[{{Jackson}(2015)}]{Jac15}
{Jackson} N (2015) {The Hubble Constant}. Living Reviews in Relativity 18,
  \doi{10.1007/lrr-2015-2}

\bibitem[{{Jakobsson} et~al(2005){Jakobsson}, {Hjorth}, {Burud}, {Letawe},
  {Lidman}, and {Courbin}}]{Jak++05}
{Jakobsson} P, {Hjorth} J, {Burud} I, {Letawe} G, {Lidman} C, {Courbin} F
  (2005) {An optical time delay for the double gravitational lens system FBQ
  0951+2635}. \aap 431:103--109, \doi{10.1051/0004-6361:20041432},
  \eprint{astro-ph/0409444}

\bibitem[{{Jauzac} et~al(2016){Jauzac}, {Richard}, {Limousin}, {Knowles},
  {Mahler}, {Smith}, {Kneib}, {Jullo}, {Natarajan}, {Ebeling}, {Atek},
  {Cl{\'e}ment}, {Eckert}, {Egami}, {Massey}, and {Rexroth}}]{Jau++16}
{Jauzac} M, {Richard} J, {Limousin} M, {Knowles} K, {Mahler} G, {Smith} GP,
  {Kneib} JP, {Jullo} E, {Natarajan} P, {Ebeling} H, {Atek} H, {Cl{\'e}ment} B,
  {Eckert} D, {Egami} E, {Massey} R, {Rexroth} M (2016) {Hubble Frontier
  Fields: predictions for the return of SN Refsdal with the MUSE and GMOS
  spectrographs}. \mnras 457:2029--2042, \doi{10.1093/mnras/stw069},
  \eprint{1509.08914}

\bibitem[{{Jee} et~al(2015{\natexlab{a}}){Jee}, {Komatsu}, and
  {Suyu}}]{JeeKomatsuSuyu2015}
{Jee} I, {Komatsu} E, {Suyu} SH (2015{\natexlab{a}}) {Measuring angular
  diameter distances of strong gravitational lenses}. \jcap 11:033,
  \doi{10.1088/1475-7516/2015/11/033}, \eprint{1410.7770}

\bibitem[{{Jee} et~al(2015{\natexlab{b}}){Jee}, {Komatsu}, {Suyu}, and
  {Huterer}}]{JeeEtal2016}
{Jee} I, {Komatsu} E, {Suyu} SH, {Huterer} D (2015{\natexlab{b}}) {Time-delay
  Cosmography: Increased Leverage with Angular Diameter Distances}. ArXiv
  e-prints \eprint{1509.03310}

\bibitem[{{Kawamata} et~al(2016){Kawamata}, {Oguri}, {Ishigaki}, {Shimasaku},
  and {Ouchi}}]{Kaw++16}
{Kawamata} R, {Oguri} M, {Ishigaki} M, {Shimasaku} K, {Ouchi} M (2016) {Precise
  Strong Lensing Mass Modeling of Four Hubble Frontier Field Clusters and a
  Sample of Magnified High-redshift Galaxies}. \apj 819:114,
  \doi{10.3847/0004-637X/819/2/114}, \eprint{1510.06400}

\bibitem[{{Keeton}(2011)}]{Kee11}
{Keeton} CR (2011) {GRAVLENS: Computational Methods for Gravitational Lensing}.
  Astrophysics Source Code Library, \eprint{1102.003}

\bibitem[{{Keeton} and {Moustakas}(2009)}]{K+M09}
{Keeton} CR, {Moustakas} LA (2009) {A New Channel for Detecting Dark Matter
  Substructure in Galaxies: Gravitational Lens Time Delays}. \apj
  699:1720--1731, \doi{10.1088/0004-637X/699/2/1720}, \eprint{0805.0309}

\bibitem[{{Keeton} and {Zabludoff}(2004)}]{K+Z04}
{Keeton} CR, {Zabludoff} AI (2004) {The Importance of Lens Galaxy
  Environments}. \apj 612:660--678, \doi{10.1086/422745},
  \eprint{astro-ph/0406060}

\bibitem[{Keeton et~al(2000)Keeton, Falco, Impey, Kochanek, Leh{\'a}r, McLeod,
  Rix, Mu{\~n}oz, and Peng}]{Keeton:2000p241}
Keeton CR, Falco EE, Impey CD, Kochanek CS, Leh{\'a}r J, McLeod BA, Rix HW,
  Mu{\~n}oz JA, Peng CY (2000) The host galaxy of the lensed quasar q0957+561.
  The Astrophysical Journal 542:74, \doi{10.1086/309517}

\bibitem[{{Kelly} et~al(2015){Kelly}, {Rodney}, {Treu}, {Foley}, {Brammer},
  {Schmidt}, {Zitrin}, {Sonnenfeld}, {Strolger}, {Graur}, {Filippenko}, {Jha},
  {Riess}, {Bradac}, {Weiner}, {Scolnic}, {Malkan}, {von der Linden}, {Trenti},
  {Hjorth}, {Gavazzi}, {Fontana}, {Merten}, {McCully}, {Jones}, {Postman},
  {Dressler}, {Patel}, {Cenko}, {Graham}, and {Tucker}}]{Kel++15}
{Kelly} PL, {Rodney} SA, {Treu} T, {Foley} RJ, {Brammer} G, {Schmidt} KB,
  {Zitrin} A, {Sonnenfeld} A, {Strolger} LG, {Graur} O, {Filippenko} AV, {Jha}
  SW, {Riess} AG, {Bradac} M, {Weiner} BJ, {Scolnic} D, {Malkan} MA, {von der
  Linden} A, {Trenti} M, {Hjorth} J, {Gavazzi} R, {Fontana} A, {Merten} JC,
  {McCully} C, {Jones} T, {Postman} M, {Dressler} A, {Patel} B, {Cenko} SB,
  {Graham} ML, {Tucker} BE (2015) {Multiple images of a highly magnified
  supernova formed by an early-type cluster galaxy lens}. Science
  347:1123--1126, \doi{10.1126/science.aaa3350}, \eprint{1411.6009}

\bibitem[{{Kelly} et~al(2016){Kelly}, {Rodney}, {Treu}, {Strolger}, {Foley},
  {Jha}, {Selsing}, {Brammer}, {Brada{\v c}}, {Cenko}, {Graur}, {Filippenko},
  {Hjorth}, {McCully}, {Molino}, {Nonino}, {Riess}, {Schmidt}, {Tucker}, {von
  der Linden}, {Weiner}, and {Zitrin}}]{Kel++16}
{Kelly} PL, {Rodney} SA, {Treu} T, {Strolger} LG, {Foley} RJ, {Jha} SW,
  {Selsing} J, {Brammer} G, {Brada{\v c}} M, {Cenko} SB, {Graur} O,
  {Filippenko} AV, {Hjorth} J, {McCully} C, {Molino} A, {Nonino} M, {Riess} AG,
  {Schmidt} KB, {Tucker} B, {von der Linden} A, {Weiner} BJ, {Zitrin} A (2016)
  {Deja Vu All Over Again: The Reappearance of Supernova Refsdal}. \apjl
  819:L8, \doi{10.3847/2041-8205/819/1/L8}, \eprint{1512.04654}

\bibitem[{{Kim} et~al(2015){Kim}, {Padmanabhan}, {Aldering}, {Allen}, {Baltay},
  {Cahn}, {D'Andrea}, {Dalal}, {Dawson}, {Denney}, {Eisenstein}, {Finley},
  {Freedman}, {Ho}, {Holz}, {Kasen}, {Kent}, {Kessler}, {Kuhlmann}, {Linder},
  {Martini}, {Nugent}, {Perlmutter}, {Peterson}, {Riess}, {Rubin}, {Sako},
  {Suntzeff}, {Suzuki}, {Thomas}, {Wood-Vasey}, and {Woosley}}]{Kim++15}
{Kim} AG, {Padmanabhan} N, {Aldering} G, {Allen} SW, {Baltay} C, {Cahn} RN,
  {D'Andrea} CB, {Dalal} N, {Dawson} KS, {Denney} KD, {Eisenstein} DJ, {Finley}
  DA, {Freedman} WL, {Ho} S, {Holz} DE, {Kasen} D, {Kent} SM, {Kessler} R,
  {Kuhlmann} S, {Linder} EV, {Martini} P, {Nugent} PE, {Perlmutter} S,
  {Peterson} BM, {Riess} AG, {Rubin} D, {Sako} M, {Suntzeff} NV, {Suzuki} N,
  {Thomas} RC, {Wood-Vasey} WM, {Woosley} SE (2015) {Distance probes of dark
  energy}. Astroparticle Physics 63:2--22,
  \doi{10.1016/j.astropartphys.2014.05.007}, \eprint{1309.5382}

\bibitem[{Klein and Roodman(2005)}]{klein2005blind}
Klein JR, Roodman A (2005) Blind analysis in nuclear and particle physics. Annu
  Rev Nucl Part Sci 55:141--163

\bibitem[{{Kneib} et~al(2011){Kneib}, {Bonnet}, {Golse}, {Sand}, {Jullo}, and
  {Marshall}}]{Kne++11}
{Kneib} JP, {Bonnet} H, {Golse} G, {Sand} D, {Jullo} E, {Marshall} P (2011)
  {LENSTOOL: A Gravitational Lensing Software for Modeling Mass Distribution of
  Galaxies and Clusters (strong and weak regime)}. Astrophysics Source Code
  Library, \eprint{1102.004}

\bibitem[{{Kochanek}(2002)}]{Koc02}
{Kochanek} CS (2002) {What Do Gravitational Lens Time Delays Measure?} \apj
  578:25--32, \doi{10.1086/342476}, \eprint{arXiv:astro-ph/0205319}

\bibitem[{{Kochanek} and {Schechter}(2004)}]{K+S04}
{Kochanek} CS, {Schechter} PL (2004) {The Hubble Constant from Gravitational
  Lens Time Delays}. Measuring and Modeling the Universe p 117,
  \eprint{astro-ph/0306040}

\bibitem[{{Kochanek} et~al(2001){Kochanek}, {Keeton}, and {McLeod}}]{KKM01}
{Kochanek} CS, {Keeton} CR, {McLeod} BA (2001) {The Importance of Einstein
  Rings}. \apj 547:50--59, \doi{10.1086/318350},
  \eprint{arXiv:astro-ph/0006116}

\bibitem[{{Koopmans}(2005)}]{Koo05}
{Koopmans} LVE (2005) {Gravitational imaging of cold dark matter
  substructures}. \mnras 363:1136--1144, \doi{10.1111/j.1365-2966.2005.09523.x}

\bibitem[{{Koopmans} and {Fassnacht}(1999)}]{K+F99}
{Koopmans} LVE, {Fassnacht} CD (1999) {A Determination of H$_{0}$ with the
  CLASS Gravitational Lens B1608+656. II. Mass Models and the Hubble Constant
  from Lensing}. \apj 527:513--524, \doi{10.1086/308120},
  \eprint{astro-ph/9907258}

\bibitem[{{Koopmans} et~al(2003){Koopmans}, {Treu}, {Fassnacht}, {Blandford},
  and {Surpi}}]{Koo++03}
{Koopmans} LVE, {Treu} T, {Fassnacht} CD, {Blandford} RD, {Surpi} G (2003) {The
  Hubble Constant from the Gravitational Lens B1608+656}. \apj 599:70--85,
  \doi{10.1086/379226}, \eprint{arXiv:astro-ph/0306216}

\bibitem[{{Koopmans} et~al(2009){Koopmans}, {Bolton}, {Treu}, {Czoske},
  {Auger}, {Barnab{\`e}}, {Vegetti}, {Gavazzi}, {Moustakas}, and
  {Burles}}]{Koo++09}
{Koopmans} LVE, {Bolton} A, {Treu} T, {Czoske} O, {Auger} MW, {Barnab{\`e}} M,
  {Vegetti} S, {Gavazzi} R, {Moustakas} LA, {Burles} S (2009) {The Structure
  and Dynamics of Massive Early-Type Galaxies: On Homology, Isothermality, and
  Isotropy Inside One Effective Radius}. \apjl 703:L51--L54,
  \doi{10.1088/0004-637X/703/1/L51}, \eprint{0906.1349}

\bibitem[{{Kundic} et~al(1997){Kundic}, {Turner}, {Colley}, {Gott}, {Rhoads},
  {Wang}, {Bergeron}, {Gloria}, {Long}, {Malhotra}, and {Wambsganss}}]{Kun++97}
{Kundic} T, {Turner} EL, {Colley} WN, {Gott} JRI, {Rhoads} JE, {Wang} Y,
  {Bergeron} LE, {Gloria} KA, {Long} DC, {Malhotra} S, {Wambsganss} J (1997) {A
  Robust Determination of the Time Delay in 0957+561A, B and a Mea surement of
  the Global Value of Hubble's Constant}. \apj 482:75--+, \doi{10.1086/304147},
  \eprint{arXiv:astro-ph/9610162}

\bibitem[{{Liao} et~al(2015){Liao}, {Treu}, {Marshall}, {Fassnacht},
  {Rumbaugh}, {Dobler}, {Aghamousa}, {Bonvin}, {Courbin}, {Hojjati}, {Jackson},
  {Kashyap}, {Rathna Kumar}, {Linder}, {Mandel}, {Meng}, {Meylan}, {Moustakas},
  {Prabhu}, {Romero-Wolf}, {Shafieloo}, {Siemiginowska}, {Stalin}, {Tak},
  {Tewes}, and {van Dyk}}]{LiaoEtal2015}
{Liao} K, {Treu} T, {Marshall} P, {Fassnacht} CD, {Rumbaugh} N, {Dobler} G,
  {Aghamousa} A, {Bonvin} V, {Courbin} F, {Hojjati} A, {Jackson} N, {Kashyap}
  V, {Rathna Kumar} S, {Linder} E, {Mandel} K, {Meng} XL, {Meylan} G,
  {Moustakas} LA, {Prabhu} TP, {Romero-Wolf} A, {Shafieloo} A, {Siemiginowska}
  A, {Stalin} CS, {Tak} H, {Tewes} M, {van Dyk} D (2015) {Strong Lens Time
  Delay Challenge. II. Results of TDC1}. \apj 800:11,
  \doi{10.1088/0004-637X/800/1/11}, \eprint{1409.1254}

\bibitem[{{Linder}(2011)}]{Lin11}
{Linder} EV (2011) {Lensing time delays and cosmological complementarity}. \prd
  84(12):123529, \doi{10.1103/PhysRevD.84.123529}, \eprint{1109.2592}

\bibitem[{{Linder}(2015)}]{Lin15}
{Linder} EV (2015) {Tailoring strong lensing cosmographic observations}. \prd
  91(8):083511, \doi{10.1103/PhysRevD.91.083511}, \eprint{1502.01353}

\bibitem[{{MacLeod} et~al(2015){MacLeod}, {Morgan}, {Mosquera}, {Kochanek},
  {Tewes}, {Courbin}, {Meylan}, {Chen}, {Dai}, and {Chartas}}]{Mac++15}
{MacLeod} CL, {Morgan} CW, {Mosquera} A, {Kochanek} CS, {Tewes} M, {Courbin} F,
  {Meylan} G, {Chen} B, {Dai} X, {Chartas} G (2015) {A Consistent Picture
  Emerges: A Compact X-Ray Continuum Emission Region in the Gravitationally
  Lensed Quasar SDSS J0924+0219}. \apj 806:258,
  \doi{10.1088/0004-637X/806/2/258}, \eprint{1501.07533}

\bibitem[{{Magain} et~al(1998){Magain}, {Courbin}, and {Sohy}}]{MCS98}
{Magain} P, {Courbin} F, {Sohy} S (1998) {Deconvolution with Correct Sampling}.
  \apj 494:472--477, \doi{10.1086/305187}, \eprint{astro-ph/9704059}

\bibitem[{{Marshall} et~al(2007)}]{Mar++07}
{Marshall} PJ, et~al (2007) {Superresolving Distant Galaxies with Gravitational
  Telescopes: Keck Laser Guide Star Adaptive Optics and Hubble Space Telescope
  Imaging of the Lens System SDSS J0737+3216}. \apj 671:1196--1211,
  \doi{10.1086/523091}, \eprint{arXiv:0710.0637}

\bibitem[{{McCully} et~al(2014){McCully}, {Keeton}, {Wong}, and
  {Zabludoff}}]{McCullyEtal2014}
{McCully} C, {Keeton} CR, {Wong} KC, {Zabludoff} AI (2014) {A new hybrid
  framework to efficiently model lines of sight to gravitational lenses}.
  \mnras 443:3631--3642, \doi{10.1093/mnras/stu1316}, \eprint{1401.0197}

\bibitem[{{McCully} et~al(2016){McCully}, {Keeton}, {Wong}, and
  {Zabludoff}}]{McCullyEtal2016}
{McCully} C, {Keeton} CR, {Wong} KC, {Zabludoff} AI (2016) {Quantifying
  Environmental and Line-of-Sight Effects in Models of Strong Gravitational
  Lens Systems}. ArXiv e-prints \eprint{1601.05417}

\bibitem[{{Meng} et~al(2015){Meng}, {Treu}, {Agnello}, {Auger}, {Liao}, and
  {Marshall}}]{Men++15}
{Meng} XL, {Treu} T, {Agnello} A, {Auger} MW, {Liao} K, {Marshall} PJ (2015)
  {Precision cosmology with time delay lenses: high resolution imaging
  requirements}. \jcap 9:059, \doi{10.1088/1475-7516/2015/09/059},
  \eprint{1506.07640}

\bibitem[{Metcalf(2005)}]{Metcalf:2005p1203}
Metcalf RB (2005) Testing Λcdm with gravitational lensing constraints on
  small-scale structure. The Astrophysical Journal 622:72,
  \doi{10.1086/427864}, (c) 2005: . The American Astronomical Society

\bibitem[{{Metcalf} and {Madau}(2001)}]{M+M01}
{Metcalf} RB, {Madau} P (2001) {Compound Gravitational Lensing as a Probe of
  Dark Matter Substructure within Galaxy Halos}. \apj 563:9--20,
  \doi{10.1086/323695}, \eprint{astro-ph/0108224}

\bibitem[{{Momcheva} et~al(2006){Momcheva}, {Williams}, {Keeton}, and
  {Zabludoff}}]{Mom++06}
{Momcheva} I, {Williams} K, {Keeton} C, {Zabludoff} A (2006) {A Spectroscopic
  Study of the Environments of Gravitational Lens Galaxies}. \apj 641:169--189,
  \doi{10.1086/500382}, \eprint{arXiv:astro-ph/0511594}

\bibitem[{{Momcheva} et~al(2015){Momcheva}, {Williams}, {Cool}, {Keeton}, and
  {Zabludoff}}]{Mom++15}
{Momcheva} IG, {Williams} KA, {Cool} RJ, {Keeton} CR, {Zabludoff} AI (2015) {A
  Spectroscopic Survey of the Fields of 28 Strong Gravitational Lenses}. \apjs
  219:29, \doi{10.1088/0067-0049/219/2/29}, \eprint{1503.02074}

\bibitem[{{More} et~al(2016){More}, {Oguri}, {Kayo}, {Zinn}, {Strauss},
  {Santiago}, {Mosquera}, {Inada}, {Kochanek}, {Rusu}, {Brownstein}, {da
  Costa}, {Kneib}, {Maia}, {Quimby}, {Schneider}, {Streblyanska}, and
  {York}}]{Mor++16}
{More} A, {Oguri} M, {Kayo} I, {Zinn} J, {Strauss} MA, {Santiago} BX,
  {Mosquera} AM, {Inada} N, {Kochanek} CS, {Rusu} CE, {Brownstein} JR, {da
  Costa} LN, {Kneib} JP, {Maia} MAG, {Quimby} RM, {Schneider} DP,
  {Streblyanska} A, {York} DG (2016) {The SDSS-III BOSS quasar lens survey:
  discovery of 13 gravitationally lensed quasars}. \mnras 456:1595--1606,
  \doi{10.1093/mnras/stv2813}, \eprint{1509.07917}

\bibitem[{{Moustakas} et~al(2008){Moustakas}, {Bolton}, {Booth}, {Bullock},
  {Cheng}, {Coe}, {Fassnacht}, {Gorjian}, {Heneghan}, {Keeton}, {Kochanek},
  {Lawrence}, {Marshall}, {Metcalf}, {Natarajan}, {Nikzad}, {Peterson}, and
  {Wambsganss}}]{Mou++08}
{Moustakas} LA, {Bolton} AJ, {Booth} JT, {Bullock} JS, {Cheng} E, {Coe} D,
  {Fassnacht} CD, {Gorjian} V, {Heneghan} C, {Keeton} CR, {Kochanek} CS,
  {Lawrence} CR, {Marshall} PJ, {Metcalf} RB, {Natarajan} P, {Nikzad} S,
  {Peterson} BM, {Wambsganss} J (2008) {The Observatory for Multi-Epoch
  Gravitational Lens Astrophysics (OMEGA)}. In: Space Telescopes and
  Instrumentation 2008: Optical, Infrared, and Millimeter, PROC SPIE, vol 7010,
  p 70101B, \doi{10.1117/12.789987}, \eprint{0806.1884}

\bibitem[{{Nierenberg} et~al(2014){Nierenberg}, {Treu}, {Wright}, {Fassnacht},
  and {Auger}}]{Nie++14}
{Nierenberg} AM, {Treu} T, {Wright} SA, {Fassnacht} CD, {Auger} MW (2014)
  {Detection of substructure with adaptive optics integral field spectroscopy
  of the gravitational lens B1422+231}. \mnras 442:2434--2445,
  \doi{10.1093/mnras/stu862}, \eprint{1402.1496}

\bibitem[{{Nightingale} and {Dye}(2015)}]{Nig++15}
{Nightingale} JW, {Dye} S (2015) {Adaptive semi-linear inversion of strong
  gravitational lens imaging}. \mnras 452:2940--2959,
  \doi{10.1093/mnras/stv1455}, \eprint{1412.7436}

\bibitem[{{Oguri}(2007)}]{Ogu07b}
{Oguri} M (2007) {Gravitational Lens Time Delays: A Statistical Assessment of
  Lens Model Dependences and Implications for the Global Hubble Constant}. \apj
  660:1--15, \doi{10.1086/513093}, \eprint{arXiv:astro-ph/0609694}

\bibitem[{{Oguri}(2015)}]{Ogu15}
{Oguri} M (2015) {Predicted properties of multiple images of the strongly
  lensed supernova SN Refsdal}. \mnras 449:L86--L89,
  \doi{10.1093/mnrasl/slv025}, \eprint{1411.6443}

\bibitem[{{Oguri} and {Marshall}(2010)}]{O+M10}
{Oguri} M, {Marshall} PJ (2010) {Gravitationally lensed quasars and supernovae
  in future wide-field optical imaging surveys}. \mnras 405:2579--2593,
  \doi{10.1111/j.1365-2966.2010.16639.x}, \eprint{1001.2037}

\bibitem[{Oguri et~al(2006)Oguri, Inada, Pindor, Strauss, Richards, Hennawi,
  Turner, Lupton, Schneider, Fukugita, and Brinkmann}]{Oguri:2006p5865}
Oguri M, Inada N, Pindor B, Strauss MA, Richards GT, Hennawi JF, Turner EL,
  Lupton RH, Schneider DP, Fukugita M, Brinkmann J (2006) The sloan digital sky
  survey quasar lens search. i. candidate selection algorithm. The Astronomical
  Journal 132:999, \doi{10.1086/506019}

\bibitem[{{Paraficz} and {Hjorth}(2010)}]{P+J10}
{Paraficz} D, {Hjorth} J (2010) {The Hubble Constant Inferred from 18
  Time-delay Lenses}. \apj 712:1378--1384, \doi{10.1088/0004-637X/712/2/1378},
  \eprint{1002.2570}

\bibitem[{{Pelt} et~al(1996){Pelt}, {Kayser}, {Refsdal}, and
  {Schramm}}]{Pelt++96}
{Pelt} J, {Kayser} R, {Refsdal} S, {Schramm} T (1996) {The light curve and the
  time delay of QSO 0957+561.} \aap 305:97, \eprint{astro-ph/9501036}

\bibitem[{{Percival} et~al(2010){Percival}, {Reid}, {Eisenstein}, {Bahcall},
  {Budavari}, {Frieman}, {Fukugita}, {Gunn}, {Ivezi{\'c}}, {Knapp}, {Kron},
  {Loveday}, {Lupton}, {McKay}, {Meiksin}, {Nichol}, {Pope}, {Schlegel},
  {Schneider}, {Spergel}, {Stoughton}, {Strauss}, {Szalay}, {Tegmark},
  {Vogeley}, {Weinberg}, {York}, and {Zehavi}}]{PercivalEtal2010}
{Percival} WJ, {Reid} BA, {Eisenstein} DJ, {Bahcall} NA, {Budavari} T,
  {Frieman} JA, {Fukugita} M, {Gunn} JE, {Ivezi{\'c}} {\v Z}, {Knapp} GR,
  {Kron} RG, {Loveday} J, {Lupton} RH, {McKay} TA, {Meiksin} A, {Nichol} RC,
  {Pope} AC, {Schlegel} DJ, {Schneider} DP, {Spergel} DN, {Stoughton} C,
  {Strauss} MA, {Szalay} AS, {Tegmark} M, {Vogeley} MS, {Weinberg} DH, {York}
  DG, {Zehavi} I (2010) {Baryon acoustic oscillations in the Sloan Digital Sky
  Survey Data Release 7 galaxy sample}. \mnras 401:2148--2168,
  \doi{10.1111/j.1365-2966.2009.15812.x}, \eprint{0907.1660}

\bibitem[{{Perlmutter} et~al(1999){Perlmutter}, {Aldering}, {Goldhaber},
  {Knop}, {Nugent}, {Castro}, {Deustua}, {Fabbro}, {Goobar}, {Groom}, {Hook},
  {Kim}, {Kim}, {Lee}, {Nunes}, {Pain}, {Pennypacker}, {Quimby}, {Lidman},
  {Ellis}, {Irwin}, {McMahon}, {Ruiz-Lapuente}, {Walton}, {Schaefer}, {Boyle},
  {Filippenko}, {Matheson}, {Fruchter}, {Panagia}, {Newberg}, {Couch}, and {The
  Supernova Cosmology Project}}]{Per++99}
{Perlmutter} S, {Aldering} G, {Goldhaber} G, {Knop} RA, {Nugent} P, {Castro}
  PG, {Deustua} S, {Fabbro} S, {Goobar} A, {Groom} DE, {Hook} IM, {Kim} AG,
  {Kim} MY, {Lee} JC, {Nunes} NJ, {Pain} R, {Pennypacker} CR, {Quimby} R,
  {Lidman} C, {Ellis} RS, {Irwin} M, {McMahon} RG, {Ruiz-Lapuente} P, {Walton}
  N, {Schaefer} B, {Boyle} BJ, {Filippenko} AV, {Matheson} T, {Fruchter} AS,
  {Panagia} N, {Newberg} HJM, {Couch} WJ, {The Supernova Cosmology Project}
  (1999) {Measurements of Omega and Lambda from 42 High-Redshift Supernovae}.
  \apj 517:565--586, \doi{10.1086/307221}, \eprint{arXiv:astro-ph/9812133}

\bibitem[{{Petters} et~al(2001){Petters}, {Levine}, and
  {Wambsganss}}]{Petters2001}
{Petters} AO, {Levine} H, {Wambsganss} J (2001) {Singularity theory and
  gravitational lensing}. Boston:Birkh\"auser (Progress in mathematical physics
  v.~21)

\bibitem[{{Planck Collaboration} et~al(2015){Planck Collaboration}, {Ade},
  {Aghanim}, {Arnaud}, {Ashdown}, {Aumont}, {Baccigalupi}, {Banday},
  {Barreiro}, {Bartlett}, and et~al.}]{Pla15}
{Planck Collaboration}, {Ade} PAR, {Aghanim} N, {Arnaud} M, {Ashdown} M,
  {Aumont} J, {Baccigalupi} C, {Banday} AJ, {Barreiro} RB, {Bartlett} JG, et~al
  (2015) {Planck 2015 results. XIII. Cosmological parameters}. ArXiv e-prints
  \eprint{1502.01589}

\bibitem[{{Poindexter} et~al(2007){Poindexter}, {Morgan}, {Kochanek}, and
  {Falco}}]{Poi++07a}
{Poindexter} S, {Morgan} N, {Kochanek} CS, {Falco} EE (2007) {Mid-IR
  Observations and a Revised Time Delay for the Gravitational Lens System
  Quasar HE 1104-1805}. \apj 660:146--151, \doi{10.1086/512773},
  \eprint{arXiv:astro-ph/0612045}

\bibitem[{{Poindexter} et~al(2008){Poindexter}, {Morgan}, and
  {Kochanek}}]{PMK08}
{Poindexter} S, {Morgan} N, {Kochanek} CS (2008) {The Spatial Structure of an
  Accretion Disk}. \apj 673:34--38, \doi{10.1086/524190},
  \eprint{arXiv:0707.0003}

\bibitem[{{Press} et~al(1992){Press}, {Rybicki}, and {Hewitt}}]{PRH92}
{Press} WH, {Rybicki} GB, {Hewitt} JN (1992) {The Time Delay of Gravitational
  Lens 0957+561. II. Analysis of Radio Data and Combined Optical-Radio
  Analysis}. \apj 385:416, \doi{10.1086/170952}

\bibitem[{{Rathna Kumar} et~al(2013){Rathna Kumar}, {Tewes}, {Stalin},
  {Courbin}, {Asfandiyarov}, {Meylan}, {Eulaers}, {Prabhu}, {Magain}, {Van
  Winckel}, and {Ehgamberdiev}}]{RK++13}
{Rathna Kumar} S, {Tewes} M, {Stalin} CS, {Courbin} F, {Asfandiyarov} I,
  {Meylan} G, {Eulaers} E, {Prabhu} TP, {Magain} P, {Van Winckel} H,
  {Ehgamberdiev} S (2013) {COSMOGRAIL: the COSmological MOnitoring of
  GRAvItational Lenses. XIV. Time delay of the doubly lensed quasar SDSS
  J1001+5027}. \aap 557:A44, \doi{10.1051/0004-6361/201322116},
  \eprint{1306.5105}

\bibitem[{{Rathna Kumar} et~al(2015){Rathna Kumar}, {Stalin}, and
  {Prabhu}}]{RK++2015}
{Rathna Kumar} S, {Stalin} CS, {Prabhu} TP (2015) {H$_{0}$ from ten
  well-measured time delay lenses}. \aap 580:A38,
  \doi{10.1051/0004-6361/201423977}, \eprint{1404.2920}

\bibitem[{{Read} et~al(2007){Read}, {Saha}, and {Macci{\`o}}}]{Rea++07}
{Read} JI, {Saha} P, {Macci{\`o}} AV (2007) {Radial Density Profiles of
  Time-Delay Lensing Galaxies}. \apj 667:645--654, \doi{10.1086/520714},
  \eprint{0704.3267}

\bibitem[{{Refsdal}(1964)}]{Ref64}
{Refsdal} S (1964) {On the possibility of determining Hubble's parameter and
  the masses of galaxies from the gravitational lens effect}. \mnras 128:307--+

\bibitem[{Riess et~al(1998)Riess, Filippenko, Challis, Clocchiatti, Diercks,
  Garnavich, Gilliland, Hogan, Jha, Kirshner, Leibundgut, Phillips, Reiss,
  Schmidt, Schommer, Smith, Spyromilio, Stubbs, Suntzeff, and
  Tonry}]{Riess:1998p21184}
Riess AG, Filippenko AV, Challis P, Clocchiatti A, Diercks A, Garnavich PM,
  Gilliland RL, Hogan CJ, Jha S, Kirshner RP, Leibundgut B, Phillips MM, Reiss
  D, Schmidt BP, Schommer RA, Smith RC, Spyromilio J, Stubbs C, Suntzeff NB,
  Tonry J (1998) Observational evidence from supernovae for an accelerating
  universe and a cosmological constant. The Astronomical Journal 116:1009,
  \doi{10.1086/300499}

\bibitem[{{Riess} et~al(2016){Riess}, {Macri}, {Hoffmann}, {Scolnic},
  {Casertano}, {Filippenko}, {Tucker}, {Reid}, {Jones}, {Silverman},
  {Chornock}, {Challis}, {Yuan}, and {Foley}}]{Rie++16}
{Riess} AG, {Macri} LM, {Hoffmann} SL, {Scolnic} D, {Casertano} S, {Filippenko}
  AV, {Tucker} BE, {Reid} MJ, {Jones} DO, {Silverman} JM, {Chornock} R,
  {Challis} P, {Yuan} W, {Foley} RJ (2016) {A 2.4\% Determination of the Local
  Value of the Hubble Constant}. ArXiv e-prints \eprint{1604.01424}

\bibitem[{{Rigault} et~al(2015){Rigault}, {Aldering}, {Kowalski}, {Copin},
  {Antilogus}, {Aragon}, {Bailey}, {Baltay}, {Baugh}, {Bongard}, {Boone},
  {Buton}, {Chen}, {Chotard}, {Fakhouri}, {Feindt}, {Fagrelius}, {Fleury},
  {Fouchez}, {Gangler}, {Hayden}, {Kim}, {Leget}, {Lombardo}, {Nordin}, {Pain},
  {Pecontal}, {Pereira}, {Perlmutter}, {Rabinowitz}, {Runge}, {Rubin},
  {Saunders}, {Smadja}, {Sofiatti}, {Suzuki}, {Tao}, and {Weaver}}]{Rig++15}
{Rigault} M, {Aldering} G, {Kowalski} M, {Copin} Y, {Antilogus} P, {Aragon} C,
  {Bailey} S, {Baltay} C, {Baugh} D, {Bongard} S, {Boone} K, {Buton} C, {Chen}
  J, {Chotard} N, {Fakhouri} HK, {Feindt} U, {Fagrelius} P, {Fleury} M,
  {Fouchez} D, {Gangler} E, {Hayden} B, {Kim} AG, {Leget} PF, {Lombardo} S,
  {Nordin} J, {Pain} R, {Pecontal} E, {Pereira} R, {Perlmutter} S, {Rabinowitz}
  D, {Runge} K, {Rubin} D, {Saunders} C, {Smadja} G, {Sofiatti} C, {Suzuki} N,
  {Tao} C, {Weaver} BA (2015) {Confirmation of a Star Formation Bias in Type Ia
  Supernova Distances and its Effect on the Measurement of the Hubble
  Constant}. \apj 802:20, \doi{10.1088/0004-637X/802/1/20}, \eprint{1412.6501}

\bibitem[{{Rodney} et~al(2016){Rodney}, {Strolger}, {Kelly}, {Brada{\v c}},
  {Brammer}, {Filippenko}, {Foley}, {Graur}, {Hjorth}, {Jha}, {McCully},
  {Molino}, {Riess}, {Schmidt}, {Selsing}, {Sharon}, {Treu}, {Weiner}, and
  {Zitrin}}]{Rod++16}
{Rodney} SA, {Strolger} LG, {Kelly} PL, {Brada{\v c}} M, {Brammer} G,
  {Filippenko} AV, {Foley} RJ, {Graur} O, {Hjorth} J, {Jha} SW, {McCully} C,
  {Molino} A, {Riess} AG, {Schmidt} KB, {Selsing} J, {Sharon} K, {Treu} T,
  {Weiner} BJ, {Zitrin} A (2016) {SN Refsdal: Photometry and Time Delay
  Measurements of the First Einstein Cross Supernova}. \apj 820:50,
  \doi{10.3847/0004-637X/820/1/50}, \eprint{1512.05734}

\bibitem[{{Rusu} et~al(2016){Rusu}, {Oguri}, {Minowa}, {Iye}, {Inada}, {Oya},
  {Kayo}, {Hayano}, {Hattori}, {Saito}, {Ito}, {Pyo}, {Terada}, {Takami}, and
  {Watanabe}}]{Rus++16}
{Rusu} CE, {Oguri} M, {Minowa} Y, {Iye} M, {Inada} N, {Oya} S, {Kayo} I,
  {Hayano} Y, {Hattori} M, {Saito} Y, {Ito} M, {Pyo} TS, {Terada} H, {Takami}
  H, {Watanabe} M (2016) {Subaru Telescope adaptive optics observations of
  gravitationally lensed quasars in the Sloan Digital Sky Survey}. \mnras
  458:2--55, \doi{10.1093/mnras/stw092}, \eprint{1506.05147}

\bibitem[{{Saha} et~al(2006){Saha}, {Coles}, {Macci{\`o}}, and
  {Williams}}]{Sah++06}
{Saha} P, {Coles} J, {Macci{\`o}} AV, {Williams} LLR (2006) {The Hubble Time
  Inferred from 10 Time Delay Lenses}. \apjl 650:L17--L20,
  \doi{10.1086/507583}, \eprint{astro-ph/0607240}

\bibitem[{{Schechter} et~al(1997){Schechter}, {Bailyn}, {Barr}, {Barvainis},
  {Becker}, {Bernstein}, {Blakeslee}, {Bus}, {Dressler}, {Falco}, {Fesen},
  {Fischer}, {Gebhardt}, {Harmer}, {Hewitt}, {Hjorth}, {Hurt}, {Jaunsen},
  {Mateo}, {Mehlert}, {Richstone}, {Sparke}, {Thorstensen}, {Tonry}, {Wegner},
  {Willmarth}, and {Worthey}}]{Sch++97}
{Schechter} PL, {Bailyn} CD, {Barr} R, {Barvainis} R, {Becker} CM, {Bernstein}
  GM, {Blakeslee} JP, {Bus} SJ, {Dressler} A, {Falco} EE, {Fesen} RA, {Fischer}
  P, {Gebhardt} K, {Harmer} D, {Hewitt} JN, {Hjorth} J, {Hurt} T, {Jaunsen} AO,
  {Mateo} M, {Mehlert} D, {Richstone} DO, {Sparke} LS, {Thorstensen} JR,
  {Tonry} JL, {Wegner} G, {Willmarth} DW, {Worthey} G (1997) {The Quadruple
  Gravitational Lens PG 1115+080: Time Delays and Models}. \apjl 475:L85--L88,
  \doi{10.1086/310478}, \eprint{astro-ph/9611051}

\bibitem[{{Schechter} et~al(2014){Schechter}, {Pooley}, {Blackburne}, and
  {Wambsganss}}]{Sch++14}
{Schechter} PL, {Pooley} D, {Blackburne} JA, {Wambsganss} J (2014) {A
  Calibration of the Stellar Mass Fundamental Plane at z \~{} 0.5 Using the
  Micro-lensing-induced Flux Ratio Anomalies of Macro-lensed Quasars}. \apj
  793:96, \doi{10.1088/0004-637X/793/2/96}, \eprint{1405.0038}

\bibitem[{{Schmidt} et~al(2012){Schmidt}, {Rix}, {Shields}, {Knecht}, {Hogg},
  {Maoz}, and {Bovy}}]{Sch++12}
{Schmidt} KB, {Rix} HW, {Shields} JC, {Knecht} M, {Hogg} DW, {Maoz} D, {Bovy} J
  (2012) {The Color Variability of Quasars}. \apj 744:147,
  \doi{10.1088/0004-637X/744/2/147}, \eprint{1109.6653}

\bibitem[{{Schneider}(1985)}]{Schneider1985}
{Schneider} P (1985) {A new formulation of gravitational lens theory,
  time-delay, and Fermat's principle}. \aap 143:413--420

\bibitem[{{Schneider}(2014)}]{Schneider2014}
{Schneider} P (2014) {Generalized multi-plane gravitational lensing: time
  delays, recursive lens equation, and the mass-sheet transformation}. ArXiv
  e-prints \eprint{1409.0015}

\bibitem[{{Schneider} and {Sluse}(2013)}]{S+S13}
{Schneider} P, {Sluse} D (2013) {Mass-sheet degeneracy, power-law models and
  external convergence: Impact on the determination of the Hubble constant from
  gravitational lensing}. \aap 559:A37, \doi{10.1051/0004-6361/201321882},
  \eprint{1306.0901}

\bibitem[{{Schneider} and {Sluse}(2014)}]{SPT}
{Schneider} P, {Sluse} D (2014) {Source-position transformation: an approximate
  invariance in strong gravitational lensing}. \aap 564:A103,
  \doi{10.1051/0004-6361/201322106}, \eprint{1306.4675}

\bibitem[{{Schneider} et~al(1992){Schneider}, {Ehlers}, and {Falco}}]{SEF92}
{Schneider} P, {Ehlers} J, {Falco} EE (1992) {Gravitational Lenses}.
  Springer-Verlag Berlin Heidelberg New York

\bibitem[{{Schneider} et~al(2006){Schneider}, {Kochanek}, and
  {Wambsganss}}]{SKW06}
{Schneider} P, {Kochanek} CS, {Wambsganss} J (2006) {Gravitational Lensing:
  Strong, Weak and Micro}. \doi{10.1007/978-3-540-30310-7}

\bibitem[{{Sharon} and {Johnson}(2015)}]{S+J15}
{Sharon} K, {Johnson} TL (2015) {Revised Lens Model for the Multiply Imaged
  Lensed Supernova "Refsdal" in MACS J1149+2223}. \apjl 800:L26,
  \doi{10.1088/2041-8205/800/2/L26}, \eprint{1411.6933}

\bibitem[{{Skidmore} et~al(2015){Skidmore}, {TMT International Science
  Development Teams}, and {Science Advisory Committee}}]{Ski++15}
{Skidmore} W, {TMT International Science Development Teams}, {Science Advisory
  Committee} T (2015) {Thirty Meter Telescope Detailed Science Case: 2015}.
  Research in Astronomy and Astrophysics 15:1945,
  \doi{10.1088/1674-4527/15/12/001}, \eprint{1505.01195}

\bibitem[{{Sonnenfeld} et~al(2011){Sonnenfeld}, {Bertin}, and
  {Lombardi}}]{SBL11}
{Sonnenfeld} A, {Bertin} G, {Lombardi} M (2011) {Direct measurement of the
  magnification produced by galaxy clusters as gravitational lenses}. \aap
  532:A37, \doi{10.1051/0004-6361/201016309}, \eprint{1106.1442}

\bibitem[{{Sonnenfeld} et~al(2015){Sonnenfeld}, {Treu}, {Marshall}, {Suyu},
  {Gavazzi}, {Auger}, and {Nipoti}}]{SonnenfeldEtal2015}
{Sonnenfeld} A, {Treu} T, {Marshall} PJ, {Suyu} SH, {Gavazzi} R, {Auger} MW,
  {Nipoti} C (2015) {The SL2S Galaxy-scale Lens Sample. V. Dark Matter Halos
  and Stellar IMF of Massive Early-type Galaxies Out to Redshift 0.8}. \apj
  800:94, \doi{10.1088/0004-637X/800/2/94}, \eprint{1410.1881}

\bibitem[{{Spergel} et~al(2015){Spergel}, {Flauger}, and {Hlo{\v z}ek}}]{SFH15}
{Spergel} DN, {Flauger} R, {Hlo{\v z}ek} R (2015) {Planck data reconsidered}.
  \prd 91(2):023518, \doi{10.1103/PhysRevD.91.023518}, \eprint{1312.3313}

\bibitem[{{Sun} et~al(2014){Sun}, {Wang}, {Chen}, and {Zheng}}]{SunEtal2014}
{Sun} YH, {Wang} JX, {Chen} XY, {Zheng} ZY (2014) {The Discovery of
  Timescale-dependent Color Variability of Quasars}. \apj 792:54,
  \doi{10.1088/0004-637X/792/1/54}, \eprint{1407.4230}

\bibitem[{{Suyu}(2012)}]{Suyu12}
{Suyu} SH (2012) {Cosmography from two-image lens systems: overcoming the lens
  profile slope degeneracy}. ArXiv e-prints \eprint{1202.0287}

\bibitem[{{Suyu} and {Halkola}(2010)}]{S+H10}
{Suyu} SH, {Halkola} A (2010) {The halos of satellite galaxies: the companion
  of the massive elliptical lens SL2S J08544-0121}. \aap 524:A94,
  \doi{10.1051/0004-6361/201015481}, \eprint{1007.4815}

\bibitem[{{Suyu} et~al(2006){Suyu}, {Marshall}, {Hobson}, and
  {Blandford}}]{Suy++06}
{Suyu} SH, {Marshall} PJ, {Hobson} MP, {Blandford} RD (2006) {A Bayesian
  analysis of regularized source inversions in gravitational lensing}. \mnras
  371:983--998, \doi{10.1111/j.1365-2966.2006.10733.x},
  \eprint{astro-ph/0601493}

\bibitem[{{Suyu} et~al(2009){Suyu}, {Marshall}, {Blandford}, {Fassnacht},
  {Koopmans}, {McKean}, and {Treu}}]{Suy++09}
{Suyu} SH, {Marshall} PJ, {Blandford} RD, {Fassnacht} CD, {Koopmans} LVE,
  {McKean} JP, {Treu} T (2009) {Dissecting the Gravitational Lens B1608+656. I.
  Lens Potential Reconstruction}. \apj 691:277--298,
  \doi{10.1088/0004-637X/691/1/277}, \eprint{0804.2827}

\bibitem[{{Suyu} et~al(2010){Suyu}, {Marshall}, {Auger}, {Hilbert},
  {Blandford}, {Koopmans}, {Fassnacht}, and {Treu}}]{Suy++10}
{Suyu} SH, {Marshall} PJ, {Auger} MW, {Hilbert} S, {Blandford} RD, {Koopmans}
  LVE, {Fassnacht} CD, {Treu} T (2010) {Dissecting the Gravitational lens
  B1608+656. II. Precision Measurements of the Hubble Constant, Spatial
  Curvature, and the Dark Energy Equation of State}. \apj 711:201--221,
  \doi{10.1088/0004-637X/711/1/201}, \eprint{0910.2773}

\bibitem[{{Suyu} et~al(2012){Suyu}, {Treu}, {Blandford}, {Freedman}, {Hilbert},
  {Blake}, {Braatz}, {Courbin}, {Dunkley}, {Greenhill}, {Humphreys}, {Jha},
  {Kirshner}, {Lo}, {Macri}, {Madore}, {Marshall}, {Meylan}, {Mould}, {Reid},
  {Reid}, {Riess}, {Schlegel}, {Scowcroft}, and {Verde}}]{Suy++12}
{Suyu} SH, {Treu} T, {Blandford} RD, {Freedman} WL, {Hilbert} S, {Blake} C,
  {Braatz} J, {Courbin} F, {Dunkley} J, {Greenhill} L, {Humphreys} E, {Jha} S,
  {Kirshner} R, {Lo} KY, {Macri} L, {Madore} BF, {Marshall} PJ, {Meylan} G,
  {Mould} J, {Reid} B, {Reid} M, {Riess} A, {Schlegel} D, {Scowcroft} V,
  {Verde} L (2012) {The Hubble constant and new discoveries in cosmology}.
  ArXiv e-prints \eprint{1202.4459}

\bibitem[{{Suyu} et~al(2013){Suyu}, {Auger}, {Hilbert}, {Marshall}, {Tewes},
  {Treu}, {Fassnacht}, {Koopmans}, {Sluse}, {Blandford}, {Courbin}, and
  {Meylan}}]{Suy++13}
{Suyu} SH, {Auger} MW, {Hilbert} S, {Marshall} PJ, {Tewes} M, {Treu} T,
  {Fassnacht} CD, {Koopmans} LVE, {Sluse} D, {Blandford} RD, {Courbin} F,
  {Meylan} G (2013) {Two Accurate Time-delay Distances from Strong Lensing:
  Implications for Cosmology}. \apj 766:70, \doi{10.1088/0004-637X/766/2/70}

\bibitem[{{Suyu} et~al(2014){Suyu}, {Treu}, {Hilbert}, {Sonnenfeld}, {Auger},
  {Blandford}, {Collett}, {Courbin}, {Fassnacht}, {Koopmans}, {Marshall},
  {Meylan}, {Spiniello}, and {Tewes}}]{Suy++14}
{Suyu} SH, {Treu} T, {Hilbert} S, {Sonnenfeld} A, {Auger} MW, {Blandford} RD,
  {Collett} T, {Courbin} F, {Fassnacht} CD, {Koopmans} LVE, {Marshall} PJ,
  {Meylan} G, {Spiniello} C, {Tewes} M (2014) {Cosmology from Gravitational
  Lens Time Delays and Planck Data}. \apjl 788:L35,
  \doi{10.1088/2041-8205/788/2/L35}, \eprint{1306.4732}

\bibitem[{{Tagore} and {Jackson}(2016)}]{TagoreAndJackson2016}
{Tagore} AS, {Jackson} N (2016) {On the use of shapelets in modelling resolved,
  gravitationally lensed images}. \mnras 457:3066--3075,
  \doi{10.1093/mnras/stw057}, \eprint{1505.00198}

\bibitem[{{Tak} et~al(2016){Tak}, {Mandel}, {van Dyk}, {Kashyap}, {Meng}, and
  {Siemiginowska}}]{TakEtal2016}
{Tak} H, {Mandel} K, {van Dyk} DA, {Kashyap} VL, {Meng} XL, {Siemiginowska} A
  (2016) {Bayesian Estimates of Astronomical Time Delays between
  Gravitationally Lensed Stochastic Light Curves}. ArXiv e-prints
  \eprint{1602.01462}

\bibitem[{{Tewes} et~al(2013{\natexlab{a}}){Tewes}, {Courbin}, and
  {Meylan}}]{TCM13}
{Tewes} M, {Courbin} F, {Meylan} G (2013{\natexlab{a}}) {COSMOGRAIL: the
  COSmological MOnitoring of GRAvItational Lenses. XI. Techniques for time
  delay measurement in presence of microlensing}. \aap 553:A120,
  \doi{10.1051/0004-6361/201220123}, \eprint{1208.5598}

\bibitem[{{Tewes} et~al(2013{\natexlab{b}}){Tewes}, {Courbin}, {Meylan},
  {Kochanek}, {Eulaers}, {Cantale}, {Mosquera}, {Magain}, {Van Winckel},
  {Sluse}, {Cataldi}, {V{\"o}r{\"o}s}, and {Dye}}]{Tew++13}
{Tewes} M, {Courbin} F, {Meylan} G, {Kochanek} CS, {Eulaers} E, {Cantale} N,
  {Mosquera} AM, {Magain} P, {Van Winckel} H, {Sluse} D, {Cataldi} G,
  {V{\"o}r{\"o}s} D, {Dye} S (2013{\natexlab{b}}) {COSMOGRAIL: the COSmological
  MOnitoring of GRAvItational Lenses. XIII. Time delays and 9-yr optical
  monitoring of the lensed quasar RX J1131-1231}. \aap 556:A22,
  \doi{10.1051/0004-6361/201220352}, \eprint{1208.6009}

\bibitem[{{The Dark Energy Survey Collaboration} et~al(2015){The Dark Energy
  Survey Collaboration}, {Abbott}, {Abdalla}, {Allam}, {Amara}, {Annis},
  {Armstrong}, {Bacon}, {Banerji}, {Bauer}, {Baxter}, {Becker},
  {Benoit-L{\'e}vy}, {Bernstein}, {Bernstein}, {Bertin}, {Blazek}, {Bonnett},
  {Bridle}, {Brooks}, {Bruderer}, {Buckley-Geer}, {Burke}, {Busha}, {Capozzi},
  {Carnero Rosell}, {Carrasco Kind}, {Carretero}, {Castander}, {Chang},
  {Clampitt}, {Crocce}, {Cunha}, {D'Andrea}, {da Costa}, {Das}, {DePoy},
  {Desai}, {Diehl}, {Dietrich}, {Dodelson}, {Doel}, {Drlica-Wagner},
  {Efstathiou}, {Eifler}, {Erickson}, {Estrada}, {Evrard}, {Fausti Neto},
  {Fernandez}, {Finley}, {Flaugher}, {Fosalba}, {Friedrich}, {Frieman},
  {Gangkofner}, {Garcia-Bellido}, {Gaztanaga}, {Gerdes}, {Gruen}, {Gruendl},
  {Gutierrez}, {Hartley}, {Hirsch}, {Honscheid}, {Huff}, {Jain}, {James},
  {Jarvis}, {Kacprzak}, {Kent}, {Kirk}, {Krause}, {Kravtsov}, {Kuehn},
  {Kuropatkin}, {Kwan}, {Lahav}, {Leistedt}, {Li}, {Lima}, {Lin}, {MacCrann},
  {March}, {Marshall}, {Martini}, {McMahon}, {Melchior}, {Miller}, {Miquel},
  {Mohr}, {Neilsen}, {Nichol}, {Nicola}, {Nord}, {Ogando}, {Palmese}, {Peiris},
  {Plazas}, {Refregier}, {Roe}, {Romer}, {Roodman}, {Rowe}, {Rykoff}, {Sabiu},
  {Sadeh}, {Sako}, {Samuroff}, {S{\'a}nchez}, {Sanchez}, {Seo},
  {Sevilla-Noarbe}, {Sheldon}, {Smith}, {Soares-Santos}, {Sobreira}, {Suchyta},
  {Swanson}, {Tarle}, {Thaler}, {Thomas}, {Troxel}, {Vikram}, {Walker},
  {Wechsler}, {Weller}, {Zhang}, and {Zuntz}}]{DESWL}
{The Dark Energy Survey Collaboration}, {Abbott} T, {Abdalla} FB, {Allam} S,
  {Amara} A, {Annis} J, {Armstrong} R, {Bacon} D, {Banerji} M, {Bauer} AH,
  {Baxter} E, {Becker} MR, {Benoit-L{\'e}vy} A, {Bernstein} RA, {Bernstein} GM,
  {Bertin} E, {Blazek} J, {Bonnett} C, {Bridle} SL, {Brooks} D, {Bruderer} C,
  {Buckley-Geer} E, {Burke} DL, {Busha} MT, {Capozzi} D, {Carnero Rosell} A,
  {Carrasco Kind} M, {Carretero} J, {Castander} FJ, {Chang} C, {Clampitt} J,
  {Crocce} M, {Cunha} CE, {D'Andrea} CB, {da Costa} LN, {Das} R, {DePoy} DL,
  {Desai} S, {Diehl} HT, {Dietrich} JP, {Dodelson} S, {Doel} P, {Drlica-Wagner}
  A, {Efstathiou} G, {Eifler} TF, {Erickson} B, {Estrada} J, {Evrard} AE,
  {Fausti Neto} A, {Fernandez} E, {Finley} DA, {Flaugher} B, {Fosalba} P,
  {Friedrich} O, {Frieman} J, {Gangkofner} C, {Garcia-Bellido} J, {Gaztanaga}
  E, {Gerdes} DW, {Gruen} D, {Gruendl} RA, {Gutierrez} G, {Hartley} W, {Hirsch}
  M, {Honscheid} K, {Huff} EM, {Jain} B, {James} DJ, {Jarvis} M, {Kacprzak} T,
  {Kent} S, {Kirk} D, {Krause} E, {Kravtsov} A, {Kuehn} K, {Kuropatkin} N,
  {Kwan} J, {Lahav} O, {Leistedt} B, {Li} TS, {Lima} M, {Lin} H, {MacCrann} N,
  {March} M, {Marshall} JL, {Martini} P, {McMahon} RG, {Melchior} P, {Miller}
  CJ, {Miquel} R, {Mohr} JJ, {Neilsen} E, {Nichol} RC, {Nicola} A, {Nord} B,
  {Ogando} R, {Palmese} A, {Peiris} HV, {Plazas} AA, {Refregier} A, {Roe} N,
  {Romer} AK, {Roodman} A, {Rowe} B, {Rykoff} ES, {Sabiu} C, {Sadeh} I, {Sako}
  M, {Samuroff} S, {S{\'a}nchez} C, {Sanchez} E, {Seo} H, {Sevilla-Noarbe} I,
  {Sheldon} E, {Smith} RC, {Soares-Santos} M, {Sobreira} F, {Suchyta} E,
  {Swanson} MEC, {Tarle} G, {Thaler} J, {Thomas} D, {Troxel} MA, {Vikram} V,
  {Walker} AR, {Wechsler} RH, {Weller} J, {Zhang} Y, {Zuntz} J (2015)
  {Cosmology from Cosmic Shear with DES Science Verification Data}. ArXiv
  e-prints \eprint{1507.05552}

\bibitem[{{Treu}(2010)}]{Tre10}
{Treu} T (2010) {Strong Lensing by Galaxies}. \araa 48:87--125,
  \doi{10.1146/annurev-astro-081309-130924}, \eprint{1003.5567}

\bibitem[{Treu and Ellis(2015)}]{T+E15}
Treu T, Ellis RS (2015) Gravitational lensing: Einstein’s unfinished
  symphony. Contemporary Physics 56(1):17--34,
  \doi{10.1080/00107514.2015.1006001},
  \urlprefix\url{http://www.tandfonline.com/doi/abs/10.1080/00107514.2015.1006001},
  \eprint{http://www.tandfonline.com/doi/pdf/10.1080/00107514.2015.1006001}

\bibitem[{{Treu} and {Koopmans}(2002{\natexlab{a}})}]{T+K02a}
{Treu} T, {Koopmans} LVE (2002{\natexlab{a}}) {The Internal Structure and
  Formation of Early-Type Galaxies: The Gravitational Lens System MG 2016+112
  at z = 1.004}. \apj 575:87--94, \doi{10.1086/341216},
  \eprint{astro-ph/0202342}

\bibitem[{{Treu} and {Koopmans}(2002{\natexlab{b}})}]{T+K02b}
{Treu} T, {Koopmans} LVE (2002{\natexlab{b}}) {The internal structure of the
  lens PG1115+080: breaking degeneracies in the value of the Hubble constant}.
  \mnras 337:L6--L10, \doi{10.1046/j.1365-8711.2002.06107.x},
  \eprint{astro-ph/0210002}

\bibitem[{{Treu} and {Koopmans}(2004)}]{T+K04}
{Treu} T, {Koopmans} LVE (2004) {Massive Dark Matter Halos and Evolution of
  Early-Type Galaxies to z \~{} 1}. \apj 611:739--760, \doi{10.1086/422245},
  \eprint{arXiv:astro-ph/0401373}

\bibitem[{{Treu} et~al(2012){Treu}, {Marshall}, and {Clowe}}]{TMC12}
{Treu} T, {Marshall} PJ, {Clowe} D (2012) {Resource Letter GL-1: Gravitational
  Lensing}. American Journal of Physics 80:753--763, \doi{10.1119/1.4726204},
  \eprint{1206.0791}

\bibitem[{{Treu} et~al(2013){Treu}, {Marshall}, {Cyr-Racine}, {Fassnacht},
  {Keeton}, {Linder}, {Moustakas}, {Bradac}, {Buckley-Geer}, {Collett},
  {Courbin}, {Dobler}, {Finley}, {Hjorth}, {Kochanek}, {Komatsu}, {Koopmans},
  {Meylan}, {Natarajan}, {Oguri}, {Suyu}, {Tewes}, {Wong}, {Zabludoff},
  {Zaritsky}, {Anguita}, {Brunner}, {Cabanac}, {Falco}, {Fritz}, {Seidel},
  {Howell}, {Giocoli}, {Jackson}, {Lopez}, {Metcalf}, {Motta}, and
  {Verdugo}}]{Tre++13}
{Treu} T, {Marshall} PJ, {Cyr-Racine} FY, {Fassnacht} CD, {Keeton} CR, {Linder}
  EV, {Moustakas} LA, {Bradac} M, {Buckley-Geer} E, {Collett} T, {Courbin} F,
  {Dobler} G, {Finley} DA, {Hjorth} J, {Kochanek} CS, {Komatsu} E, {Koopmans}
  LVE, {Meylan} G, {Natarajan} P, {Oguri} M, {Suyu} SH, {Tewes} M, {Wong} KC,
  {Zabludoff} AI, {Zaritsky} D, {Anguita} T, {Brunner} RJ, {Cabanac} R, {Falco}
  EE, {Fritz} A, {Seidel} G, {Howell} DA, {Giocoli} C, {Jackson} N, {Lopez} S,
  {Metcalf} RB, {Motta} V, {Verdugo} T (2013) {Dark energy with gravitational
  lens time delays}. ArXiv:13061272 \eprint{1306.1272}

\bibitem[{{Treu} et~al(2016){Treu}, {Brammer}, {Diego}, {Grillo}, {Kelly},
  {Oguri}, {Rodney}, {Rosati}, {Sharon}, {Zitrin}, {Balestra}, {Brada{\v c}},
  {Broadhurst}, {Caminha}, {Halkola}, {Hoag}, {Ishigaki}, {Johnson}, {Karman},
  {Kawamata}, {Mercurio}, {Schmidt}, {Strolger}, {Suyu}, {Filippenko}, {Foley},
  {Jha}, and {Patel}}]{Tre++16}
{Treu} T, {Brammer} G, {Diego} JM, {Grillo} C, {Kelly} PL, {Oguri} M, {Rodney}
  SA, {Rosati} P, {Sharon} K, {Zitrin} A, {Balestra} I, {Brada{\v c}} M,
  {Broadhurst} T, {Caminha} GB, {Halkola} A, {Hoag} A, {Ishigaki} M, {Johnson}
  TL, {Karman} W, {Kawamata} R, {Mercurio} A, {Schmidt} KB, {Strolger} LG,
  {Suyu} SH, {Filippenko} AV, {Foley} RJ, {Jha} SW, {Patel} B (2016)
  {''Refsdal'' Meets Popper: Comparing Predictions of the Re-appearance of the
  Multiply Imaged Supernova Behind MACSJ1149.5+2223}. \apj 817:60,
  \doi{10.3847/0004-637X/817/1/60}, \eprint{1510.05750}

\bibitem[{{Vanderriest} et~al(1982){Vanderriest}, {Felenbok}, {Schneider},
  {Wlerick}, {Bijaoui}, and {Lelievre}}]{Van82}
{Vanderriest} C, {Felenbok} P, {Schneider} J, {Wlerick} G, {Bijaoui} A,
  {Lelievre} G (1982) {The photometry of 0957 plus 561 - Detection of
  short-period variations}. \aap 110:L11--L14

\bibitem[{{Vanderriest} et~al(1989){Vanderriest}, {Schneider}, {Herpe},
  {Chevreton}, {Moles}, and {Wlerick}}]{Van89}
{Vanderriest} C, {Schneider} J, {Herpe} G, {Chevreton} M, {Moles} M, {Wlerick}
  G (1989) {The value of the time delay Delta t(A, B) for the 'double' quasar
  0957+561 from optical photometric monitoring}. \aap 215:1--13

\bibitem[{{Vegetti} and {Koopmans}(2009)}]{V+K09}
{Vegetti} S, {Koopmans} LVE (2009) {Statistics of CDM substructure from strong
  gravitational lensing: quantifying the mass fraction and mass function}.
  \mnras, in press \eprint{0903.4752}

\bibitem[{{Vegetti} et~al(2014){Vegetti}, {Koopmans}, {Auger}, {Treu}, and
  {Bolton}}]{Veg++14}
{Vegetti} S, {Koopmans} LVE, {Auger} MW, {Treu} T, {Bolton} AS (2014)
  {Inference of the cold dark matter substructure mass function at z = 0.2
  using strong gravitational lenses}. \mnras 442:2017--2035,
  \doi{10.1093/mnras/stu943}, \eprint{1405.3666}

\bibitem[{{Vuissoz} et~al(2007){Vuissoz}, {Courbin}, {Sluse}, {Meylan},
  {Ibrahimov}, {Asfandiyarov}, {Stoops}, {Eigenbrod}, {Le Guillou}, {van
  Winckel}, and {Magain}}]{Vui++07}
{Vuissoz} C, {Courbin} F, {Sluse} D, {Meylan} G, {Ibrahimov} M, {Asfandiyarov}
  I, {Stoops} E, {Eigenbrod} A, {Le Guillou} L, {van Winckel} H, {Magain} P
  (2007) {COSMOGRAIL: the COSmological MOnitoring of GRAvItational Lenses. V.
  The time delay in SDSS J1650+4251}. \aap 464:845--851,
  \doi{10.1051/0004-6361:20065823}, \eprint{astro-ph/0606317}

\bibitem[{{Vuissoz} et~al(2008){Vuissoz}, {Courbin}, {Sluse}, {Meylan},
  {Chantry}, {Eulaers}, {Morgan}, {Eyler}, {Kochanek}, {Coles}, {Saha},
  {Magain}, and {Falco}}]{Vui++08}
{Vuissoz} C, {Courbin} F, {Sluse} D, {Meylan} G, {Chantry} V, {Eulaers} E,
  {Morgan} C, {Eyler} ME, {Kochanek} CS, {Coles} J, {Saha} P, {Magain} P,
  {Falco} EE (2008) {COSMOGRAIL: the COSmological MOnitoring of GRAvItational
  Lenses. VII. Time delays and the Hubble constant from WFI J2033-4723}. \aap
  488:481--490, \doi{10.1051/0004-6361:200809866}, \eprint{0803.4015}

\bibitem[{{Walsh} et~al(1979){Walsh}, {Carswell}, and {Weymann}}]{WCW79}
{Walsh} D, {Carswell} RF, {Weymann} RJ (1979) {0957 + 561 A, B - Twin
  quasistellar objects or gravitational lens}. \nat 279:381--384,
  \doi{10.1038/279381a0}

\bibitem[{{Warren} and {Dye}(2003)}]{W+D03}
{Warren} SJ, {Dye} S (2003) {Semilinear Gravitational Lens Inversion}. \apj
  590:673--682, \doi{10.1086/375132}, \eprint{astro-ph/0302587}

\bibitem[{{Weinberg} et~al(2013){Weinberg}, {Mortonson}, {Eisenstein},
  {Hirata}, {Riess}, and {Rozo}}]{Wei++13}
{Weinberg} DH, {Mortonson} MJ, {Eisenstein} DJ, {Hirata} C, {Riess} AG, {Rozo}
  E (2013) {Observational probes of cosmic acceleration}. \physrep 530:87--255,
  \doi{10.1016/j.physrep.2013.05.001}, \eprint{1201.2434}

\bibitem[{{Wong} et~al(2011){Wong}, {Keeton}, {Williams}, {Momcheva}, and
  {Zabludoff}}]{Won++11}
{Wong} KC, {Keeton} CR, {Williams} KA, {Momcheva} IG, {Zabludoff} AI (2011)
  {The Effect of Environment on Shear in Strong Gravitational Lenses}. \apj
  726:84, \doi{10.1088/0004-637X/726/2/84}, \eprint{1011.2504}

\bibitem[{{Wucknitz}(2002)}]{Wuc02}
{Wucknitz} O (2002) {Degeneracies and scaling relations in general power-law
  models for gravitational lenses}. \mnras 332:951--961,
  \doi{10.1046/j.1365-8711.2002.05426.x}, \eprint{arXiv:astro-ph/0202376}

\bibitem[{{Wucknitz} et~al(2004){Wucknitz}, {Biggs}, and {Browne}}]{WBB04}
{Wucknitz} O, {Biggs} AD, {Browne} IWA (2004) {Models for the lens and source
  of B0218+357: a LENSCLEAN approach to determine $H_0$}. \mnras 349:14--30,
  \doi{10.1111/j.1365-2966.2004.07514.x}, \eprint{arXiv:astro-ph/0312263}

\bibitem[{{Xu} et~al(2016){Xu}, {Sluse}, {Schneider}, {Springel},
  {Vogelsberger}, {Nelson}, and {Hernquist}}]{XuEtal2016}
{Xu} D, {Sluse} D, {Schneider} P, {Springel} V, {Vogelsberger} M, {Nelson} D,
  {Hernquist} L (2016) {Lens galaxies in the Illustris simulation: power-law
  models and the bias of the Hubble constant from time delays}. \mnras
  456:739--755, \doi{10.1093/mnras/stv2708}, \eprint{1507.07937}

\bibitem[{{Xu} et~al(2009){Xu}, {Mao}, {Wang}, {Springel}, {Gao}, {White},
  {Frenk}, {Jenkins}, {Li}, and {Navarro}}]{Xu++09}
{Xu} DD, {Mao} S, {Wang} J, {Springel} V, {Gao} L, {White} SDM, {Frenk} CS,
  {Jenkins} A, {Li} G, {Navarro} JF (2009) {Effects of dark matter
  substructures on gravitational lensing: results from the Aquarius
  simulations}. \mnras pp 1108--+, \doi{10.1111/j.1365-2966.2009.15230.x},
  \eprint{0903.4559}

\end{thebibliography}



\end{document}